\documentclass[12pt]{article} 
\usepackage[hyperfootnotes=false]{hyperref}
     \usepackage{amsmath}
      \usepackage{amssymb}
  \usepackage{graphicx}
  \usepackage{xcolor}
  \newlength{\abstractwidth}
 \usepackage{mciteplus} 
\usepackage{subcaption}
 \usepackage{bbold}

  \newcommand{\be}{\begin{equation}}
  \newcommand{\bea}{\begin{eqnarray}}
  \newcommand{\eea}{\end{eqnarray}}
  \newcommand{\beq}{\begin{equation}}
  \newcommand{\ee}{\end{equation}}
  \newcommand{\eeq}{\end{equation}}

  \newcommand{\half}{{1\over 2}}

\def\la{\label}

\def\32{{3 \over 2 } }
\def\sign{ {\rm sgn} }

  \def\ba{\begin{eqnarray}}
  \def\ea{\end{eqnarray}}

 \def\simleq{\; \raise0.3ex\hbox{$<$\kern-0.75em
      \raise-1.1ex\hbox{$\sim$}}\; }
 \def\simgeq{\; \raise0.3ex\hbox{$>$\kern-0.75em
      \raise-1.1ex\hbox{$\sim$}}\; }

\def\nref#1{(\ref{#1})}

\def\ket#1{\left|#1\right\rangle}

\begin{document}

\begin{titlepage}
  \bigskip

  \bigskip\bigskip

  \bigskip

\begin{center}
 
\centerline
{\Large \bf {Eternal traversable wormhole }}
 \bigskip

 \bigskip
{\Large \bf { }} 
    \bigskip
\bigskip
\end{center}

  \begin{center}

 \bf {Juan Maldacena$^1$  and Xiao-Liang Qi$^{2,1}$   }
  \bigskip \rm
  
\bigskip
 $^1$Institute for Advanced Study,  Princeton, NJ 08540, USA  

 \rm 
 \bigskip
   $^2$Stanford University,  CA 94305, USA\\
\rm
 \bigskip

  \bigskip \rm
\bigskip
 
\rm

\bigskip
\bigskip

  \end{center}

 \bigskip\bigskip
  \begin{abstract}

  We construct a nearly-$AdS_2$ solution describing an eternal traversable wormhole. The solution contains negative null energy generated by quantum fields under the influence of an external coupling between the two boundaries. In parallel,  we discuss  two SYK systems coupled by a relevant interaction. The physics of the two cases is very similar. They both share a ``gravitational'' subsector which is identical.  The solution within this subsector sets the stage for dynamics which is almost conformal invariant.  We study this system in detail, both in gravity and in the SYK model. The coupled SYK models have an interesting phase diagram at finite temperature, displaying the usual Hawking-Page transition between the thermal AdS phase at low temperature and the black hole phase at high temperature. Interestingly, these two phases are continuously connected in the microcannonical ensemble.  

 \medskip
  \noindent
  \end{abstract}
\bigskip \bigskip \bigskip

\vspace{1cm}

\begin{center}
{\it Dedicated to the memory of Joe Polchinski,  \\ an outstanding colleague in many dimensions. } 
\end{center}

\vspace{2cm}

  \end{titlepage}

   \tableofcontents


\section{Introduction and motivation}

 $AdS_2$ is a very simple two dimensional spacetime that has two asymptotic boundaries. 
  Furthermore, $AdS_2$, viewed as a global spacetime, has two boundaries that are causally connected. 
 We can send a signal from one boundary to the other. It behaves like a  traversable wormhole! See figure \ref{Coordinates}(a). 
 
  In this paper we consider physical situations where a spacetime very similar  to $AdS_2$ arises, in the framework of nearly-$AdS_2$ gravity 
  \cite{Almheiri:2014cka,Jensen:2016pah,Maldacena:2016upp,Engelsoy:2016xyb}.
  It involves a solution that balances classical and quantum effects. In order to get the computation under control one needs a large number
  of quantum fields. These quantum fields are in a state with negative null energy because we set up an interaction between the two boundaries. 
  This is an interaction that looks non-local in the bulk, but could arise locally in a higher dimensional ambient space. 
  This is like the interaction that makes wormholes traversables \cite{Gao:2016bin}. In this case, we get a traversable wormhole that is static and time independent, 
  an eternal traversable wormhole. 
   
  We also analyze a closely related problem in the SYK model \cite{Sachdev:1992fk,KitaevTalks,Kitaev:2017awl}.
   We consider two identical SYK models coupled by a simple bilinear term. 
  When the coupling is small, the low energy physics of the model is nearly conformal and its features can be described by the reparametrization mode 
  described in \cite{KitaevTalks,Kitaev:2017awl,Maldacena:2016hyu,Jevicki:2016ito}.  In fact, the SYK model and the nearly-$AdS_2$ spacetime share a common subsector that is associated to gravitational physics. For this common subsector we can perform an analysis that is valid for both cases. 
  This means that the basic mechanism that renders the wormhole traversable is common to both, and relies on the physics of the emergent spontaneously and explicitly broken time reparametrization symmetry. 
  This common description is valid when the interaction between the two systems is relatively small, so that its effects become significant at the energy scales where the SYK model is nearly scale invariant. It is expected to be valid for any quantum mechanical system with an emergent approximate conformal symmetry in the IR. 
  This version of coupled SYK model is also interesting purely from the quantum mechanical point of view. We get a gapped system, which in itself is not surprising, but we get an excitation spectrum that is largely controlled by an SL(2) symmetry that is broken in a controlled way. The net result is that one can predict 
  the spectrum of excitations of  the model. The energy levels are set by the dimensions of operators of a single SYK model. 
   A similar feature is present in higher dimensional CFT$_d$s where the spectrum of operators of the theory is related to the spectrum of energies of the 
   CFT$_d$ on an $S^{d-1}\times$(time). In this case we have a nearly-CFT$_1$ and an $S^0$ consists of two points, which are the two copies of SYK. 
   So, our construction can be a way to realize a form of the state/operator map in a nearly-CFT$_1$. Of course, here 
   we need to add an extra interaction between the two sides. 
   So there are some similarities and differences with the higher dimensional situation. In this paper we spell them out in detail.
    {It should be noted that various models with interplay of SYK interaction and bilinear terms have been studied in the literature, such as\cite{banerjee2017solvable,chen2017competition,Azeyanagi:2017drg,song2017strongly,chowdhury2018translationally}. Gravity solutions describing weakly
    interacting CFTs were discussed in \cite{Bachas:2017rch}.}

  In the SYK model, we can also consider stronger couplings, which we can also analyze by solving the large $N$ Schwinger-Dyson equations. 
  These equations can be solved numerically. They can also be solved analytically in the large $q$ limit ($q$ is defined in \nref{SingleSYK}). 
  
  In writing this paper we have tried to separate a bit the gravity and the SYK discussions so that it can be read also by readers who are not familiar with one or the other. 
  So the reader should feel free to skip some sections. 
  
  In section \ref{SecAdST}, we review $AdS_2$ and nearly-$AdS_2$ gravity.  We explain the setup that leads to a solution where we preserve the global time translation symmetry of $AdS_2$. We also describe the effective action that describes gravitational effects in this theory. 
  
  In section \ref{SYKSection} we review a single SYK model and describe the system consisting of  two coupled SYK models.
  We describe how the reparametrization mode encodes some important aspects of the physics. This action has the same form as the one describing the 
  gravitational modes of nearly-$AdS_2$ gravity. 
  
   In section \ref{LowSec} we analyze the common effective action that appeared in the previous two sections. 
   We find the equations that determine the size of the energy gap of the system and the overall scale of the spectrum. 
   We also discuss some aspects of the quantization of this action. We describe how the approximate SL(2) symmetry is realized on the spectrum. 
     We also connect the ground state of this coupled system to the thermofield double state of the two decoupled systems. It turns out
     that both states are very close to each other and we quantify how close they are. 
     
     In section \ref{SYKAny} we return to the analysis of the two coupled SYK models, but now beyond the low energy limit. This is done numerically for $q=4$ 
      and analytically 
     in the large $q$ limit. We describe some aspects of the thermodynamics of the coupled model. 
    In the large $q$ limit we also show  that the ground state of the coupled model is equal  to the thermofield double state of the decoupled model for a specific
    temperature. Finally we also discuss the solution in the case that the microscopic couplings of the two SYK models are not exactly equal.

            \section{ Nearly $AdS_2$ gravity with a global time isometry }
  
  \la{SecAdST}
  
\subsection{Global $AdS_2$ } 

\begin{figure}[h]
\begin{center}
\includegraphics[scale=.5]{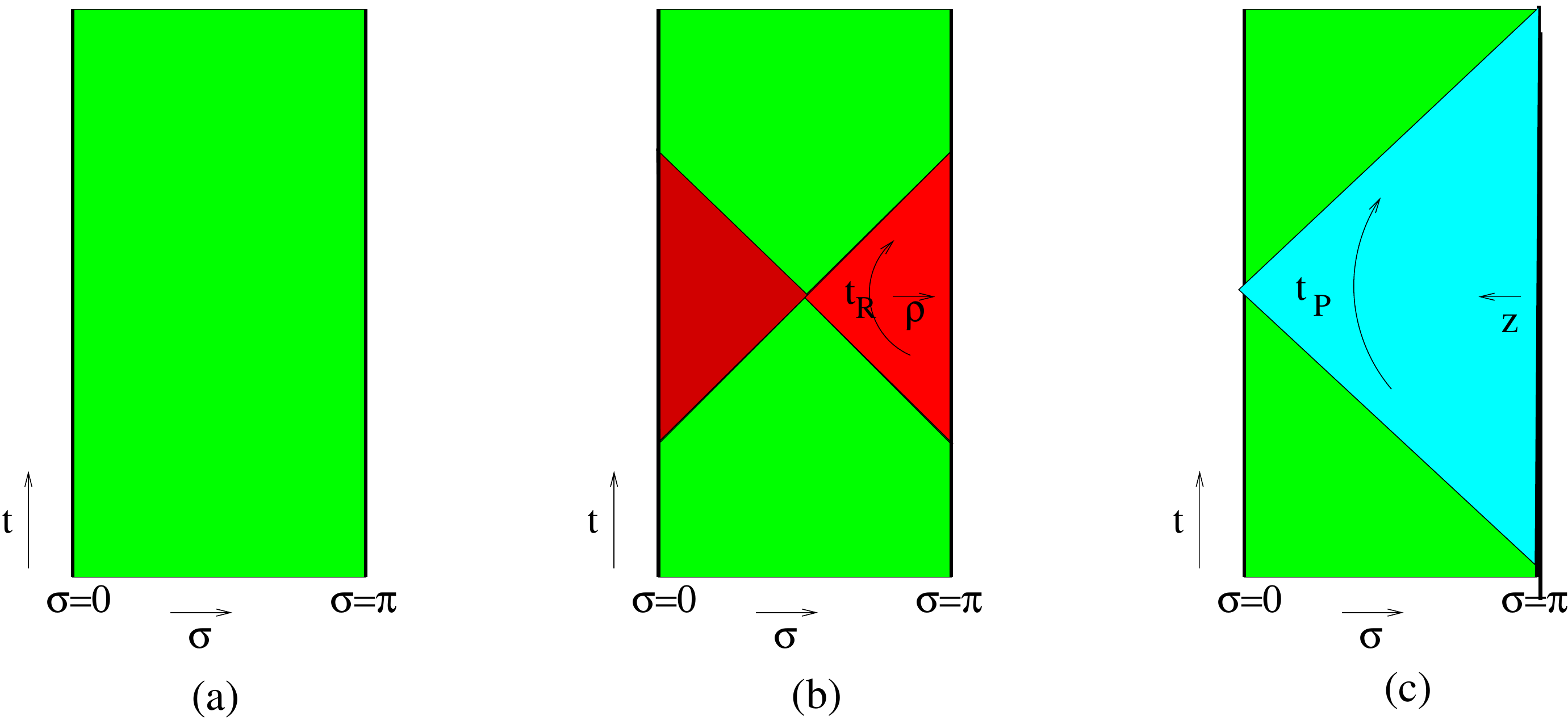} 
\caption{  (a) Full $AdS_2$ Penrose diagram. It has two boundaries, one at $\sigma =0$ and one at $\sigma = \pi$, see 
\nref{GlobCoord}. (b) The right triangle is covered by the Rindler or Thermal coordinates in \nref{PoinRind}. (c) The full triangle is covered by the Poincare coordinates \nref{PoinRind}.     }
\label{Coordinates}
\end{center}
\end{figure}

We start by recalling a few properties of $AdS_2$ and setting some notation. 
Two dimensional anti-de-Sitter space can be written in terms of global coordinates that cover the whole space-time 
\be \la{GlobCoord}
ds^2 = { - dt^2 + d\sigma^2 \over \sin^2 \sigma } ~,~~~~~~\sigma \in [ 0 , \pi ] 
\ee
Other coordinate systems that are also popular are 
\be  \la{PoinRind}
{\rm Poincare:} ~~ds^2 = { - dt_P + dz^2 \over z^2 } ~,~~~~~~~{\rm Rindler/Thermal: } ~~ds^2 = { - dt_R^2 \sinh^2 \rho + d\rho^2 } 
\ee 
These coordinate systems do not cover the full spacetime, see figure \ref{Coordinates}. 
The full spacetime has an $SL(2,R)$ group of isometries, but the above coordinate systems manifest only one of them, a different generator for each of the coordinate systems. This generator corresponds to the time translation symmetry for each of the choices of the time coordinate.

We can also describe $AdS_2$  in terms of global coordinates $Y^M$ with the constraint 
$ -(Y^{-1}) ^2 - (Y^0)^2 + (Y^1)^2 =-1$ (or, more precisely, its  universal cover). It is also possible to describe the boundary in terms
of   projective coordinates $X^M$ with the constraint $X.X =0$ and the identification $X^M \sim \lambda X^M$. We can introduce the boundary analogs of the 
previous time coordinates  via 
\bea  \la{BdyTimes}
 & ~&e^{ i t_r}  = X^{-1} + i X^0 ~,~~~~{\rm for } ~~~X^1 =1 ~,~~~~~e^{ i t_l} = X^{-1} + i X^0 ~,~~~~{\rm for } ~~~ X^1=-1
 \cr
 & ~&   { X^0 \over X^{-1} + X^1 } = t_P ~,~~~~~~ ~,~~~~~ e^{ t_R} = X^{1} + X^0 ~,~~~ {\rm for} ~~X^{-1}=1
  \eea
  Here $t_l$, $t_r$ are the global time coordinates along each of the two boundaries, which are selected via the ``gauge'' choice $X^1 = \pm 1$ respectively. 
  These equations    enable us to find the relations between these times, for example, 
  \be \la{RepTwo}
   { X^0 \over X^{-1} + X^1 } = t_P = \tan{ t_r \over 2 } = - { 1 \over \tan{ t_l \over 2 } }  = \tanh { t_R \over 2 } 
   \ee
   These relations also arise by relating the bulk coordinates in \nref{GlobCoord} \nref{PoinRind} and then moving close to the boundary.

   \subsection{Nearly $AdS_2$ gravity } 
   
   Purely $AdS_2$ asymptotic boundary conditions are not consistent in a theory of quantum gravity with finite energy excitations
   \cite{Maldacena:1998uz}  (see \cite{Galloway:2018dak} for a recent related discussion). 
   The next best possibility is to consider Nearly-$AdS_2$ boundary conditions \cite{Almheiri:2014cka,Maldacena:2016upp,Jensen:2016pah,Engelsoy:2016xyb}, 
   which are described by Jackiw-Teitelboim gravity 
  \cite{Jackiw:1984je,Teitelboim:1983ux} 
  \be \la{ActJT}
   S =  { \phi_0 \over 2 }  \left[  \int R  + 2 \int_{\rm Bdy} K \right] +   { \half } \left[  \int  \phi ( R + 2 ) + 2 \phi_b \int_{\rm Bdy} K \right] 
   + S_{\rm matter}[\chi, g] 
   \ee
   with boundary conditions that fix the boundary value of the metric and the dilaton, 
   \be \la{BCon}
   ds|_{\rm Bdy}  = {  d u \over \epsilon} ~,~~~~~ \phi|_{\rm Bdy} = \phi_b = { \phi_r \over \epsilon}
   \ee  and taking $\epsilon $ to zero,
   see \cite{Maldacena:2016upp} for more details.  
    The term proportional to  $\phi_0$ is a topological term that will not contribute to the main solutions we will describe here, 
   which have the topology of a strip or a cylinder, so we will mostly ignore it. $\chi$ denotes the matter fields, and we assumed that $\phi$ does not appear in the matter action. This action is a good approximation to the dynamics of nearly extremal black holes \cite{Almheiri:2014cka,Nayak:2018qej,Kolekar:2018sba}.
  
    In this theory, the only gravitational mode can be viewed as living at the boundary of the space. 
   Since the metric is set to be exactly $AdS_2$, the gravitational dynamics comes purely from the location of the physical boundary in that rigid $AdS_2$ space. 
   There are a couple of different ways to think about this location. One is to first solve the dynamics of matter,  then compute the dilaton from the 
   (metric) equation of motion in \nref{ActJT}, and finally
    find the location where the dilaton has the desired boundary value \cite{Almheiri:2014cka}.
    Alternatively, we can reduce the problem 
   explicitly to a dynamical system at the boundary. There are different parametrizations for this dynamical system. A  
    conventient one involves picking   the basic dynamic variable as a time reparametrization with an action which is the Schwarzian \cite{Maldacena:2016upp,kitaevIASChaosWorkshop}
   \be \la{SchAct}
   S =   -\phi_r \int \{ t_P(u),u \} du 
   \ee
   where $u$ is the boundary time. Here we can view $t_P(u)$ as a map between the physical proper time, or boundary time, and the interior Poincare time $t_P$. 
   We can consider a spacetime with two boundaries, and then we get \nref{SchAct} for each of the two boundaries. 
   When we consider this problem, we have an   $SL(2)$ symmetry acting as $t_P \to {( a t_P + b ) /( c t_P + d )} $. This symmetry should be treated as a {\it gauge} 
   symmetry. This means  we impose that the associated charges vanish as a gauge constraint\footnote{This $SL(2)$ symmetry, which is not broken and is not physical, should not be confused with 
   the physical $SL(2)$ operation on $u$ which is explicitly broken by \nref{SchAct} to simply $u$-translations.} \cite{Maldacena:2016upp}. 
We see that $\phi_r$, which is a quantity of dimensions of length, defined via \nref{BCon}, sets the strength of the coupling for the gravitational action 
\nref{SchAct}.
   
\begin{figure}[h]
\begin{center}
\includegraphics[scale=.5]{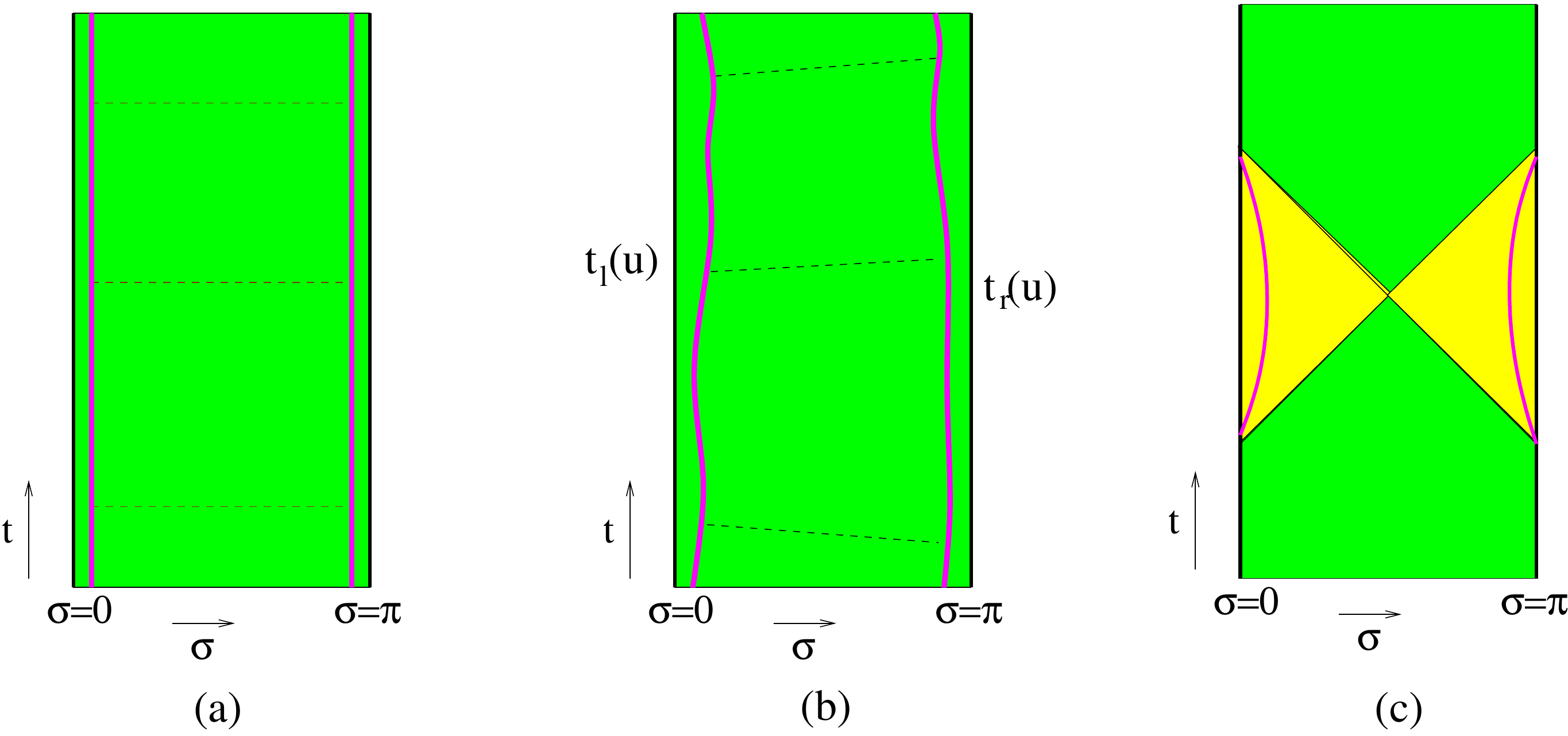} 
\caption{  (a)  Trajectories of the physical boundaries (in magenta) 
for the Nearly-$AdS_2$ geometry with a global time isometry. These trajectories are
 the lines where the dilaton acquires its boundary value. 
 It can be obtained by introducing an interaction between the two boundaries.  (b) We can describe the fluctuations of the boundary trajectories
  in terms of a pair of functions,  $t_l(u)$ and $t_r(u)$, mapping (rescaled) proper time $u$  along the trajectory to the  global $AdS_2$ time coordinate $t$. The 
  dotted lines can be viewed as insertions of the interaction Hamiltonian. They join points with the same value of $u$ on both boundaries.  (c)
 The physical boundaries for a Nearly-$AdS_2$ geometry with thermal isometry. Here the two boundary trajectories cover only a finite range of global time and we cannot send a signal between the two trajectories. }
\label{BdyTrajectories}
\end{center}
\end{figure}

         A simple Lorentzian solution of \nref{ActJT} is 
   \be \la{TFDSol}
   t_P = \tanh{ t_R(u) \over 2 }= \tanh{   \pi u \over \beta }  ~,~~~~~{\rm or}~~~~t_R(u) = { 2 \pi \over \beta } u 
   \ee
    which can be interpreted as a thermal black hole configuration with free energy, energy and entropy given by \cite{Maldacena:2016upp}
    \be \la{EnThr}
    F = - { \phi_r \over 2 }{  ( 2 \pi)^2  \over \beta^2} ~,~~~~ E = - \phi_r \{ t_P, u \}=  { \phi_r \over 2 }{  ( 2 \pi)^2  \over \beta^2}~,~~~~
    S = { \phi_r (2 \pi)^2 \over \beta } 
    \ee
      The full Lorentzian solution has two boundaries and the total gauge SL(2) charges vanish. 
  For this solution, the two boundary trajectories correspond to lines of constant $\rho$ in the Rindler/Thermal coordinates \nref{PoinRind}, and we cannot send signals between the boundaries. See figure \nref{BdyTrajectories}(c).

   In this paper we are interested in  creating a situation where the boundaries correspond to lines of constant $\sigma$ in the 
 global coordinates \nref{GlobCoord}. 
  In this configuration we would have a   $t$-translation invariant dilaton which grows towards both boundaries. 
   There is no solution of this kind  in the pure JT gravity with no matter \nref{ActJT}.
    Furthermore, there is no solution of this kind if we assume that we have matter which obeys the integrated null energy condition in the bulk. 
  In fact,  one of the equations of motion for the metric\footnote{\nref{NegEn} is the two dimensional version of the Raychaudhuri equation.}
   implies ($x^\pm = t\pm \sigma$) \cite{Maldacena:1998uz}
   \be \la{NegEn}
- \partial_+    (\sin^2 \sigma   \partial_+ \phi ) =   T_{++} \sin^2\sigma  \longrightarrow    - 2 \phi_r =- \left.  (\sin^2\sigma \partial_+\phi)\right|_{-\infty}^{+\infty} = 
  \int_{-\infty}^{\infty} dX^+ T_{X^+ X^+}   
  \ee
  where we have integrated the left expression 
 along a null line, with $X^+$ being   the affine coordinate along the null line, namely $dX^+ = dx^+/\sin^2\sigma$. 
    We see that the left hand side is  negative if $\phi $ is growing towards both boundaries as ${ \phi_r \over 
    |\sigma-\sigma_{\rm bdy}|}$. 
    On the other hand the right hand side is 
    non-negative if the integrated  null energy condition holds. 
    This problem is a special case of the general 
    topological censorship result \cite{Galloway:1999bp,Galloway:1999br} that forbids traversable wormholes and non-trivial topology in asymptotically flat or $AdS_D$, $D>2$,
     spaces (assuming the integrated null energy condition)\footnote{ One could imagine the following method for violating the null energy condition. First we consider a bulk 
    CFT in $AdS_2$. Since it is  a CFT, we get the same results as if we had it on a flat strip. On a flat  strip,  the energy is negative an equal to 
    $-c/24$ \cite{Bloete:1986qm}.  Nevertheless, when we go back to $AdS_2$ we have a contribution from the stress tensor from the conformal anomaly that cancels this out, so that we indeed have zero null energy in $AdS_2$. This has to be the case since the $AdS_2$ stress tensor should be SL(2) invariant. See appendix \ref{GenBulkCFT} for more discussion.}. 
   
   This result can be avoided by introducing an interaction that directly couples the two boundaries \cite{Gao:2016bin}. Therefore we now 
  add a boundary interaction of the form 
   \be \la{OOInt}
   S_{int} =
    g \sum_{i=1}^N \int du O^i_L(u) O^i_R(u) 
   \ee
   where $O^i$ are a set of $N$  operators with dimension $\Delta$. We consider a theory with $N$ operators because we will take $N$ to be large so as 
   to enhance the effects of these operators. $g$ has dimensions of [energy]$^{2\Delta -1}$. The operators are the field operators in the bulk evaluated at the positions of the boundary. So our bulk theory has at least $N$ matter fields. 
   This is an interaction across the two boundaries, similar to the one considered in \cite{Gao:2016bin}. 
   For the reasons explained in \cite{Gao:2016bin},  this interaction produces 
   negative energy in the bulk, see also \cite{Susskind:2014yaa,Maldacena:2017axo,Susskind:2017nto,Kourkoulou:2017zaj,Almheiri:2018ijj}. 
   Notice that \nref{NegEn} implies that a finite amount of negative energy is enough to resolve the problem.
In appendix \ref{FBF} we compute this negative energy explicitly for the case of a massless free fermion in the bulk, and in  \ref{DilProf} we compute the profile of the dilaton
with this negative energy source. 
Here we will analyze the same problem using the effective action for the boundary graviton. 

   When $g$ is sufficiently small, we can approximate the effects of the interaction \nref{OOInt} by replacing 
      \be \la{Expec}
       \langle e^{ i  g  \sum_i \int   du   O^i_L(u) O^i_R(u) } \rangle \sim e^{ i g  \sum_i \int dt  \langle  O^i_L(u) O^i_R(u) \rangle }   
       \ee
     in the path integral.  This amounts to resumming a series of ladder type diagrams of the form indicated in figure \ref{BdyTrajectories}(a). 
     These diagrams dominate in the large $N$ small $g$ limit, with $Ng$ kept fixed. 
     We can further couple  this to the gravity modes by performing a reparametrization  of the left and right times. 
  These   are a map between the proper boundary time $u$ and the global  times $t_l(u)$, $t_r(u)$ at the two boundaries \nref{BdyTimes}. 
   The system is then described  by the  effective action  
   \be
   \la{ActGr} 
   S = \int d u \left[    -\phi_r  \left\{  \tan{ t_l(u) \over 2} ,u \right\}    -\phi_r\left \{  \tan{ t_r(u) \over 2} ,u \right\}  +
  { g N \over 2^{ 2 \Delta } }   \left( {t'_l(u) t'_r(u) \over \cos^2 { t_l(u) -t_r(u) \over 2 } 
   } \right)^{  \Delta } \right] 
   \ee
   where we normalized the 
    correlator for $O$ so that it takes the form    $ \langle O(t^1_P) O(t^2_P)\rangle =   |t_P^1-t^2_P|^{-2\Delta}$ in Poincare coordinates.  
     We have reparametrized $t_P$ 
   by $t_l$ and $t_r$ indicated in \nref{RepTwo}. This correlator in terms of $t_l$ and $t_r$ has the form we expect for an $AdS_2$ correlator in global coordinates 
   where $t_l$ and $t_r$ are the boundary global times at the left and right boundaries respectively. 
   This is a way to describe the position of the physical boundary, see figure \ref{BdyTrajectories}(b).
       
   We postpone the analysis of this action until we obtain the same action from two coupled copies of the SYK model in the next section.     
       
    \section{ Two coupled SYK models } 
    \la{SYKSection}
    
\subsection{Review of the SYK model } 

The SYK model has a Hilbert space generated by $N$ Majorana fermions $\psi^i$ with a Hamiltonian of the form \cite{Sachdev:1992fk,KitaevTalks}
\bea \la{SingleSYK}
H &=& (i) ^{ q/2} \sum_{1 \leq j_1 \leq j_2  \cdots \leq  j_q } J_{j_1 j_2 \cdots j_q} \psi^{j_1} \psi^{j_2} \cdots \psi^{ j_q }  ~,
\cr
 & ~& \langle J_{j_1 \cdots j_q }^2 \rangle =
  { 2^{ q -1}  {\cal J } ^2 (q-1)! \over q N^{q-1} } ~~~ ({ \rm no ~sum } ) 
\eea
where the couplings are drawn from a random gaussian distribution with the mean indicated above. The factor of $i$ is necessary to get a hermitian
Hamiltonian when $q/2$ is odd. 

At large $N$ the model can be solved by writing an equation for the fermion two point function $G(\tau_1 , \tau_2) = { 1 \over N } 
\sum_j \langle \psi^j(\tau) \psi^j(0) \rangle $. 
This equation has the form 
\be \la{SDeqn}
\partial_{\tau_1} G(\tau_1,\tau_3) - \int d\tau_2 \Sigma(\tau_1,\tau_2) G(\tau_2,\tau_3) = \delta(\tau_{13}) ~,~~~~~~ 
\Sigma(\tau_1,\tau_2) = \frac{\mathcal{J}^2}q [ 2G(\tau_1,\tau_2)]^{q-1} 
\ee
These equations  follow from the euclidean action 
\be \la{FreeSG}
- S_E/N =  \log {\rm Pf}( \partial_\tau - \Sigma )  - \half  \int d\tau_1 d\tau_2 \left[ \Sigma(\tau_1,\tau_2) G(\tau_1,\tau_2) - { {\cal J}^2 \over 2 q^2  } [ 2 G(\tau_1,\tau_2)]^q 
\right] 
\ee
This equations \nref{SDeqn} can be solved in Euclidean time and   can be used to evaluate the free energy using \nref{FreeSG}. In general it is only possible to solve the equations numerically. 
However, it is possible to solve them analytically at large $q$. It is also possible to solve them in general at low temperatures, but not too low, 
$ 1 \ll \beta {\cal  J} \ll N$, in terms of a scaling  ansatz of the form (for $ 1 \ll {\cal J} \tau_{12} \ll \beta { \cal J} $) 
\be \la{GCorr}
G(\tau_1,\tau_2) =  c_\Delta \sign{ \tau_{12} } {1 \over | {\cal  J}  \tau_{12} |^{ 2 \Delta } } ~,~~\Delta = { 1 \over q } ~,~~~~c_\Delta   = \half  \left[ 
  ( 1- 2 \Delta)  { \tan \pi \Delta \over \pi \Delta}  \right]^\Delta
\ee

At low energies the model has an emergent reparametrization symmetry changing $G$ to $ G_f = [ f'(u_1) f'(u_2) ]^\Delta G(f(u_1) ,f(u_2) ) $ were $G$ is given 
in \nref{GCorr}. This symmetry is  
  explicitly broken by $1/{\cal J}$ effects leading to a Schwarzian action \cite{Maldacena:2016hyu,Kitaev:2017awl}. 
\be
 S = - {N \alpha_S \over {\cal  J}  } \int du \{ f(u) , u \}   
 \ee
We can then get the  thermal correlators and free energy by setting  $f = \tan{ \pi u \over \beta } $. The coupling constant $\alpha_S$ is determined by four-point function calculation of the SYK model, which goes as  $\alpha_S \sim  \frac{1}{4 q^2}$ in large $q$ limit. (See Ref. \cite{Maldacena:2016hyu}).
 
  We are interested in the thermofield double state of the SYK model, which is a state in the Hilbert space of two SYK sites, defined as   follows 
 \be 
 \ket{TFD_\beta}=Z_\beta^{-1/2}e^{-\beta\left(H_L+H_R\right)}\ket{I}
 \ee
 Here $Z_\beta$ is the thermal partition function at inverse temperature $\beta$. $\ket{I}$ is the thermofield double state at infinite temperature, which is a maximally entangled state between two systems. $H_L$ is the SYK Hamiltonian applied to the left system, and $H_R$ is the Hamiltonian of the right system, defined so that 
  $\left(H_L-H_R\right)\ket{I}=0$. For the SYK model with fermions $\psi^i_L,~\psi^i_R$, we can define $\ket{I}$ by
 \be \la{Idef}
  \left(\psi^j_L+i\psi^j_R\right)\ket{I}=0,~\forall j
 \ee
 This  determines $\ket{I}$ as the ground state of complex fermions with annihilation operators $f_j= \frac1{\sqrt {2}}\left(\psi^j_L+i\psi^j_R\right)  $. With this convention, for the SYK model we obtain $H_R=(-1)^{q/2}H_L$ which describes a SYK model with equal or opposite coupling.  The thermofield double state $\ket{TFD_\beta}$ satisfies $\left(H_L-H_R\right)\ket{TFD_\beta}=0$ by construction. The reduced density matrix of each subsystem (left or right) is the thermal density matrix with temperature $\beta^{-1}$, and entropy $S_0+\frac{N(2\pi)^2\alpha_S}{\beta\mathcal{J}}$. $S_0$ is a constant zero temperature entropy.  
    
 
\subsection{Two coupled SYK models } 

In this article we concentrate on a model where we start with two decoupled SYK models, both described by  same couplings, up to a sign for odd $q/2$,  
$J_{j_1 \cdots j_q} ^L = (-1)^{q/2} J^R_{j_1 \cdots j_q} $. 
A situation like this arises when we consider the thermofield double of the original system.  
So we  denote by $\psi^i_L$ and $\psi^i_R$ the fermions of the two copies the SYK model. 
But now we  introduce a coupling between the two sides of the form 
\be \la{Hint} 
H_{\rm total} = H_{\rm L, SYK} + H_{\rm R,SYK} + H_{\rm  int} ~,~~~~~~H_{\rm int} = i \mu \sum_j    \psi_L^j \psi_R^j 
\ee

Note that the state $|I\rangle$, defined in \nref{Idef}, is  also the ground state of $H_{int}$. For small $\mu$ we expect that the system described by the 
total Hamiltonian \nref{Hint} can be described as the nearly conformal system associated to two copies of the usual SYK model \nref{SingleSYK}
 plus a relevant deformation associated
to the $\mu$ coupling. We expect that the system  flows  in the IR to a gapped phase described
by a ground state $|G\rangle$ of the combined system. 

In this paper we will study this gapped phase in detail. We will show that at small coupling $\mu$, or at large $q$ for any coupling, the 
ground state $|G\rangle$ is very close to the thermofield double state $|TFD\rangle$ of the decoupled system with a particular inverse temperature $\beta(\mu)$, that is determined by the mass $\mu$.  
For small mass $\mu$, this could be 
expected for the following reason. We expect that the system will develop an approximate conformal symmetry at energy scales less than 
${\cal J}$. Then we can analyze the effects of the coupling as a perturbation to the approximately conformal system. The state that will minimize the interacting Hamiltonian 
\nref{Hint} will try to maximize the left-right correlations in order to make $H_{\rm int}$ as small as possible. The thermofield double state is a pure state of the combined,
but decoupled, left and right systems that has a relatively large value for the left-right correlators. As we decrease the temperature of the TFD, 
 we decrease the expectation value of $H_L$ and $H_R$, we also decrease $H_{\rm int}$, but at a different rate. There will be a temperature for the TFD where the energy 
 is minimized.  Furthermore, in the conformal approximation we can imagine a conformal map between the circle that prepares the TFD to two parallel lines, which 
 would naively prepare the same state. This is analogous to the usual map between the sphere and the Euclidean cylinder for higher dimensional CFTs. This map is not
 strictly valid in our case, because the conformal symmetry is not exact, but it leads us to expect that the TFD double could be related to the ground state of a closely related system, whose preparation involves Euclidean time evolution over an infinite time. Furthermore, the interaction term in \nref{Hint} selects a state with relatively large correlations between the left and right degrees of freedom, as we have in the TFD state.
     In the rest of the paper we will present  precise and detailed arguments 
 for the relationship between $|G\rangle $ and $|TFD\rangle$. Similar phenomena have been studied in $(1+1)$-d CFTs\cite{li2008entanglement,qi2012general}.


\begin{figure}[h]
\begin{center}
\includegraphics[scale=.38]{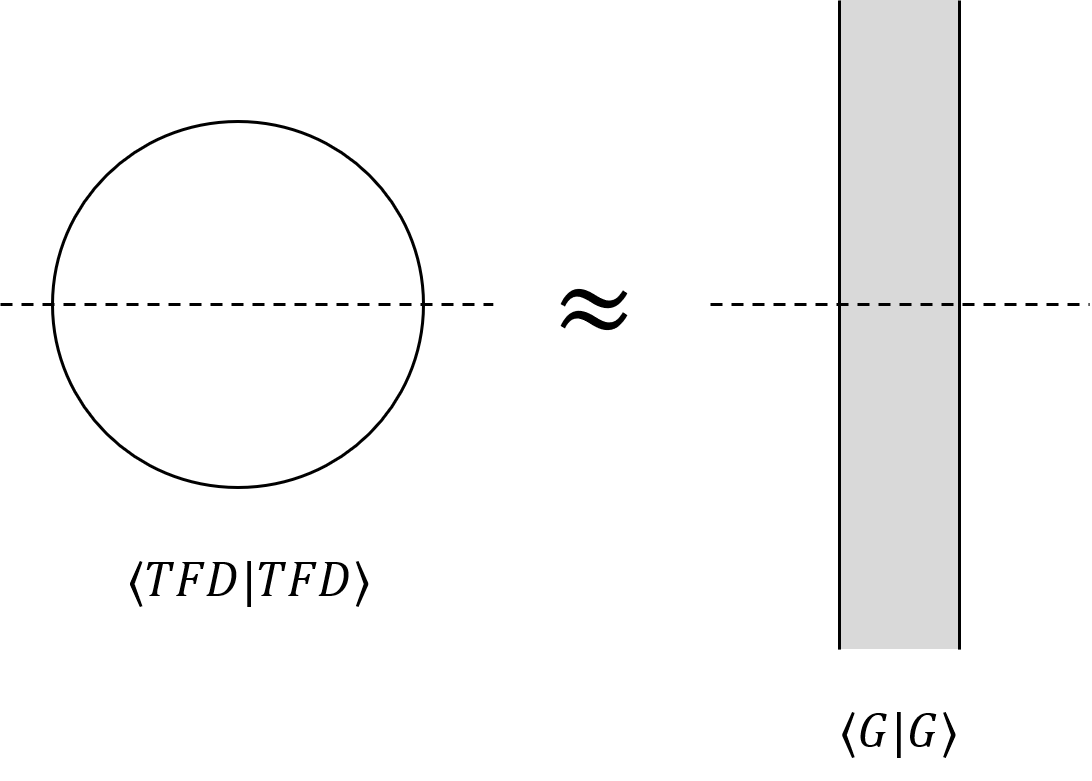} 
\caption{Schematic comparison of the Euclidean path integral that prepares the $|TFD\rangle$ state and the ground state of coupled system $|G\rangle$. The thermal circle is related to two parallel lines by a conformal transformation. The conformal symmetry is weakly broken. The competition of $H_L+H_R$ and $H_{\rm int}$ controls the symmetry breaking and selects the $|TFD\rangle$ with a certain temperature as the ground state.
}
\label{TFDvsG}
\end{center}
\end{figure}
 
In the two decoupled systems the $|TFD\rangle $ state undergoes a non-trivial time evolution when we evolve forwards in time for both the left and right times by the same
amount. On the other hand $|G\rangle$ is invariant under time  evolution by the coupled Hamiltonian \nref{Hint}. 
We had mentioned that the system develops an approximate conformal symmetry that is explicitly broken in the IR. In the case of the decoupled system we 
preserve  a boost-like symmetry corresponding to the usual symmetry of TFD states (forward evolution in $t_R$ and backward in $t_L$). On the other hand 
the state $|G\rangle $ preserves a different element of $SL(2)$, which corresponds to forward global time translations on the two times. Moreover, we will see that some 
features of the spectrum of excitations of the system around the ground state are still governed by the approximate SL(2) symmetry, which is a unique feature of the $(0+1)$-d case and does not occur in higher dimensional CFT's with a relevant coupling.


\subsection{ Low energy analysis} 

In this subsection we find the properties of the ground state when the mass is very small $\mu \ll J$ so that we can use the low energy solution. 
At leading order, when we ignore the effects of conformal symmetry breaking, we have conformal invariant correlators. If we assume the ground state is close to $|TFD\rangle$, the approximate conformal symmetry allows us to obtain the left-right correlator by reparameterization of the single side conformal correlator, which takes the form 
\be \la{ConfCoP}
\langle \psi_L(t_l)  \psi_R(t_r)  \rangle =  c_{\Delta } { i \over [ 2 {\cal J } \cos{ t_l - t_r \over 2 } ]^{2 \Delta } }   
\ee
We can get this from \nref{GCorr} via a reparametrization $t_P = \tan{t_r \over 2} $, $t_P= -{ 1/\tan{ t_l \over 2 } } $, with $t_P$ being Lorentzian time, as in
\nref{BdyTimes}. The $i$ is necessary since the operator $\psi_L(t_l)  \psi_R(t_r)$  is antihermitian. 
Here we are simply choosing a particular time parametrization of the two boundaries. 
In order to determine whether these correlators are close to a solution, we need to take into account the physics of the reparametrization mode. 
This can be done by  introducing an arbitrary reparametrization $t_l(u)$ and $t_r(u)$ of the above correlator, \nref{ConfCoP}. From each single SYK model we get a Schwarzian action for this mode. In addition, the  interaction term in \nref{Hint} leads to an  additional term. This  additional term can be obtained by taking the expectation value of the fields in 
\nref{Hint}, exponentiating as in \nref{Expec}, and performing a reparametrization. 
This implies that we can write down a Lorentzian action of the form
\be \la{fLfR}
S  = N \int du \left\{ - { \alpha_S \over {\cal J }  } \left( \{ \tan{ t_l(u)\over 2 } , u \} + \{ \tan{ t_r(u)\over 2 } , u \} \right) + 
\mu { c_\Delta  \over (2 {\cal J} )^{  2 \Delta } } \left[ { t'_l(u) t'_r(u) \over \cos^2 { t_l(u) - t_r(u) \over 2 } } \right]^{   \Delta } \right\}
\ee
Note that the action has an overall factor of $N$. 
This form of the low energy action is the same as what we got in gravity \nref{ActGr}. So, in the next section we will analyze this low energy 
action, coupled to conformal matter, and describe several physical consequences. 

  Let us say a few more words on the rationale behind the approximation \nref{fLfR}. As we reviewed in \nref{FreeSG}, we can describe the 
two decoupled systems around the thermofield double state in terms of an effective action. This effective action contains a soft mode, or a low action mode, that 
can be viewed as  pseudo-Goldstone bosons for an spontaneously and explicitly broken reparametrization symmetry  \cite{KitaevTalks}. The 
action in the $G,\Sigma$ space develops a shallow valley \cite{KitaevTalks}. 
Motion along the valley is parametrized in terms of the variables in \nref{ActGr}. The last term in  \nref{ActGr} is the projection of the  the interaction term  
$H_{\rm int}$ in the coupled Hamiltonian \nref{Hint}  to  this valley.  This extra term modifies the location of the minimum along this valley.
 When we approximate the dynamics by 
\nref{ActGr} we are neglecting the change of the state along the ``hard'' directions and  taking into account only  its change along she ``shallow'' directions. 
In other words, we can imagine we have a particle in a shallow valley and we are turning on a small external force. The new equilibrium position will be approximately given
by another point in this valley, obtained after evaluating the external potential along the valley. 
 The description of this new ``point'' is what we will study in detail in the next section.

We should also note that we could have started instead with an interaction Hamiltonian with a more general relevant coupling term, such as
$H_{int} = gN^{1-p}  (i  \psi_L^j \psi^j_R)^p $. In this case, for $p< q$, we also obtain an effective action like \nref{fLfR} but with $\Delta \to p \Delta$ and 
a different prefactor for the interaction term.

\section{ Low energy analysis of the coupled theory } 
\la{LowSec}

In this section we will study several physical properties that arise both in the case of gravity with the two boundary coupling 
\nref{OOInt} as well as in the case of two coupled SYK models described by \nref{Hint}. In both cases, for relatively small interaction strength we 
get a description that involves the same action \nref{fLfR} and \nref{ActGr}. Just to set notation for this section, let us write it yet once more 
\be \la{fLfRres}  
S  = N \int d \tilde u \left\{  -   \left( \{ \tan{ t_l(\tilde u)\over 2 } ,\tilde  u \} + \{ \tan{ t_r(\tilde u)\over 2 } , \tilde u \} \right) + 
{ \eta   }  \left[ { t'_l(\tilde u) t'_r(\tilde u) \over \cos^2 { t_l(\tilde u) - t_r(\tilde u) \over 2 } } \right]^{   \Delta } \right\}
\ee
The relation between parameters in the two previous actions and \nref{fLfRres} is 
 \be \la{RelPar}
 \tilde u \equiv { \cal J \over \alpha_S} u = { N \over \phi_r } u ~,~~~~~~ \eta \equiv  { \mu \alpha_S \over \cal J } 
 { c_\Delta 
 \over ( 2 \alpha_S)^{2 \Delta}}  = 
 { g \over 2^{2\Delta} } \left( { N \over \phi_r} \right)^{2\Delta -1} 
 \ee
 We  have defined a rescaled time, $\tilde u$. 
In the SYK model $\tilde u$ is basically time measured in units of $1/{\cal J }$.

   This action should be supplemented by $SL(2)$ constraints stating that the total $SL(2)$ charges vanish \cite{Maldacena:2016upp}.  

\subsection{Classical solution}

$N\gg 1$  governs the approach to the classical limit. We imagine that $N$ is larger than any other parameter that appears in the problem. 
 We can first look for a simple  solution by making a linear ansatz 
\be
   \la{CandSol} t_r(\tilde u) =t_l(\tilde u) = t' \tilde u  ~,~~~~~~{\rm with } ~~~~t' \equiv { d t \over d \tilde u } =    {\rm constant}
   \ee
   Note that we defined $t'$ as a derivative with respect to $\tilde u$. To turn into a derivative with respect to $u$ we
   have to use \nref{RelPar}. 
\nref{CandSol} is a solution of the equations of motion for \nref{fLfRres} for any value of $t'$. This looks a bit surprising at first since we expected to 
have a single solution. In these formulas $t'$ is a constant. 

We should however  recall that the action \nref{fLfRres}  
 should be supplemented by   constraints stating that the total $SL(2)$ charges vanish \cite{Maldacena:2016hyu}.  
Let us explain this in more detail. 
  The action 
   \nref{fLfRres} has a global SL(2) symmetry generated by 
   \be 
     \delta t_l = \epsilon^0  + \epsilon^+ e^{ i t_l} + \epsilon^-  e^{ - i t_l } ~,~~~~~~~\delta t_r = 
     \epsilon^0  - \epsilon^+ e^{ i t_r} - \epsilon^- e^{ - i t_r }
     \ee
     These are simply the  $SL(2)$ symmetries of the coordinate $X^M$  expressed in terms of $t_l$ or $t_r$ defined through \nref{BdyTimes}. 
 We can compute the associated charges  by using the Noether procedure on  \nref{fLfRres}, see appendix \ref{SLTwo}. 
 Here
      we will quote the answer only for the charge $Q_0$ evaluated 
     on the configuration \nref{CandSol}. This  gives \footnote{The expression for the charge
      can be simply derived from \nref{ActGr} by dropping all terms with derivatives higher than the first derivative, since they will not contribute when $t'$ is a constant. 
      Note that $\{ \tan{t_l(u) \over 2} , u\} = {t_l'}^2/2 $ up to terms with higher derivatives. Then the Noether charge is just simply the usual one associated to 
     $t$-shifts, the conserved total ``momentum'' along the $t$ direction.}
     \bea \la{Qalph}
     Q_0 &=& - 2  t'  + 2 \eta  \Delta  {t'}^{2\Delta -1} = 0  
     \\
      (t'  )^{2(1-\Delta ) } &=& \eta  \Delta ~,~~~~~~~ \left( { 1 \over {\cal J } } { dt \over du } \right)^{2 (1-\Delta) } = { \mu \Delta \over 2 {\cal J } 
      \alpha_S} { 2 c_{\Delta } \over 2^{ 2 \Delta } } \la{tPrime}
     \eea
     where we imposed that the charge is zero and obtained the value of $t'$. We have also expressed the answer in 
     terms of SYK parameters and the original time, $u$. 
      This value determines the relation between the boundary time $\tilde u$ and the 
     bulk time $t$.  
 
 The validity of the action, \nref{fLfRres},  as a good approximation to the original SYK requires that 
 \be \la{Valcl}
 t' \ll 1 ~,~~~~~~ \eta \ll 1 
 \ee
 so that the low energy approximation leading to \nref{fLfRres} is valid. Looking at \nref{tPrime} we see
 that we need   $ 0 < \Delta < 1$.
 In addition, in order for the classical approximation to \nref{fLfRres} to be valid we also need 
 \be
 \la{ClassApV} 
 N t' = N ( \eta \Delta)^{ 1 \over 2 ( 1 - \Delta ) } \gg 1 ~,
 \ee
 since the inverse of this quantity is the effective dimensionless coupling of the Schwarzian theory. 
 For large enough $N$ we can ensure both \nref{ClassApV} and  \nref{Valcl} by taking 
a  small enough $\eta$, but not too small. 
 
   Once we know $t'$ we can go from the conformal matter correlators to the physical ones\footnote{To get the operators as a function of the unrescaled 
   boundary time $u$ we write ${\cal O}(u) =  \left( { d \tilde u \over du } \right)^{\Delta } {\cal O }(\tilde u) $ using \nref{RelPar}.} 
   \be \la{Enbu}
   \langle {\cal O } (t_l) {\cal O}(t_r) \rangle =   \left[ { 1 \over\cos{ t_l - t_r \over 2 }} \right]^{2\tilde \Delta }   \longrightarrow 
   \langle {\cal O} (\tilde u_1) { \cal O } (\tilde u_2 ) \rangle =    \left[ { t' \over\cos{t' ( \tilde u_1 - \tilde u_2 ) \over 2 } } \right]^{2\tilde \Delta } 
     \ee
     This gives the physical value of left-right correlation functions. In the gravity theory, these can be the two point functions of some matter fields that propagates
     in the bulk, which could be the same or different, than the matter field that gives rise to the interaction across the boundaries \nref{OOInt}. In the SYK model 
     it could be the elementary fermion $\psi^k$ or any composite operator that we obtain after taking an operator product of those. For that reason we have indicated
     that its conformal weight, $\tilde \Delta$, need not be the same as that of the fields that give rise to the explicit coupling between the two sides.

     The value of $t'$, \nref{tPrime}, determines the energy scale of bulk excitations, or equivalently, the energy scale of the conformal excitations of the SYK model.
      An operator with dimension $\tilde \Delta $, or a bulk field with the corresponding mass, 
       has energies $E_t = \tilde \Delta + n$ with respect to global bulk time.  This form of the spectrum is fixed by $SL(2)$ symmetry.   Then, with respect to (rescaled) boundary  time $\tilde u$ they 
     have energies  
     \be \la{ConfEn}
      E_{\tilde u} = t' E_t = t' ( \tilde \Delta + n ) ~,~~~~~~~ E_u = { d t \over d u } ( \tilde \Delta + n ) 
      \ee
      where the subindex in $E_{\tilde u}$ or $E_{u}$ states whether the energy is conjugate to $\tilde u$ or $u$ time translations. The difference is a simple rescaling \nref{RelPar}. Here ${ d t \over du } $ is just a constant, independent of $u$. 
     Therefore we see that $t'$ sets the scale of the energy gap of the system. 
     More precisely, for the systems we have been describing the energy gap is 
     \be \la{GapDef}
      E_{{\rm gap}, \, \tilde u} = t' \Delta ~, ~~~~~~~ E_{{\rm gap}, \, u} = { {\cal J } \over \alpha_S} t' \Delta = { N \over \phi_r} t' \Delta  = { dt \over d u } \Delta 
      \ee
     This part of the spectrum is invariant under physical $SL(2)$ transformations of boundary time. 
     This is an interesting property. If we looked at the coupled SYK hamiltonian \nref{Hint}, we see that the mass term dominates at low energies and we would have 
     expected a generic gapped system, with arbitrary energy spacings. Here we get a pattern of energy spacings that reflect an $SL(2)$ symmetry. This is the type of spectrum we expect from a state/operator correspondence. As we discuss below the gravitational degree of freedom does not follow this pattern. 
     
\begin{figure}[h]
\begin{center}
\includegraphics[scale=.4]{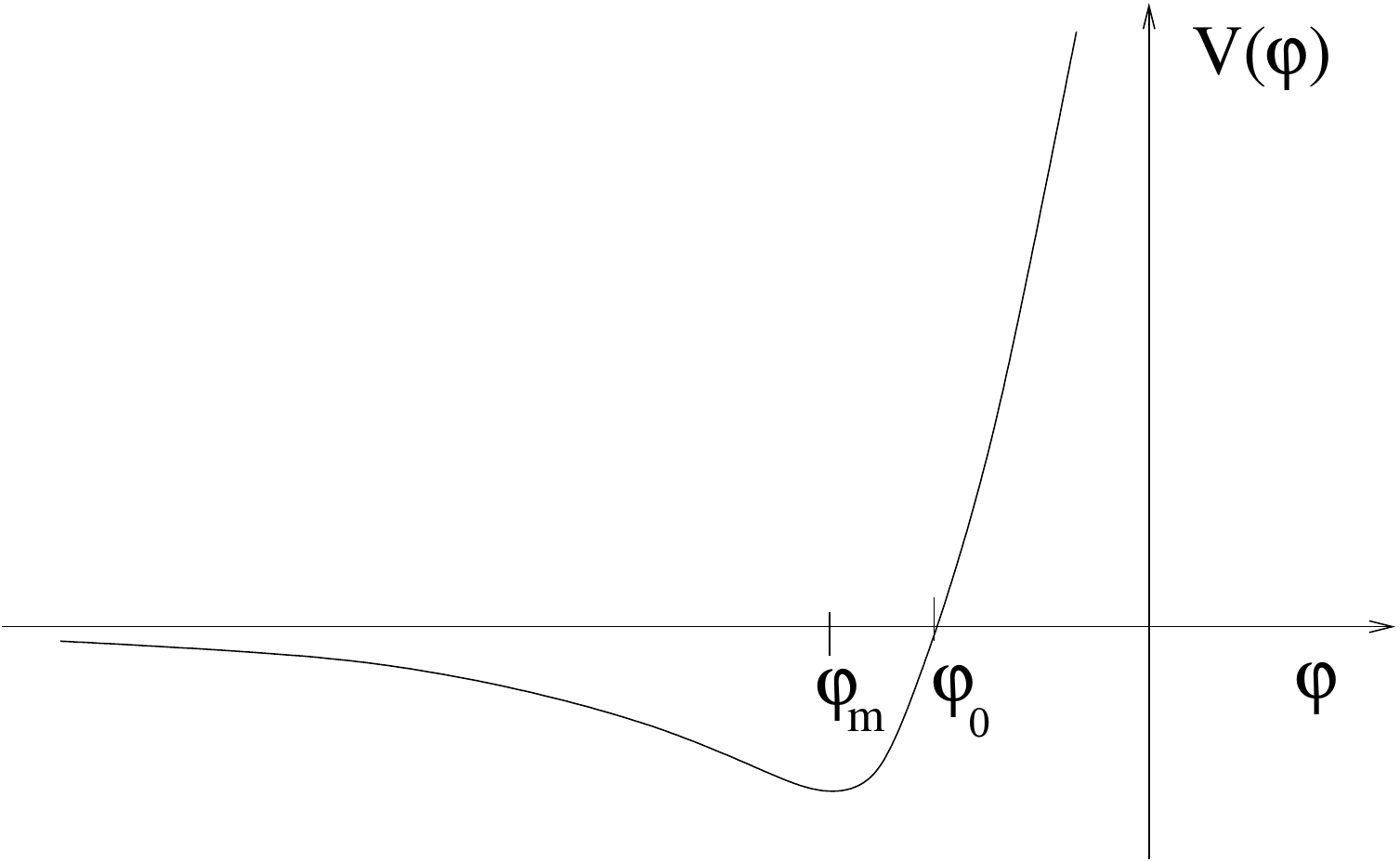} 
\caption{ Efective potential for the single degree of freedom associated to the gravitational mode, see \nref{PotV}. Here $\varphi_m$ is the minimum and 
$\varphi_0$ is where it crosses zero.    }
\label{Potential}
\end{center}
\end{figure}

     It is interesting to wonder what all the physical solutions of \nref{fLfRres} are. In appendix 
     \ref{SLTwo} we show that all solutions of \nref{fLfRres}  with zero $Q_\pm $  SL(2) charges 
     can be gauge transformed to solutions where the left and right times are equal. 
     Namely, we can restrict our attention to  
      solutions of the form $t_r(\tilde u) = t_r(\tilde u)= t(\tilde u)$, but with a general $\tilde u$ 
     depedence. This ansatz sets to zero the $Q_\pm$  SL(2) charges, see appendix \ref{SLTwo}.  As before, the equations come from demanding that 
     $Q_0 =0$, which, written in terms of $\varphi = \log t'(\tilde u)$, reads (see appendix \ref{SLTwo}),  
     \be \la{phieq}
     0 =Q_{0 }  = 2 N  e^{-\varphi} \left[   - e^{ 2\varphi }  -  \varphi''  +  \eta  \Delta  e^{ 2 \Delta \varphi } \right] 
     \ee
     We see that this looks like the equations of motion of a non-relativistic particle in a potential  with the 
     action   
     \be \la{PotV}
       S =N  \int d \tilde u   \dot \varphi^2 - V(\varphi) ~,~~~~~~~~  V =   e^{ 2 \varphi} - \eta  e^{2 \Delta \varphi}
      \ee 
     see figure \ref{Potential}. The solution we found in \nref{tPrime} corresponds to the minimum of the potential, $ e^{ 2 ( 1 -\Delta) \varphi_m} =\eta \Delta$. 
     It is possible to check that solutions of \nref{phieq}   also solve the equations of motion, which are simply  a
     time derivative of  \nref{phieq}. The oscillations around the minimum of the potential give rise to a harmonic oscillator degree of freedom with frequency 
     $\omega_{\tilde u} = t' \sqrt{ 2 (1-\Delta) } $, or physical energies 
     \be
     E_ {\tilde u} = t' \sqrt{ 2 (1-\Delta) } ( n + \half ) ~,~~~~~~~~ E_u = { d t \over du }  \sqrt{ 2 (1-\Delta) } ( n + \half )  
     \la{OscEn}
     \ee
    This harmonic oscillator corresponds to the physical gravitational excitations of  the system. 
    This single quantum mechanical degree of freedom corresponds to the ``boundary graviton''. It encodes the gravitational dynamics and backreaction of the system.  Even though we call it a ``boundary graviton'' it is a global mode that refers to both boundaries, the left and the right boundary and cannot be associated to just one of them. 
    The fact that we get a minimum and an ordinary harmonic oscillator around the minimum is telling us that the solution we found in \nref{tPrime} is stable. 
    Furthermore, the energy scale is also set by $t'$. These excitations, as opposed to the ones in \nref{ConfEn} do not organize into an SL(2) multiplet. This is 
    the leading feature of the spectrum associated to the breaking of the $SL(2)$ symmetry. Note also that the energy \nref{OscEn} is larger than \nref{GapDef} for
    the range of $\Delta$s that we are considering. This justifies the statement that \nref{GapDef} is the actual energy gap of the system. 
            
    It is also interesting to compute the conserved energy associated to the lagrangian in \nref{fLfRres}. This is the Noether charge under $\tilde u$ translations. 
    We find 
    \bea \la{EnGenN}
    E_{\tilde u }/N &=&  - \{ \tan{ t_l(\tilde u)  \over 2 } , \tilde u \} -   \{ \tan{ t_r(\tilde u) \over 2 } , \tilde u \}  + \eta (2 \Delta -1) \left[ { t'_l(\tilde u)  t'_r (\tilde u) \over { \cos{ (t_l - t_r ) \over 2 } } } \right]^{2\Delta }
    \\
    &=&   - ( 2 \varphi'' - { \varphi'}^2 + e^{ 2 \varphi})   - \eta (1-2 \Delta)    e^{ 2 \Delta \varphi}  = 
    ( {\varphi'}^2 + e^{ 2 \varphi} ) - \eta  e^{ 2 \Delta \varphi }  \la{EnVph}
    \eea 
    where we used the equations of motion. 
    This is the   energy for a particle in a potential \nref{PotV}. In comparing with the SYK Hamiltonian \nref{Hint}, we note that only the terms with a $ - \eta e^{ 2 \Delta \varphi} $ in \nref{EnGenN} corresponds to $\langle H_{\rm int} \rangle$. The rest of the terms corresponds to $\langle H_L + H_R \rangle$. 
    In particular, for the minimum at $   e^{ 2 (1-\Delta) \varphi_m }   = (  t')^{ 2 - 2\Delta } = \eta \Delta $,
    \nref{tPrime},  we get the energy of the ground state 
    \be \la{GrstEn}
    E_{G \, \tilde u }/N = -  { ( 1-\Delta) \over \Delta } e^{ 2 \varphi_m} =  -     { ( 1-\Delta) \over \Delta }    \left( \eta  \Delta \right)^{ 1 \over 1 -\Delta } 
   ~,~~~~~~~E_{G \, u } = - N { (1-\Delta) \over \Delta } { \alpha_S \over {\cal J } } \left( { dt \over d u} \right)^2
    \ee
  where we have also given the expression in terms of SYK parameters \nref{RelPar}. 
  The fact that we obtained a single degree of freedom is in agreement with analysis of 
   the standard Rindler-Thermal-whormhole in \cite{Kuchar:1994zk}, which found 
  a two dimensional phase space, corresponding to the energy and the relative time shift.

    The $\varphi$ dynamical system \nref{PotV}  also has unbounded solutions. 
     An unbounded solution has  $\varphi  \sim  -   \gamma  \tilde u $ for large times, which implies that 
    $t'(u) \propto  e^{- \gamma \tilde u }  $ for large times. This means that the global time $t $ has a maximal range and that the boundary particle trajectory approaches the $AdS$ boundary, since $\sigma = \epsilon t'$. In the gravity case, these solutions develop a horizon. 
      This is also what happens when
    $\eta =0$, where we get the solutions  corresponding to the TFD \nref{TFDtsol}. 
    
    What we discussed so far  applies to  the case where 
    we do not add any further bulk excitations. 
    If we also add bulk particle excitations, then we need to reanalyze the problem,
     now imposing a different value for the SL(2) charges. 
    In particular if we have a bulk excitation with   global energy $E_t^{\rm bulk}$, and it is at rest in $AdS_2$\footnote{What we really want is
    that it is the   lowest energy state in the corresponding SL(2) representation.},  then we expect that the charge $Q_0$ should be 
    set to $Q_0= E_t^{\rm bulk}$ instead to zero in \nref{phieq}. If we treat the bulk classically, then, if the particle is at rest, then the other charges remain zero, $Q_\pm =0$.
    When we integrate to get \nref{PotV} we see that we get an extra term in the potential
    \be \la{VShift}
     V \to V + E_t^{\rm bulk} e^{\varphi }/N 
     \ee
      The extra term is suppressed in the classical limit (for large $N$). So if the extra added energy 
    is of order one, such as the energy added by the states in \nref{ConfEn}, then we have a small correction.
    This does not  change the physics much. The shift in the minimum of the potential implies that we get 
    an extra energy of order $E_{\tilde u} \to E_{\tilde u} + E_t e^{\varphi_m} $ which is the correct, expected new boundary energy, see appendix \ref{SLTwo}. 
    If we add a large amount of energy in the bulk, then we could remove the possibility of having bound states, as we will see in the next section. 
      
      
   \subsection{Quantum version} 
        \la{SecQuantum} 
        
    We can consider the quantum version of \nref{fLfRres}. In principle, we should find the right
    integration  measure for our problem, etc. Instead, we notice that the Liouville-type  action  in 
    \nref{PotV} was already shown to describe correctly the TFD (for $\eta=0$) \cite{Bagrets:2016cdf,Bagrets:2017pwq,Stanford:2017thb}. So, to treat our  case,
     we simply add the effects of the new interaction term. 
    Therefore  we want to quantize action 
    \be \la{QMAct}
     S = N \int d\tilde u   [ { \dot \varphi}^2 - e^{ 2 \varphi}   + \eta    e^{ 2 \Delta \varphi } ]
     \ee
    Perhaps a better argument 
    for \nref{QMAct} is the following.  As argued in appendix \nref{SLTwo} any 
    solution of \nref{fLfRres} that obeys the constraints can be gauge-transformed to a solution of \nref{QMAct} and 
    futher obeys the gauge condition $t_l(0)= t_r(0) =0$.    
    We can compute the symplectic form for the original problem \nref{fLfRres} on this space of solutions.
  One can check that this symplectic form, at $u=0$,  is the same as the one we get from \nref{QMAct}. 
   
       After shifting $\varphi \to \tilde \varphi =  \varphi - \log N $ we get  a 
      Schroedinger equation for $\psi(\tilde \varphi)$ of the form 
     \be \la{Scheq}
     N E_{\tilde u}  \psi = -{  \kappa^2 \over 4 } \psi   =  - { 1 \over 4} { \psi''   } + [ e^{ 2 \tilde \varphi} - \tilde \eta e^{ 2 \Delta \tilde \varphi} ] \psi   ~,~~~~~\tilde \eta \equiv  \eta N^{ 2 - 2 \Delta } 
     \ee
     
     {\bf Exact solution for $\Delta =\half $ } 
     
     For $\Delta =1/2$ we can solve this eigenvalue problem and we find the quantization condition\footnote{
The solution that decay at $\varphi \to \infty $ is $\psi \sim e^{ \kappa \varphi  - 2 e^\varphi } U( \half - \tilde \eta + \kappa, 1 + 2 \kappa;  4e^\varphi )$, with $U$ a confluent hypergeometric function.   Imposing that it decays as $\varphi \to -\infty$ we get \nref{QuantCo}.}
     \be \la{QuantCo}
     \kappa_n = \tilde \eta-\half - n > 0 ~,~~~~~n=0,1,\cdots, \hat N  ~,~~~~\tilde \eta = N \eta~,~~~~ {\rm for} ~~~\Delta =\half 
     \ee
    which leads to a finite number of states 
    \be \la{NbS}
    \hat N_{\rm bound ~states}=\lfloor \tilde \eta + \half \rfloor \sim  \eta N  
    \ee
    Note that via \nref{ClassApV} $\eta N$ also controls how classical the system is. 
    In the quantum analysis here we can go to small $\eta N$ and we can see that for $\eta N < \half$ 
    we actually have no bound state, so that we 
     do not have a solution of the kind we are discussing, and we only get the black hole like configurations. 
     The case $\Delta =\half$ that we are discussing here is such that the boundary coupling $g$ is dimensionless and 
     it is essentially equal to $\eta$ here, see appendix \ref{FBF}.

So far we are only discussing the  
     states of the ``graviton mode''. If, in addition, we have some extra fields, with their own bulk 
     energy $E_t$,  (and zero $q_\pm $ SL(2) charges) 
      then we get an extra term in the potential as in \nref{VShift}, 
  which results in an effective shift (for $\Delta =\half$)
     $\eta \to \eta - E_t/N$. This   reduces the number of bound states available for the gravity modes.

    {\bf Approximate discussion for general $\Delta $ } 
       
   For general $\Delta$ we could not solve the equation \nref{Scheq} exactly, but the fact that we get a finite number of states 
   is  also true. For large $\tilde \eta$ we can get an estimate on the number of states by a WKB analysis 
   \bea
   \hat N &\sim & { 1\over 2 \pi } \int p dq = {  N 4  \over 2 \pi }  4 \int_{-\infty}^{\varphi_0} d\varphi \sqrt{ \eta e^{ 2\Delta \varphi} - e^{ 2 \varphi} } = 
   { 2 N e^{\varphi_0} \over \pi } \int_0^1 dz \sqrt{ z^{ -2(1- \Delta) } -1 }  
 \cr
 \hat N &\sim & {  N  \over\sqrt{ \pi} } \frac{\Gamma \left(\frac{\Delta }{2-2 \Delta }\right)}{ \Gamma
   \left(\frac{1}{2-2 \Delta }\right) } \eta^{ 1 \over 2 ( 1 - \Delta ) }  =  ( N t') {  1  \over\sqrt{ \pi} } \frac{\Gamma \left(\frac{\Delta }{2-2 \Delta }\right)  }{ \Gamma
   \left(\frac{1}{2-2 \Delta }\right)     \Delta^{ 1 \over 2 - 2 \Delta  }  }
    ~,~~~~~~{\rm where } ~~~e^{2 (1-\Delta) \varphi_0}  = \eta ~~~~~~~~
   \eea
   We see that the number continues to be propotional to the same parameter $(N t')$ that governs 
   the classicality of the system, playing the role of $1/\hbar$ \nref{ClassApV}.

   { \bf Comment on the case with  a small number of fields  in the interaction } 
   
   In this paper we have mostly concentrated on the case where we have $N$ fields in the bulk, or 
   $N$ operators in the interaction term. 
   If we had a smaller number, say $k$, then  have 
   the same  action as in in \nref{fLfRres} but with the replacement 
   \be
   \la{RepNk} 
   \eta \to { k \over N } \eta 
   \ee
   We imagine keeping $k$ fixed and of order one as we take $N$ large. 
   This makes $\eta $ even smaller, which is good in \nref{Valcl}, but in \nref{ClassApV} we get 
   now 
   \be
   N t' \propto N^{ 1 - 2 \Delta \over 2 ( 1 - \Delta ) } 
   \ee
   So that for the system to have a classical wormhole, with  a large number of bound states,  we need that $\Delta < \half $. In this case the interaction is relevant. 
In the gravity case, this requires that the bulk fields obey the alternate boundary condition, whether they are bosons  \cite{Klebanov:1999tb} or fermions
   \cite{Cacciapaglia:2008bi}. 
  It would be nice to study this case further, to check the precise regime of validity of such configurations, but we will not do it in this paper. 
   
     \subsection{ Getting the thermofield double from a quench } 
     \la{MatchingTFD} 
     
     In this subsection we consider a configuration where we prepare the ground state of the coupled system, with non-zero $\eta$,
      and then, at $\tilde u=0$ we switch off $\eta$. So we have $\eta(\tilde u) = \eta \theta(-\tilde u)$. 
      We can view this as a quench where we set the coupling between the two systems to zero at time equal to zero. 
      We argue that the state we get using  the low energy approximation   \nref{fLfRres} (with $\eta \to \eta(\tilde u) $) is precisely the thermofield double
      of two decoupled systems.

     We will argue that the solution for positive times has the same form as the thermofield double described in  \nref{TFDSol}, which in our variables becomes 
     \be \la{TFDtsol}
     t(\tilde u) = 2 \arctan\left[ \tanh\left({ \pi \tilde u \over \tilde \beta}\right) \right] ~,~~~~~~~
      \varphi = \log t' = \log\left[ {2 \pi \over \tilde \beta \cosh { 2 \pi \tilde u \over\tilde  \beta } } 
     \right] ~,~~~~~{\tilde u } > 0 ~, 
     \ee
     (see \nref{RepTwo}). We can check that this is a solution of the theory with $\eta=0$.  At $\tilde u$ equal to zero it matches 
      the constant $\varphi$ solution given by 
     \nref{tPrime}, which sits at the minimum of the potential in figure \ref{Potential}, provided that 
     \be \la{TemMa}
      t' = { 2 \pi \over \tilde  \beta }  ~,~~~ {\rm or} ~~~~~~ { d t \over d u } = { 2 \pi \over \beta } 
      ~,~~~~~~~{\rm where} ~~~~ \tilde \beta = { \cal J \over \alpha_S} \beta = { N \over \phi_r} \beta ~,
      \ee
      where $\tilde \beta$ is the rescaled inverse temperature. 
      We have indicated how to translate 
      to SYK or gravity parameters.  This also gives a new physical interpretation for $t'$.  Up to a factor of $2\pi$, 
      $t'$  is an effective (rescaled) temperature. It is the temperature of the TFD state that most closely resembles the vacuum of the coupled system. 
      Notice that the physical temperature of the coupled system is zero, when we discuss the ground state of the coupled system, as we are doing here. 
      
\begin{figure}[h]
\begin{center}
\includegraphics[scale=.4]{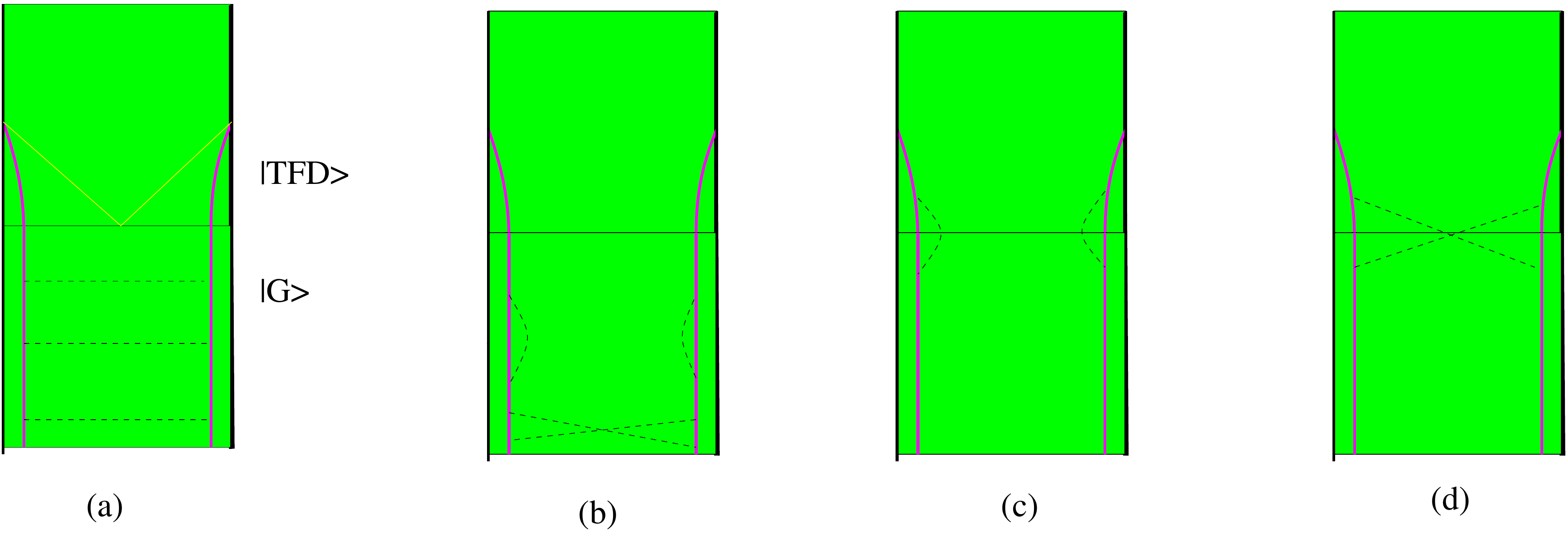} 
\caption{  (a) We turn off the interaction hamiltonian at $u=0$. Then the boundary trajectories look like those of the thermofield double state after that time. We see that at zero time, the ground state of the coupled system is very close to the thermofield double state of the decoupled system. We also display the diagrams that are summed over when we make the approximation 
in   \nref{Expec}. (b) some of the diagrams not included in \nref{Expec}.  
 (c) and (d) are the two   diagrams that compute the leading corrections to the overlap of the two states
 in section \ref{SecOver}.  }
\label{Overlap}
\end{center}
\end{figure}

     In summary, the solution sits at a constant value of $\varphi$ given by \nref{tPrime} for $u< 0$. Immediately 
      after $u=0$, the $\eta$ term in the potential \nref{PotV} disappears, $\varphi'$ is zero, but $\varphi''$ is non-zero. This implies that 
      the third derivative of $t(u)$, $t'''(u)$,  jumps at $u=0$, while all lower order derivatives are continuous. 
      This precise value of the jump can also be obtained by analyzing the original equations that come from   \nref{fLfRres}   around
      $\tilde u=0$, with an $\eta$ that is time dependent. See figure \ref{Overlap}(a). 
      
      Note that the energy after the transition is 
      \be \la{EnThAf}
      E_{\tilde u }   =N  t'^2 = N { (2 \pi)^2 \over \tilde \beta^2 } 
      \ee
      which is equal to the internal energy of the two copies of the thermal state, see \nref{EnThr}. 
      This is not the same as the energy \nref{GrstEn}, which is measured with respect to a different Hamiltonian, one with $\eta \not =0$. 
      On the other hand, if we subtract the expectation value of this extra term, we get the same as in \nref{EnThAf}
      \be
      E_G   + g N  \langle O_L O_R \rangle = E_G  + \eta N   {t'}^{\, 2 \Delta} = N  t'^{\,2}
      \ee
      after we use \nref{tPrime} and \nref{EnVph}.
      
      In addition, we can imagine that we take the TFD at inverse temperature $\check \beta $ as a variational ansatz for the interacting Hamiltonian. Then the 
      variational energy can be written as  the sum of two terms 
      \be \la{VarEn}
      E_{\rm var} /N = \left( {2 \pi \over {\check \beta } } \right)^2 + \langle H_{\rm int} \rangle =  \left( {2 \pi \over {\check \beta } } \right)^2 - 
      \eta \left( { 2 \pi \over \check \beta  } \right)^{2\Delta } 
      \ee
      The first is the energy of the two decoupled systems, and the second is the expectation value of the interaction term. 
      Minimizing the energy over the choice of $\check \beta $ we find that we get the same equations as before, $\check \beta = \tilde \beta$, 
       (see Eqs. \nref{TemMa} and \nref{tPrime}). The variational energy \nref{VarEn} at the minimum is the same as the ground state energy \nref{GrstEn}.

      Notice that this is also saying that the entanglement entropy between the left and right systems, for the ground state of the coupled system, 
       is equal to the entropy in the TFD situation, 
      which is the black hole entropy, or the entropy of a single SYK model at temperature given by \nref{TemMa}, \nref{EnThr}. 
                
      \subsection{Computing the leading correction to the overlap } 
      \la{SecOver}
      
      In this subsection we compute the leading correction to the statement in the previous section. We will see that the state resulting from the quench, 
      resulting from setting the interaction to zero at $\tilde u =0$, is not exacly the TFD. We will do so by computing the leading corrections to the inner product 
      between the ground state of the interacting model and the thermofield double. These go beyond the approximation described by \nref{fLfRres}. 
      
      If we denote by $|G\rangle$ the ground state of the coupled system and by $|TFD\rangle$ the usual thermofield double state, then we can compute
       the overlap between the two states 
      \be \la{Ovlp}   \Omega = { |\langle TFD| G\rangle |^2 \over \langle TFD | TFD\rangle \langle G| G \rangle } 
      \ee
      In the approximation where we describe the system using \nref{fLfRres}, and we treat it as in subsection \nref{MatchingTFD}, we find that the overlap is one, so that the
      two states are equal. 
      The reason is that we have an action that is local in time, and we have a solution which is piecewise the same as the solutions we had for the states in the denominator, 
      therefore we get a cancellation of terms between the numerator and the denominator. 
      %
       In this approximation,  we are treating the interaction as in \nref{Expec}, 
       which corresponds to summing diagrams likes the one in figure \ref{Overlap}(a). 
       The treatment in subsection \nref{MatchingTFD} corresponded to  coupling the gravity   modes by performing a reparametrization of these 
       correlation functions. 
       In this approximation we are neglecting the particle creation that occurs when we suddenly turn off the coupling at $u=0$. 
       We can include these effects by considering the effect of further diagrams, such as the ones in
        figure \ref{Overlap}(b). These diagrams are not local in time, they  involve an integral over two times. 
        When we turn off the coupling,  some of these diagrams 
       do not cancel between numerator and the denominator in \nref{Ovlp}. In fact, using the overall time translation symmetry, we see that the diagrams that do not cancel are precisely the two diagrams in figure \ref{Overlap}(c,d). 
       These diagrams appear in the denominator but not in the numerator. 
       With our interaction involving $N$ fields such as $\sum_i O^i_L O^i_R$,
        such diagrams give only a single factor of $N$. 
        Note that the effect of these diagrams is {\it not} includded in the action \nref{fLfRres}, they are higher order effects 
        in the small coupling that couples the two system. Here we simply evaluate those effects. We get a slightly different answer depending on whether the operators that couple the two boundaries are bosons or fermions.  
       Their final contribution looks like 
       \bea 
       \log \Omega  &=& 
       - N  \eta^2 {t'}^{ 4 \Delta}    \int_{0}^\infty d\tilde u \int_{-\infty}^0 d\tilde u' \left[ \pm { 1 \over [\cosh (t' (\tilde u -\tilde  u')/2 ]^{4 \Delta}  }+  { 1 \over [\sinh (t' (\tilde u - \tilde u')/2 ]^{4 \Delta} }  \right]
  \notag
       \\ &=& - N  \eta^2   {t'}^{ 4 \Delta -2 }   h_\pm(\Delta)= - N {t'}^2 { h_\pm(\Delta) \over \Delta^2 } 
       %
    %
      \la{LogF}  \\ \notag 
       &~& ~~~~~~{\rm where} ~~~h_\pm(\Delta ) = 4 \int_0^\infty dx x [ \pm \cosh^{-4\Delta} x + \sinh^{-4\Delta}x ] ~~~~~~~~~
       \eea
       where the $+$ corresponds to the case that the operators on each boundary are bosons, while the $-$ corresponds to the case that they are fermions (as in the two coupled SYK models we discussed in section \ref{SYKSection}). 
       The formula \nref{LogF} 
       includes the effects of particle creation. When we can create particles the probablity that we end up with the unexcitated TFD state is smaller, and for that reason we get   $ \Omega < 1$.  As $\eta $ becomes small, this correction becomes smaller. Notice that it is still of order $N$. 
       The integral in \nref{LogF} converges for $ 0 < \Delta < \half$. As $\Delta \to \half$ we find that \nref{LogF} diverges. $\Delta =\half$ corresponds to a conformally coupled field in the bulk and the boundary interaction is marginal. Turning off the coupling suddenly then leads to divergent particle production in the UV. 
       For the case of fermions $h_-(\Delta ) \sim 7 \zeta(3) \Delta $ as $\Delta \to 0$, while for bosonic operators $ 2 h_+(\Delta) \sim  1/\Delta^2$ for $\Delta \to 0$.

       \subsection{Finite temperature  } 
        
        \la{SecFTlow}
       
    We can now consider the coupled system at finite temperature. This finite physical temperature should not be confused with the effective temperature of the 
       thermofield double that we discussed above near \nref{TemMa}. We will denote the physical inverse temperature of the coupled system by $\tilde \beta_{ph}$.
       This is the period of the physical (rescaled as in \nref{RelPar}) time $\tilde u$, $\tilde u \sim \tilde u + \tilde \beta_{ph}$. 
       
       \subsubsection{Lower temperatures}
       \la{LowerTemp}

       In the $AdS_2$ picture, we expect that for very low temperatures we should be obtaining a solution with a periodically indentified euclidean $AdS_2$,  where we indentify the Euclidean $AdS_2$ global time coordinate, $t\sim t+ \beta'$. We will adjust $\beta'$ momentarily.  We imagine that we can impose a similar condition on the conformal solution of the SYK model. 
       We then impose  a periodicity condition on the Schwarzian variables 
       \be t_l(\tilde u + \tilde \beta_{ph} ) = t_l(\tilde u) + \beta' ~,~~~~~~ t_r(\tilde u + \tilde \beta_{ph}) = t_r(\tilde u ) + \beta' 
       \ee
       We continue to consider solutions with $t_l = t_r$ and with constant derivative, which requires $t_l = t_r = { \beta' \over \tilde \beta_{ph} } \tilde u $. 
       We can evaluate the euclidean action on these solutions to find 
       \be \la{ActVa}
         - S_E/N  = \tilde \beta_{ph} \left[ -(t')^2 + \eta (t')^{2\Delta } \right] + \log Z_{\rm bulk}(\beta')  ~,~~~~~~~~t' = { \beta' \over \tilde \beta_{ph} } 
       \ee
       where the first term comes from the classical Schwarzian action \nref{fLfRres} and the last term is the partition function of a single bulk field, where we imagine that
       we have $N$ bulk fields. In the SYK this would be the partition function over the conformal sector obtained as the appropriate sum over 
        representations with the thermal weights given by $\beta'$. 
         We can now view $\beta'$ as a variational parameter and minimize the action \nref{ActVa} with respect to $\beta'$ to obtain the equation
       \bea
       0 &=& - 2 t' + 2 \eta \Delta (t')^{ 2 \Delta -1} - \epsilon(\beta')  ~,~~~~\epsilon(\beta') = - \partial_{\beta'} \log Z_{\rm bulk}
       \eea
       This is also the equation we would obtain in the Lorentzian theory when we set the $SL(2)$ charge $Q_0$ to zero. Here $\epsilon(\beta')$ is the
        finite temperature energy of the bulk fields at  inverse temperature $\beta'$. At low temperatures,  $\Delta \beta'\gg 1$, the energy is approximately 
        $\epsilon \sim \Delta e^{ -\Delta \beta' }$, so that
        \bea
        0  &=& -2 t' + 2 \eta \Delta (t')^{ 2 \Delta -1} -\Delta e^{ -\Delta \beta'}  ~,~~~~~~~ t' = { \beta' \over \tilde \beta_{ph} }  \la{finTequl}
       \eea
    
\begin{figure}[h]
\begin{center}
\includegraphics[scale=.55]{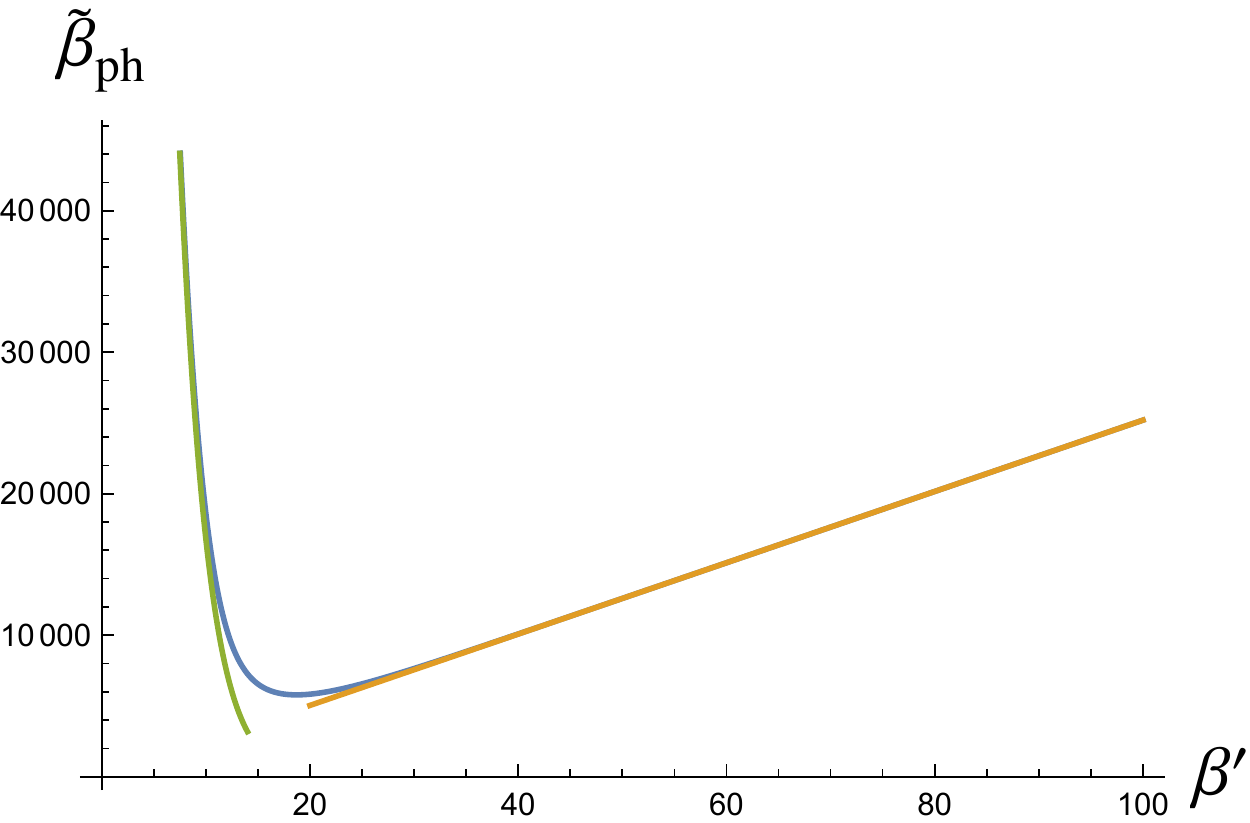}  ~~~~~~~~~~\includegraphics[scale=.55]{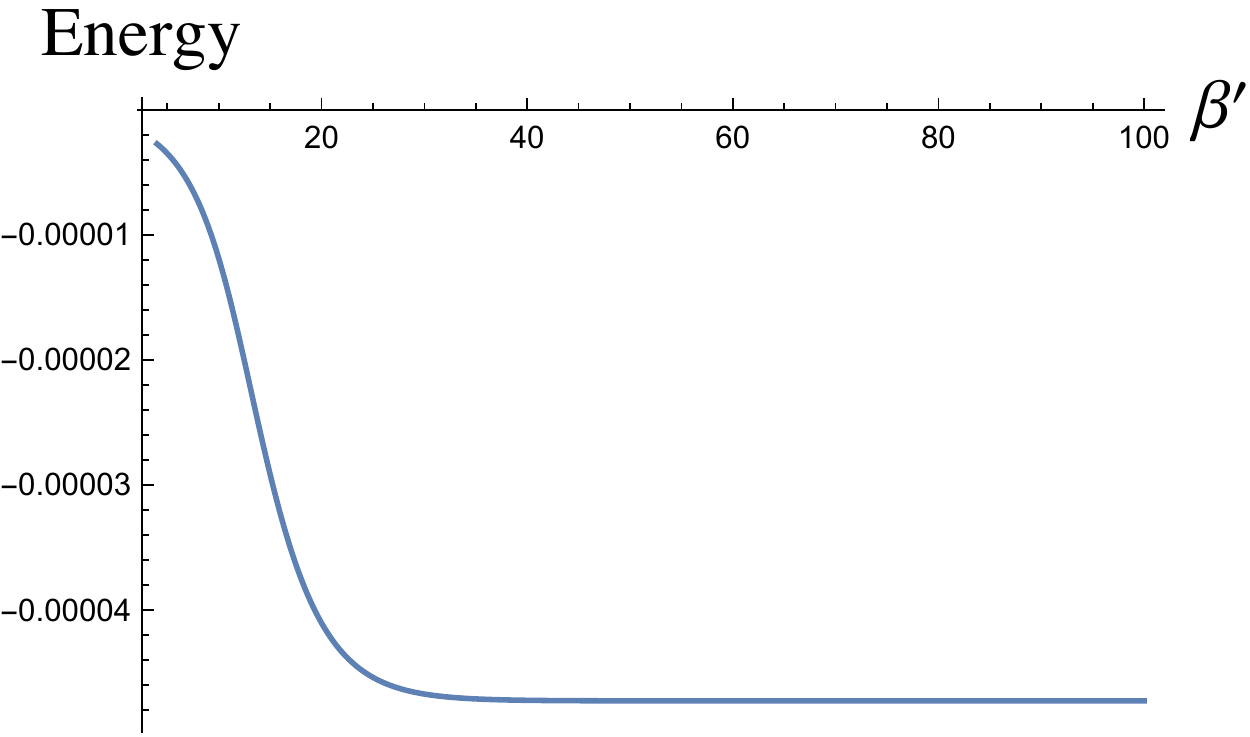} 
\\
~~(a) ~~~~~~~~~~~~~~~~~~~~~~~~~~~~~~~~~~~~~~~~~~~~~~~~~~~~~~~~~~~~~(b)
\caption{ (a) We plot the relation between the physical temperature $\tilde \beta_{ph}$ and $\beta'$, which is 
the period of the global euclidean time coordinate $t$.  $\beta'$  
can be viewed as the periodic identification of the global Euclidean time coordinate in the $AdS_2$ space, in units of the $AdS_2$ radius. In blue we have the numerical solution of \nref{finTequl} for 
$\Delta =1/4$ and $\eta = 10^{-3}$. In orange we plot the approximate solution $\beta'_+$ \nref{Fsolt} and in green the approximate solution  $\beta'_-$ 
\nref{SecSol}. We see that there exists a minimum value for $\tilde \beta_{ph}$ (or  maximum value for the temperature) for the solution to exist. For very 
low values of $\beta'$ we stop trusting the approximations leading to \nref{finTequl}. (b) We plot the energy as a function of the parameter $\beta'$. Notice that 
for a given $\tilde \beta_{ph}$ we have two values of $\beta'$ and two energies. The energy increases monotonically as $\beta'$ decreases. The 
asymptotic value for large $\beta'$ is just the zero temperature energy \nref{GrstEn}.   }
\label{LowTemperature}
\end{center}
\end{figure}

        From \nref{finTequl} we can solve for $\beta'$. The interesting aspect is that there is more than one solution, and that the solution exists only for $\tilde \beta_{ph}$ sufficiently large. See figure \ref{LowTemperature}(a).

  We can find approximate solutions to \nref{finTequl}  as follows.   Here we will analyze only the case with $\Delta < \half $. We also  need small $\eta$, \nref{Valcl}. 
       In the first solution, we neglect the third term in \nref{finTequl} and reobtain 
       \nref{Qalph} \nref{tPrime} 
       \be \la{Fsolt}
       (t')^{ 2 (1-\Delta) } = \eta \Delta \longrightarrow  \beta'_+ = t'  \tilde \beta_{ph} = ( \eta \Delta)^{ 1 \over 2 (1-\Delta) } \tilde  \beta_{ph} 
       \ee
       In the second solution we neglect the first term in \nref{finTequl} and obtain 
       \be \la{SecSol}
       (2 \eta)^{ 1\over (1- 2 \Delta ) } \tilde \beta_{ph} = \beta'_- e^{ - \Delta \beta'_-/(1- 2 \Delta) } 
       \ee
       We have denoted the  two approximate solutions by $\beta'_+$ and $\beta'_-$. In order to be consistent with the approximations we need  $\eta \tilde \beta_{ph}^{1-2\Delta} \ll 1$ for the second 
       solution \nref{SecSol} and we used this approximation for the second expression.
       Therefore we have two solutions within the regime 
       \be 
       \eta ^{ - { 1 \over ( 2 - 2 \Delta) } } \ll \tilde \beta_{ph} \ll  \eta^{ - { 1 \over ( 1 - 2 \Delta) } } 
       \ee
       Of course the first solutions \nref{Fsolt} exists also for arbitrarily large $\tilde \beta_{ph}$. 
       The two solutions merge in the lower range of $\tilde \beta_{ph}$, but for the precise value    we need to 
       consider the full equation and solve it numerically, which is shown in figure \ref{LowTemperature}(a). 
       It can be checked by taking the second derivative of the action \nref{ActVa} with respect to $\beta'$ that 
       that the $\beta'_+$ branch is stable and the $\beta'_-$ in unstable in the cannonical ensemble. It is interesting to compute the energy along the 
       whole curve, using \nref{EnGenN}, or $E/N = -t'^2 - \eta(1-2\Delta){t'}^{2\Delta }$, with $t'$ obeying \nref{finTequl}. We find that it varies monotonically, see figure \ref{LowTemperature}. This suggests that if we have an isolated system and we increase the energy 
       in the microcannonical ensemble, then we can explore the solutions with small $\beta'$ which were unstable in the canonical ensemble.

     \subsubsection{Higher temperatures} 
       
      \la{HigherTemp}
       
\begin{figure}[h]
\begin{center}
\includegraphics[scale=.55]{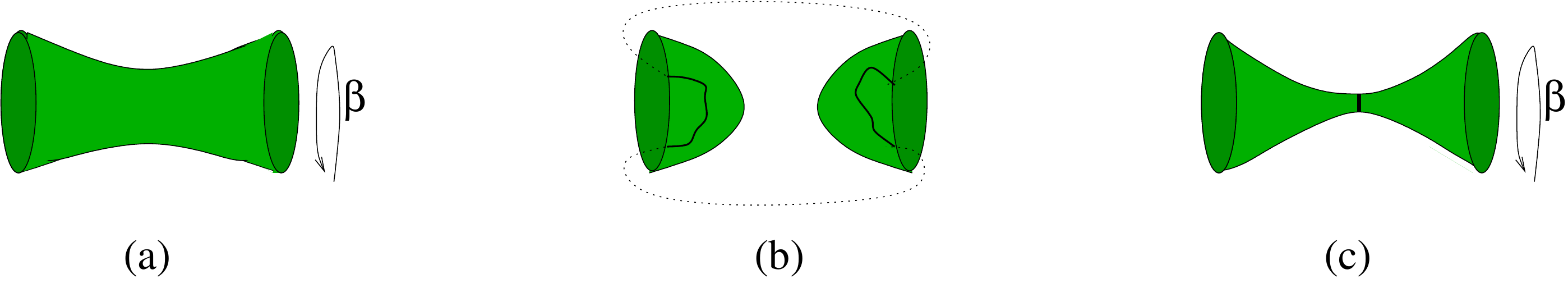} 
\caption{  Here we display three branches for a given temperature. The physical temperature is the same for all three, and it is represented by the size of the circle at boundary.  
 (a) In the low temperature phase we have a geometry that connects the two sides. The thermal circle is not contractible, and there are almost no bulk excitations. 
(b) The high temperature phase contains two separate geometries, each with a contractible euclidean time circle. 
 The leading coupling between two is given by the diagram displayed, which is proportional to 
$\mu^2$. The dotted lines indicate that the times are equal. The solid lines indicate two point functions. Finaly in (c) we display the phase that is unstable in
the cannonical ensemble that we discussed in section \ref{LowerTemp}, which can be interpreted as $AdS$ with global time identified and a small amount of thermal 
excitations in the bulk. The topology is the same as for (a), but the minimal size of the Euclidean time circle in the interior is smaller, so that we  are thermally 
exciting   the bulk modes a bit (represented by the black  line at the neck). 
The physical temperature from the boundary point of view is the same as in (a).      }
\label{TwoPhases}
\end{center}
\end{figure}

       If we consider the coupled system at sufficiently high temperatures, we expect that the Euclidean space solution corresponds to two separate Euclidean black holes. This is {\it not } described by the action \nref{fLfRres}, but it can 
       still be described using conformal methods applied to the two decoupled black holes. We have the results corresponding to two decoupled thermal systems, each computed using the answers in the nearly conformal approximation. These are corrected to leading order by the diagrams in figure \ref{TwoPhases}(b). 
   We then find    
       \be \la{HigTg}
       \log Z = 2 S_0 + N  { (2 \pi)^2  \over \tilde \beta_{ph} } + N  {  \eta^2    { \tilde  \beta_{ph}}^{ 2- 4 \Delta  }   } \int_0^1 d x \left[ { \pi \over \sin x \pi} \right]^{4\Delta } 
       + \cdots 
       \ee  
       where $S_0$ is the ``ground state'' entropy of each SYK model or  $AdS_2$ factor. In $AdS_2$
       $S_0 = 2 \pi \phi_0$ and it arises from the topological term in \nref{ActJT} when the Euclidean 
       solution has the topology of the disk, as in figure \ref{TwoPhases}(b). There is no contribution
        when the topology is that of a cylinder, as in figure \ref{TwoPhases}(a). 
       We see that the correction in \nref{HigTg}  is small for small $  \eta$. 
       On the other hand, the low temperature form of the partition function is dominated by the ground state energy \nref{GrstEn}
       and is of the form 
       \be \la{LowTg}
       \log Z = - N \beta_{ph} E_{G \, u}  +  N \exp\left(  - \beta_{ph} \Delta { dt \over du } \right) + \cdots 
       \ee
       where we have also included the leading correction from weakly exciting the particles mentioned in \nref{ConfEn} and we assume that there are $N$ of
       them, as in the SYK model. Both \nref{HigTg} and \nref{LowTg} assume that we are at temperatures much smaller than ${\cal J}$. We have also subtraced a common 
       contribution to the ground state energy, which in the SYK model is proportional to $ - N {\cal J }$.

       In the cannonical ensemble, we expect to get a first order phase transition between these two phases. The most notable aspect of this high temperature phase is 
       its entropy $2S_0$. While in the low temperature phase we have small entropy but we have negative energy \nref{GrstEn}. 
       We expect the transition temperature, $T_c$,  to be approximately given by the equation  
       \be \la{Nft}
        2 S_0 = - E_G/T_c  ~,~~~~~ \beta_c = { 2 S_0 \over - E_G } 
        \ee
  It is possible to check that the temperature we obtain in \nref{Nft} is well within  the regime of approximation of both 
  \nref{HigTg} and \nref{LowTg}, at least for $\Delta $ not
  too small\footnote{ 
  This is checked as follows. First to check \nref{LowTg} we compute 
  $\tilde \beta_{c } \Delta t' \propto { \Delta^2 \over t' } $ which is very large and so the exponential corrections in \nref{LowTg} are small. 
  Similarly, we can check that $\eta^2 { \tilde \beta}^{ 2 - 4 \Delta } \propto ( t'/\Delta)^{ 4 \Delta }$, which is small if \nref{Valcl} holds. For very small $\Delta$ this 
  last expression becomes of order one, when we scale parameters as in secion \ref{LqZT}.}.
    In other words, \nref{Nft}, is correct in the limit of fixed $\Delta$ and very small $\eta$. 
   We will see in section \ref{FinTLq}, that for very large $q$  the transition happens beyond the reach of the approximation in \nref{HigTg}. 
       
       Then we conclude that, for small $\eta$, 
        we can describe the physics on both sides of the transition by using Schwarzian type variables. It is important to notice that the Schwarzian variables we use at 
        low or high temperatures  are
       not the same! The are based on different underlying conformal states for the rest of the fields. In the gravity description, they are based on different underlying 
       geometries, as figure \ref{TwoPhases}. 
                   
          On the other hand,  we  can consider the microcannonical ensemble and start increasing the energy above the vacuum.  
          As we do so, we introduce  bulk particles and  we need
          to go beyong the pure gravity, or pure Schwarzian, approximation. This can be done in the SYK model by solving the large $N$ equations of the coupled system.
           We will dicuss this in more detail in section \ref{SYKAny}.

                \subsection{Some properties of correlation functions } 
                
                  Here we consider the two point functions of a  field ${\cal O}$ that propagates in $AdS_2$. We can also think of it as one of the fields 
                  that we are using in the interaction that produces the solution. In the SYK model we can think of ${\cal O}$ as one of the basic fermions, or any other 
                  conformal operator with definite conformal dimension in the conformal limit. 
                  
\begin{figure}[h]
\begin{center}
\includegraphics[scale=.4]{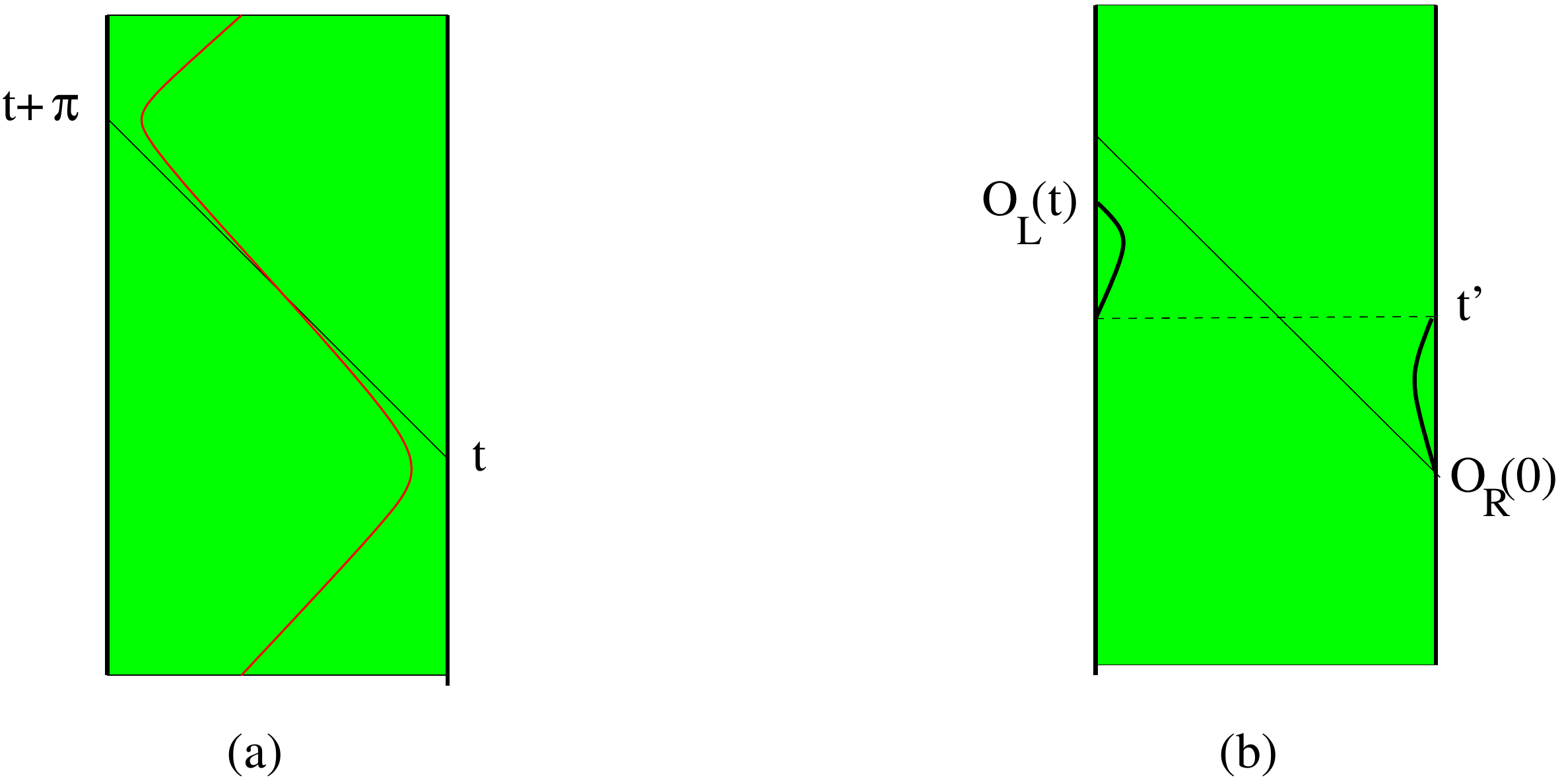} 
\caption{ (a) Trajectory of a particle moving in $AdS_2$. In the conformal approximation, given  by \nref{CorrAdS}, there is a 
singularity at $t_r- t_l = \pi $. (b) Diagram contributing to the commutator between left and right operators  before the operator 
on the left  crosses the bulk  light cone of the operator on the right. The black lines in the bulk represent propagators. Here we drew $AdS$ diagrams but the same 
properties hold for the two coupled SYK models. 
 }
\label{Singularities}
\end{center}
\end{figure}

                  The two point functions for this field on the same side and on different sides are given by 
                  \be \la{CorrAdS}
                  \langle {\cal O}_L(\tilde u_1)  {\cal O}_L(\tilde u_2) \rangle =e^{ - i \pi \Delta } 
                   \left[ {   t' t' \over  \sin^2  { t' (u_1-u_2 - i \epsilon ) \over 2 }  } \right]^{   \Delta } ~,~~~~~~~~
                  \langle {\cal O}_L(\tilde u_1)  {\cal O}_R(\tilde u_2) \rangle = \left[ {   t' t' \over   \cos^2  { t' (\tilde u_1- \tilde u_2 ) \over 2 }  } \right]^{   \Delta }  
                  \ee
                  in Lorentzian signature. Note that the left-left correlator has singularities at $t' \tilde 
                  u_{12} =0, 2 \pi, 4 \pi $. While the left right correlator has singularities
                  at $t' \tilde u_{12} = \pi , \pi + 2 \pi , \cdots $. These singularities correspond to the ones expected for light-rays that start from one of the operators and 
                  bounce back and forth in $AdS_2$, see figure \ref{Singularities}(a). They should be regulated by the appropriate $i \epsilon$ prescription. Furthermore, all these singularities,  are
                    regulated by UV effects, namely effects that take us away from exact conformal symmetry, already to the leading order in the large $N$ limit.  In the SYK model these kick in at values of $\tilde u$ which differ by an order one amount (or an order $1/{\cal J} $ in $u$) from the naively singular one.  
                      We will see this explicitly when we analyze the SYK model at large $q$ in section 
                     \ref{LqZT}.  Finally, gravitational 
                  back reaction effects should also smooth out these singularities for the following reason. The singularities arise when we exchange high energy particles. This exchange back-reacts on the gravitational degrees of freedom, which moves the boundary trajectories and smears out the insertion of the other particles. Similar effects were also discussed in the analysis of traversable wormholes in  \cite{Maldacena:2017axo}.

                  Notice that the left-right correlators reflect the fact that we can send signals from one system to the other. Of course this is not surprising since we are coupling them. However, as in the traversable wormhole situation, it is interesting how the the signal is moving between the boundaries. The interaction is setting up an eternal-traversable wormhole. A particle starts near one boundary, living mostly in the first system and then moves to the other side, living mostly in the second system. This is 
                  somewhat reminiscent of two weakly coupled oscillators where the excitation moves between one oscillator to the other, except that here it is doing this through the bulk geometry.  Notice that the same property holds in the two coupled SYK models!

  Note also the following feature. Let us concentrate on the coupled SYK models in \nref{Hint}. Let us first set the SYK couplings to zero, 
  ${\cal J } =0$, but keeping a non-zero mass-like coupling $\mu$. In this case, a fermion on one side will move to the other side after a time
  of order $\Delta u_0=1/\mu$. This is just a simple free fermionic harmonic oscillator. Now, let us  turn on  a relatively large value of ${\cal J} \gg \mu$ so 
  that the low energy discussion in this section applies. 
  Then the time to go accross is $\pi$ in time units measured by $t$.  Via $t'$, \nref{tPrime}, this translates into a boundary time of the 
  order of $ \Delta u \propto 1/t'  $. Comparing this time to the time $\Delta u_0$ in the free theory, we get 
  $\frac{\Delta u}{\Delta u_0} \propto    ({ \mu \over {\cal J } })^{ 1-2\Delta    \over 2 (1-\Delta ) } $, which vanishes in small $\mu$ limit if $\Delta < 1/2 $. This means that the 
 self  interactions, within each SYK model,  accelerate the transfer of information from the left to the right system, relative to the non-interacting system. Of course, for this increase in the information transfer rate we need that the microscopic random couplings of the two models are correlated.

                  Note that the second correlator in \nref{CorrAdS} is such that 
                  the commutator (or anti-commutator, for fermionic operators)  between the left and right fields is exactly zero 
                  for times ${ | t' (u - u')|   } < { \pi } $. In other words, 
                  \be \la{CommZe}
                     \langle [ O_L( \tilde u) , O_R(0) ]_{\pm} \rangle \propto  \sin( 2 \pi \Delta) \left[ { t' \over - \cos( { t'  \tilde u \over 2 } ) } \right]^{2 \Delta } \theta( t' \tilde u -\pi ) ~,~~~~~~~0 < t' \tilde u < 3\pi  ~,~~~~~
                  \ee 
                  This is true to leading order in the $ \eta $ expansion. 
                  However, there are some diagrams that contribute to the commutator which are of the form displayed in figure \ref{Singularities}(b). 
                  This gives an extra term in the commutator (or anticommutator for fermion fields)  (for dimensionless times $t< \pi$) of the form 
                  \be \la{CommEta}
                  \langle [ O_L(t) , O_R(0) ]_{\pm} \rangle \propto \eta {t'}^{ 4 \Delta -1}  ( \sin { 2 \pi \Delta } )^2 \int_0^t {d \tau  \over [ \sin{ \tau \over2 } \sin {(t- \tau) \over 2 } ]^{2\Delta } } 
                ~,~~~ 0 < t < \pi 
                  \ee
                  here we assume that ${\cal O} $ is one of the fields appearing in the interaction term. We see that
                   there is an extra suppression due to an extra factor relative to \nref{CommZe} of  $\eta \,  {t'}^{2 \Delta -1} \propto t'  \ll 1 $  (see \nref{Valcl}). 
                  However, it is the leading non-zero contribution in the time range in \nref{CommEta}
                  
 On the other hand, for larger times, we already get a non-zero commutator \nref{CommZe}. 
        
 \subsection{The three generators of SL(2) } 
 
 We have mostly discussed here the ``gravitational'' degrees of freedom. We have also mentioned  ``bulk'' degrees of freedom that transform in 
 non-trivial representations of SL(2). These are present both in the gravity case and in the SYK situation. These degrees of freedom transform 
 under SL(2) representations with generators $q_a$ that act purely on them. 
 When these degrees of freedom are added to the ``gravitational'' sector \nref{fLfRres} we need to modify the constraints so that now they read 
 \be
 Q_a + q_a =0
 \ee
 where $Q_a$ are the SL(2) charges of \nref{fLfRres}. We discuss a bit more how this works in appendix \ref{SLTwo}. 
 
 We have the  time translation symmetry of the full sytem, generated by the total Hamiltonian $H_{\rm tot} = H_L + H_R + H_{\rm int}$ in the 
 coupled SYK systems. This symmetry acts, up to a redshift factor,  like the generator $q_0$ on the conformal degrees of freedom, 
 \be
 H_{\rm tot} - E_{G \, u } \sim  { dt \over du } q_ 0 + o (1/N)   \la{QzDef}
 \ee
 where we have subtracted the ground state energy. It 
  also follows from the discussion in section \nref{MatchingTFD}, where we matched the ground state to the TFD state, 
  that at $u=0$ the difference of Hamiltonians
 \be \la{Boostc}
 H_R - H_L |_{u=0}  \sim { dt \over du } { (q_+ + q_-) \over 2 } + o (1/N) 
 \ee
 acts like the boost generator at $u=0$. This statement has also corrections due to the particle creation discussed in section \nref{SecOver}, which we will ignore, 
 assuming that we take a limit of very large $N$ and very small $\eta$.  
 The operator on the left is time dependent in the full coupled system.  It does not commute with $H_{\rm tot}$. 
 It turns out that the third SL(2) generator corresponds to a similar operator but at another time (see appendix \ref{SLTwo} for more details)
 \be
 H_R - H_L |_{u = { \pi \over 2 }{ du \over d t } } = e^{ i T H_{\rm tot} } (H_L - H_R )e^{ - i T H_{\rm tot} }    \sim { dt \over du } { (q_+ - q_-) \over 2i } + o (1/N)
    ~,~~~~~~T = { \pi \over 2 } { d u \over dt }  \la{Transl}
       \ee        
  The time shift corresponds to a shift by $\pi/2$ in the IR time $t$. Formula \nref{Transl} is dervied as follows. 
 We could run the 
  argument leading to  \nref{Boostc} but at time $t=\pi/2$. The operator on the right hand side is the boost generator around the bulk point 
  $\sigma =0, ~ t= \pi/2$ in coordinates \nref{Coordinates}. An alternative local expression for the third generator 
   (not equal to \nref{Transl}), arises from taking the commutator between \nref{QzDef} and \nref{Boostc}. 
  
  In conclusion, we have identified three operators \nref{QzDef}, \nref{Boostc} and \nref{Transl} that are completely well defined in the 
  boundary theory. We have argued that these operators act like the three generators
   of SL(2) on the conformal infrared degrees of freedom of the theory. We have derived this indirectly. It would be nice to derive this more
   directly in the SYK model. In particular, one would like to understand their commutation relations, and their $1/N$ corrections. 
   However, one can easily verify that in coupled SYK model the microscopic operators $H_{\rm tot}-E_{Gu}$ and $H_R-H_L$, and their commutator, do not form a closed SL(2) algebra. This is expected since the SL(2) symmetry only emerges for low energy states. It woud be interesting to learn how to compute the effective commutation relations in the low energy subspace so as to verify the approximate SL(2) symmetry. We will not do it in this work.

    \section{ The two coupled SYK models beyond the low energy limit } 

\la{SYKAny}

 \subsection{ Large $N$ equations } 
 
 In this section we study the large $N$ equations for the fermion two point function for the coupled system. 
 We study them in Euclidean time.  The effective action with collective variables $G$ and $\Sigma$ can be easily generalized to the coupled system with Hamiltonian \nref{Hint}.  One difference is that we now have left-left, left-right, etc,  correlators, $G_{LL}(\tau_1,\tau_2)$, $G_{LR}(\tau_1,\tau_2)$, etc. 
 The effective action is 
 \begin{eqnarray} \label{EfAct}
 -S_E/N&=&\log{\rm Pf}\left(\partial_\tau\delta_{ab}-\Sigma_{ab}\right)-\frac12\int d\tau_1d\tau_2\sum_{a,b}\left[\Sigma_{ab}(\tau_1,\tau_2)G_{ab}(\tau_1,\tau_2)-s_{ab}\frac{ {\cal J}^2}{2 q^2}[ 2 G_{ab}(\tau_1,\tau_2)]^q\right] + \nonumber\\
 & &+ { i \mu \over 2}  \int d\tau_1 \left[ - G_{LR}(\tau_1,\tau_1) + G_{RL}(\tau_1,\tau_1) \right] 
 \end{eqnarray}
 Here $a,b=L,R$ denotes the two sides. Note that the functions obey the antisymmetry condition $G_{ab}(\tau_1,\tau_2) = - G_{ba}(\tau_2,\tau_1) $. 
 Here $s_{ab}$ is a sign, $s_{LL} = s_{RR} =1$, $s_{LR} = s_{RL} =(-1)^{q/2} $, which arises because for odd $q/2$ the signs of the couplings in the left Hamiltonian
 are the opposite than those on the right Hamiltonian (with the same absolute value). 
 The equations we get from \nref{EfAct} are very similar to the ones in \nref{SDeqn}.  
  If we think of the $L,R$ and $\tau_1,\tau_2$ indices as one combined index, then the equations have the same structure as in \nref{SDeqn}. 
  The only difference is that there is an additional term, $\mu\delta(\tau_{1}-\tau_2 )$ in the expression for the left-right self energy $\Sigma_{LR}$. 
  Just to be more explicit, we can write some of the equations 
  \bea \la{SDCoup}
 & ~&  \partial_{\tau_1} G_{LL} -\Sigma_{LL} * G_{LL } -\Sigma_{LR} * G_{RL} = \delta
 \cr
 & ~&  \partial_{\tau_1} G_{LR} -\Sigma_{LL} * G_{LR } -\Sigma_{LR} * G_{RR} = 0 
 \cr
 & ~&  \Sigma_{LL} = { { \cal J}^2 \over q } ( 2  G_{LL})^{q-1} ~,~~~~~~~\Sigma_{LR} =(-1)^{q/2} { { \cal J}^2 \over q } (2  G_{LR})^{q-1} - i \mu \delta(\tau_{12} ) ~, ~~
 \eea
 where $*$ denotes a convolution, as in \nref{SDeqn}. 
 The other equations can be similarly listed. $\delta$ in the first equation represents $\delta(\tau_1-\tau_2)$. 
 
  For any solution, the energy can be computed by 
 \be \la{EGen}
 { E \over N } = \left[ { 1 \over q } \partial_{\tau_1} G_{LL} + { 1 \over q } \partial_{\tau_1} G_{RR} + i \mu \left(1 - { 2 \over q } \right) G_{LR} \right]_{\tau_{12} = 0^+}
 \ee
This formula is derived by noticing that 
\be
 N \partial_{\tau_1} G_{LL}|_{\tau_{12} = 0+} = \sum_i \langle \partial_\tau \psi_L^i \psi_L^i \rangle = 
\sum_i \langle [ H,\psi^i_L ] \psi^i_L \rangle =  \langle q H_L + H_{int} \rangle
\ee
 where the operators are all at the same time. The factor of $q$ comes from the fact that a given individual coupling in \nref{SingleSYK} contributes to $q$ terms in the sum over $i$. We  have a similar equation for $G_{RR}$. We can also express the expectation value of the 
 interaction Hamiltonian in terms of $G_{LR}$ to get to \nref{EGen}. Note that \nref{EGen} is an exact formula for the energy when we think of $N G$ as the sum of all
 the fermion correlators\footnote{By ``exact'', we mean that it is valid for finite $N$ and for definite values of the random couplings of the model.}.
 
 In order to solve the equations is it convenient to consider the system at finite temperature. Then the euclidean time is periodic. We can also assume that we have an 
 ansatz where all functions depend only on the difference of times. Then we expect that $G_{LL}(\tau)$ remains positive and is symmetric around $\tau = \beta/2$ and
 obeys $G_{LL}(0)=\half$. On the other hand, we expect that $G_{LR}(\tau)$ is purely imaginary and is antisymmetric around $\tau = \beta/2$.   This property is
 consistent with the fact that it should be anti-periodic under $\tau \to \tau + \beta$. The fact that it 
 is imaginary follows simply the the factors if $i$ in \nref{SDCoup}.
 
 The  equations \nref{SDCoup}  can be analyzed numerically or analytically for large $q$. We discuss some numerical results in the next section. 
 
 \subsection{Numerical analysis } \label{sec:numerics}
 
  To gain more intuition, we solve the Schwinger-Dyson equation (\ref{SDCoup}) numerically by iteration.\footnote{There are numerical subtleties in the iteration procedure, as has been pointed out in Appendix G of Ref. \cite{Maldacena:2016hyu}. Here we use the same weighted iteration as in Ref. \cite{Maldacena:2016hyu}.} For numerical purposes, it is easier to work at finite temperature, where the imaginary time is periodic   $\tau=\tau+\beta$. Then $\tau$ is discretized as $\tau=\beta \frac{n}{M},~n=0,1,2,...,M$, where the integer $M$ determines the UV cutoff of frequency $\omega_{\rm max}=\frac{\pi}\beta M$.
   In our calculation we take $M=10^5$. To  describe correctly the continuous time physics, the discretization time scale $\frac{\beta}M$ needs to be much smaller than ${\cal J}^{-1}$ and $\mu^{-1}$, which requires $M\gg \beta {\cal J},~\beta \mu$. 

\begin{figure}[htbp]
\begin{center}
\includegraphics[scale=.25]{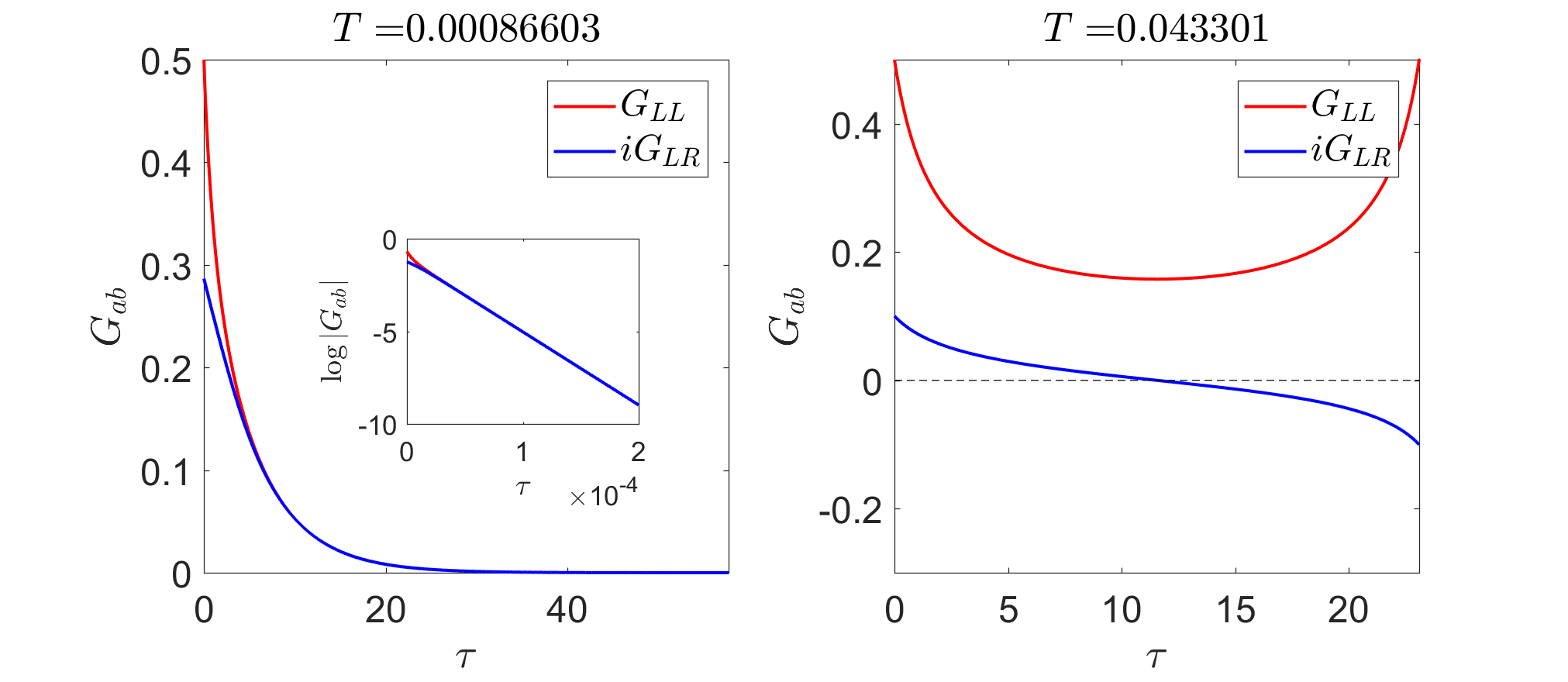} 
~~~~~~~~~~~~~~~~~~~~~~~~~(a)~~~~~~~~~~~~~~~~~~~~~~~~~~~~~~~~~~~~~~~(b) 
\caption{Numerical solution of $G_{ab}(\tau)$ at (a) low temperature and (b) higher temperature. The calculations are done for $\mu=0.075,~\mathcal{J}=1,~q=4$. (The low temperature case in (a) is also anti-periodic in imaginary time, but we have only shown the short time part for clarity.) Since $G_{LR}$ is purely imaginary, the imaginary part is plotted. The inset of panel (a) is a log plot of $\log\left|G_{ab}\right|$ which shows that the two-point function decays exponentially in time.}
\label{SDEfig1}
\end{center}
\end{figure}

\begin{figure}[htbp]
\begin{center}
\includegraphics[scale=.25]{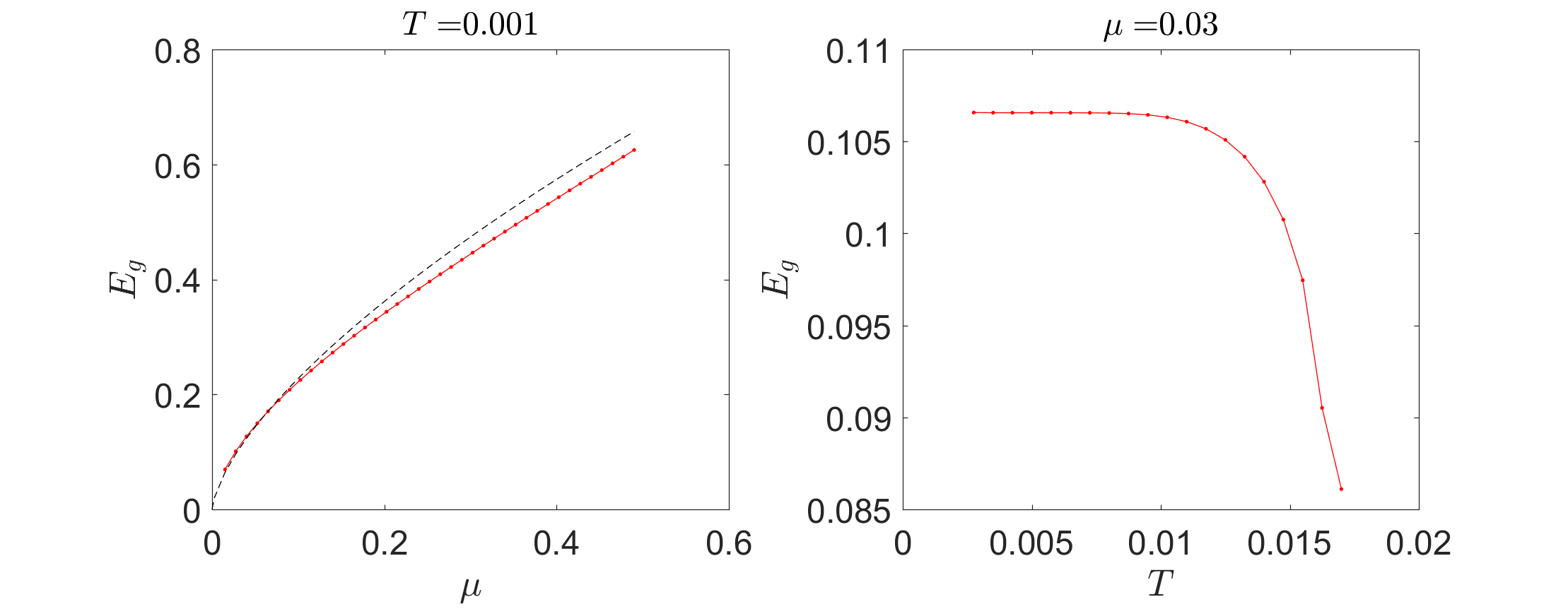} 
~~~~~~~~~~~~~~~~~~~~~~~~~~~(a)~~~~~~~~~~~~~~~~~~~~~~~~~~~~~~~~~~~~~~~~~~~~(b) 
\caption{The (effective) energy gap $E_{\rm gap}$ defined by exponential fitting $G_{ab}(\tau)\propto e^{-E_{\rm gap}\tau}$, as a function of (a) coupling $\mu$ and (b) temperature $T$. The dashed line in (a) shows a fitting to the power law behavior $E_{\rm gap}\propto \mu^{2/3}$. The calculation is done for $\mathcal{J}=1,~q=4$.}
\label{SDEfig2}
\end{center}
\end{figure}

\begin{figure}[htbp]
\begin{center}
\includegraphics[scale=.28]{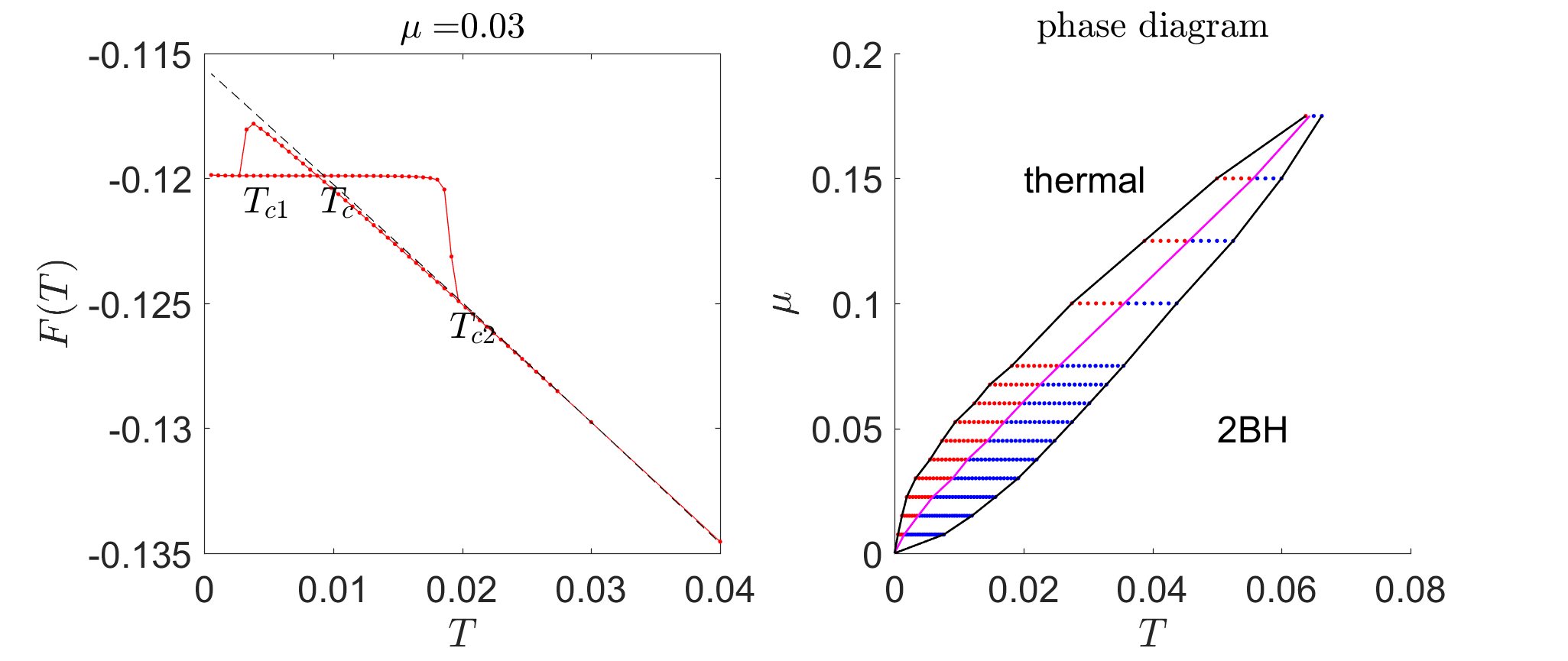} 
~~~~~~~~~~~~~~~~~(a)~~~~~~~~~~~~~~~~~~~~~~~~~~~~~~~~~~~~~~~~~~~~~~~~(b) 
\caption{(a) The free energy of the saddle point solution. Starting from the highest temperature, we decrease the temperature to lowest value and then increase it again back to the highest value. The solution at each temperature step is used as the initial condition of the next step. Two saddle points coexist in the region $T\in[T_{c1},T_{c2}]$, and the free energy of the two saddle points cross each other at some temperature $T_c$ in this region. The black dashed line is the free energy of decoupled two SYK sites, for comparison. (b) The phase diagram in $\mu-T$ plane obtained by calculating the free energy hysteresis curves for different values of $\mu$. The red (blue) dots are data points where the free energy of the high temperature saddle point is higher (lower) than that of the low temperature one. The black solid lines are the lower and upper critical temperatures $T_{c1},T_{c2}$, and the pink line is the thermodynamic transition temperature $T_c$. The calculation is done for $q=4, \mathcal{J}=1$. }
\label{SDEfig3}
\end{center}
\end{figure}

\begin{figure}[htbp]
\begin{center}
\includegraphics[scale=.28]{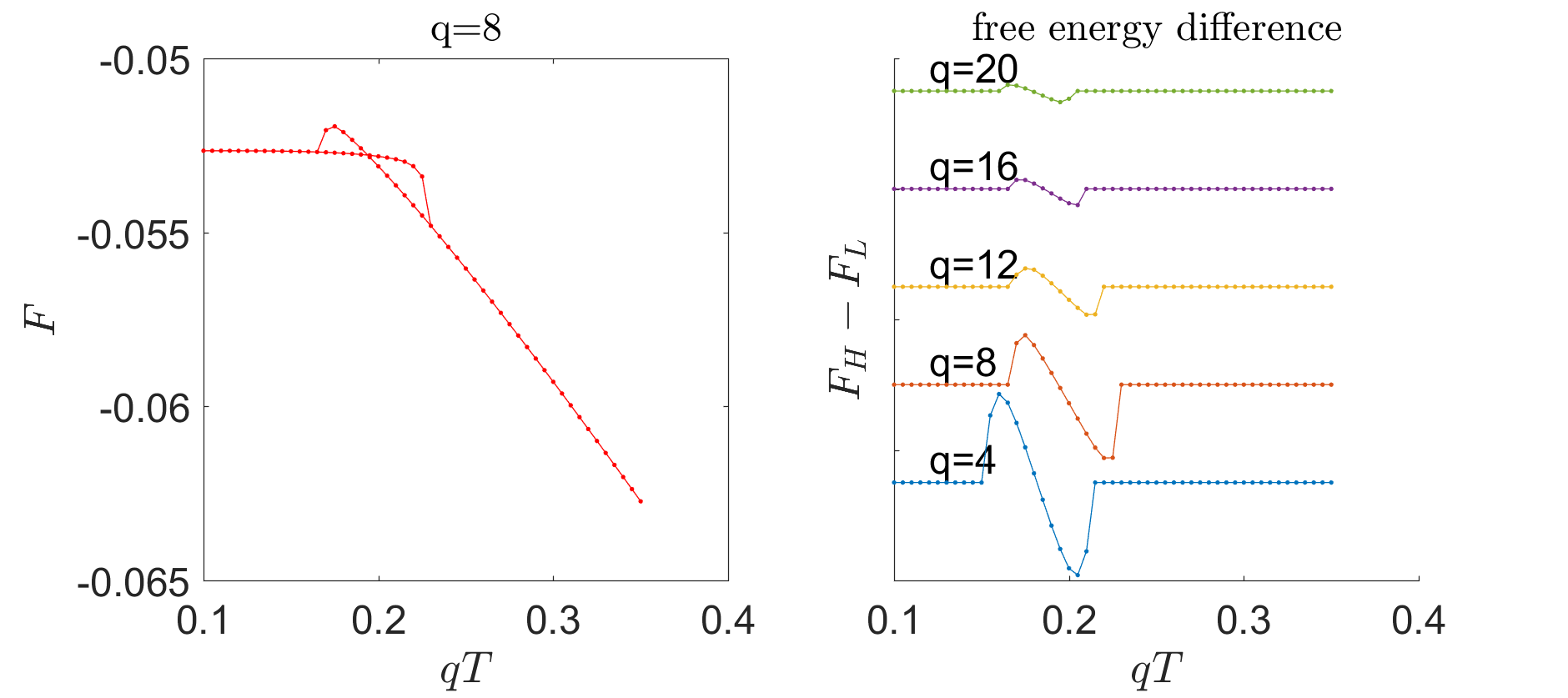} 
~~~~~~~~~~~~~(a)~~~~~~~~~~~~~~~~~~~~~~~~~~~~~~~~~~~~~~~~~~(b) 
\caption{(a) The hysteresis curve for $q=8$. (b) The difference between the free energy of the two saddle points $\Delta F=F_H-F_L$ for different $q$. The curves are offset by a constant for clarity. It should be noted that the temperature is rescaled by $q$ since the transition occurs at different temperature region for different $q$. The calculations are done for $\hat{\mu}=\frac{\mu}q=0.5$. }
\label{SDEfig4}
\end{center}
\end{figure}

The solution of $G_{ab}(\tau)$ for two different temperatures is shown in figure \ref{SDEfig1}. At low temperature, we obtain an exponentially decaying two-point function $G^{ab}(\tau)\propto e^{-E_{\rm gap}\tau}$ (with the same exponent $E_{\rm gap}$ for both $G^{LL}$ and $G^{LR}$. This confirms that the coupled SYK model has a gap $E_{\rm gap}$ above the ground state. Figure \ref{SDEfig2} (a) shows the gap $E_{\rm gap}$ as a function of coupling $\mu$. In the region of small $\mu$, the result is consistent with the dimensional analysis 
\be
E_{\rm gap} \propto \mu^{\frac1{2-2\Delta}}=\mu^{2/3},
\ee
which we discussed earlier in Eqs. (\ref{tPrime}) \nref{GapDef}. For larger $\mu$, we see a cross-over to linear dependence $E_{\rm gap} \propto \mu+{\rm constant}$, which is
what we expect when the interaction term in \nref{Hint} dominates.

Since the numerics is setup for finite temperature, we can also study the temperature dependence of two-point functions and thermodynamic properties. 
Figure \ref{SDEfig2}(b) shows how the exponential decaying factor $E_{\rm gap}$  decreases slowly with temperature.\footnote{At finite temperature, the two-point function at long time is actually a sum of exponentially decaying and exponentially increasing contributions, due to the finite periodicity in time. To obtain the temperature dependence of effective gap $E_{\rm gap}$ we used the long time behavior of $G_{ab}$, which will be discussed later in Eq. (\ref{Long}).}

At higher temperature, for $\mu$ that is not too large, we observe a first order phase transition, as can be seen from the behavior of Gibbs free energy, see figure \ref{SDEfig3}(a). To see the phase transition, we start from high temperature and decrease the temperature gradually. We use the solution $G_{ab}(T)$ at  temperature $T$ as the initial condition of the iteration for the next step with a lower
 temperature $T-\Delta T$. When we reach the lowest temperature, we start increasing the temperature and use $G_{ab}(T)$ at each step as the initial condition of next step with temperature $T+\Delta T$. The hysteresis curve we obtain in figure \ref{SDEfig3}(a) suggests that within a range of temperature $T\in[T_{c1},T_{c2}]$, the free energy has two local minima. The annealing from higher temperature makes the solution stay in the high temperature local minimum, until the minimum almost
 disappears at temperature $T_{c1}$ and $G_{ab}$ hops to the other minimum. Similarly at temperature $T_{c2}$ the low temperature local minimum disappears. The same features were observed in similar massive deformations 
 of SYK-like models in \cite{Azeyanagi:2017drg}. 
 
 The results in  figure \ref{SDEfig3} are in agreement with the general discussion in section \ref{HigherTemp}. Namely, we see 
   that the low temperature phase has constant energy and the high temperature phase has constant entropy. They cross where those properties still hold.  
  
    Varying $\mu$ and $T$ leads to a two-dimensional phase diagram, as is shown in figure \ref{SDEfig3}(b). The phase diagram suggests that the first order transition exists for arbitrarily small $\mu$ but goes away at large enough $\mu$. The first order phase transition here can be considered as an analog of the Hawking-Page transition\cite{Hawking:1982dh,Witten:1998zw}
 between the thermal gas in AdS$_2$ geometry and the black hole solution. In higher dimensions, the boundary is connected, while in two-dimensions the boundary is disconnected and the black hole solution consists of two AdS$_2$ black holes, as we have discussed in section \ref{SecFTlow} and figure \ref{TwoPhases}. 
 Compared with the higher temperature Einstein gravity case, the current theory has a large number of bulk fields, of order $N$. 
 Therefore it was not immediately obvious that a first order phase transition, instead of a smooth crossover, should exist between the two distinct geometries. 
In fact, as we will discuss later, the two phases are continuously connected in the microcannonical ensemble (at least at large $q$). 
 It is an interesting question what is the generic physical reason for such phase transition. 
 
The results above are all obtained for $q=4$, and the generalization to higher $q$ is straightforward. We observe qualitatively the same physics for all $q$: the exponentially decaying low temperature two-point function, the super-linear growth of gap $E_{\rm gap}(\mu)$ at small $\mu$, and the first order phase transition for a region of $\mu\in(0,\mu_{\rm max}]$. As an example, figure \ref{SDEfig4} shows the hysteresis curve for different $q$. In the next section, we study the large $q$ limit analytically, which provide further understanding to the form of two-point function, energy gap and  phase diagram. We will compare the numerics with analytic results there. 

 \subsection{Large $q$ analysis. Zero temperature } 
 \la{LqZT}
 
 In this section we study the coupled system at large $q$ as in \cite{Maldacena:2016hyu}. At large $q$ we write the correlators as 
 \be \la{GqExp}
 G_{LL} = \half \sign(\tau) ( 1 + { 1 \over q } g_{LL} + \cdots ) ~,~~~~~~~G_{LR} = { i \over 2 } ( 1 + { 1 \over q } g_{LR} + \cdots ) 
 \ee
 It is convenient to scale $\mu = \hat \mu/q$, keeping $\hat \mu$ fixed when $q\to \infty$. 
 We can take then derivatives of \nref{SDCoup} and expand in $1/q$ to obtain 
 \be \la{LqE}
 \partial_\tau^2 g_{LL} = 2 {\cal J }^2 e^{g_{LL} } ~,~~{\rm for}~~~~\tau > 0~~~ ~;~~~~~~~~ \partial_{\tau}^2 g_{LR} = - 2 {\cal J }^2 e^{g_{LR} } - 2 \hat \mu \delta(\tau)  
 ~,~~{\rm for ~any~}\tau \ee
 Surprisingly we see that the two equations decouple. The minus sign in front of $e^{g_{LR}}$ arises from the factors of $i$ in $G_{LR}$ in \nref{GqExp}\footnote{
For odd $q/2$, this also includes the extra sign in \nref{SDCoup}.}. 
 These two functions are related by the boundary conditions, which are 
 \be \la{BCzT} 
  g_{LL}(0)=0 ~,~~~~~~~ \partial_\tau g_{LR}(0) = -\hat \mu ~,~~~~~~ {g_{LL} -g_{LR} \to 0} ~,~~~~~{\rm as }~~\tau \to \infty 
 \ee
 The first condition comes from $G_{LL}(0)=\half$, the second comes from demanding that we reproduce the $\delta$ function in 
 \nref{LqE}. The last condition applies to the zero temperature situation and is explained in detail  in appendix \ref{LargeqApp}.
 
  An alternative derivation of the Liouville equations (\ref{LqE}) is by  performing first a 
  large $q$ expansion of the effective action (\ref{EfAct}). One can show that to the order of $\frac1{q^2}$ the effective action is equivalent to a Liouville action:
  \begin{eqnarray}
\frac1NS_{\rm eff}&=&\frac1{4q^2}\int_{\tau_1>\tau_2} d\tau_1 d\tau_2\left(\partial_{\tau_1}g_{LL}(\tau_1,\tau_2)\partial_{\tau_2}g_{LL}(\tau_1,\tau_2)-\partial_{\tau_1}g_{LR}(\tau_1,\tau_2)\partial_{\tau_2}g_{LR}(\tau_1,\tau_2)\right)\nonumber\\
& &-\frac{\mathcal{J}^2}{q^2}\int_{\tau_1>\tau_2} d\tau_1d\tau_2\left(e^{g_{LL}(\tau_1,\tau_2)}+e^{g_{LR}(\tau_1,\tau_2)}\right)-\frac{\hat{\mu}}{q^2}\int d\tau g_{LR}(\tau,\tau)\label{LiouvilleAction}
\end{eqnarray}
Such effective action for a single SYK model has been discussed in Appendix B of Ref. \cite{cotler2017black}. For completeness we include the derivation for the coupled model in Appendix \ref{app:Liouville}. In writing \nref{LiouvilleAction} we have neglected possible long-time contributions of the determinants.
  The Liouville action is defined on a half plane $\tau_1\geq \tau_2$, with the boundary condition
\be
g_{LL}(\tau,\tau)=0,~\left.\left(\partial_{\tau_1}-\partial_{\tau_2}\right)g_{LR}(\tau_1,\tau_2)\right|_{\tau_2=\tau_1}=2\left|\hat{\mu}\right| 
\ee
The effective action approach is useful, since it provides a general  derivation of the Liouville equations of motion. It  
applies to different cases such as ground state and finite temperature, and more general situations when the couplings are time-dependent. The different cases are described by the same effective action, with different boundary conditions.

\begin{figure}[h]
\begin{center}
\includegraphics[scale=.3]{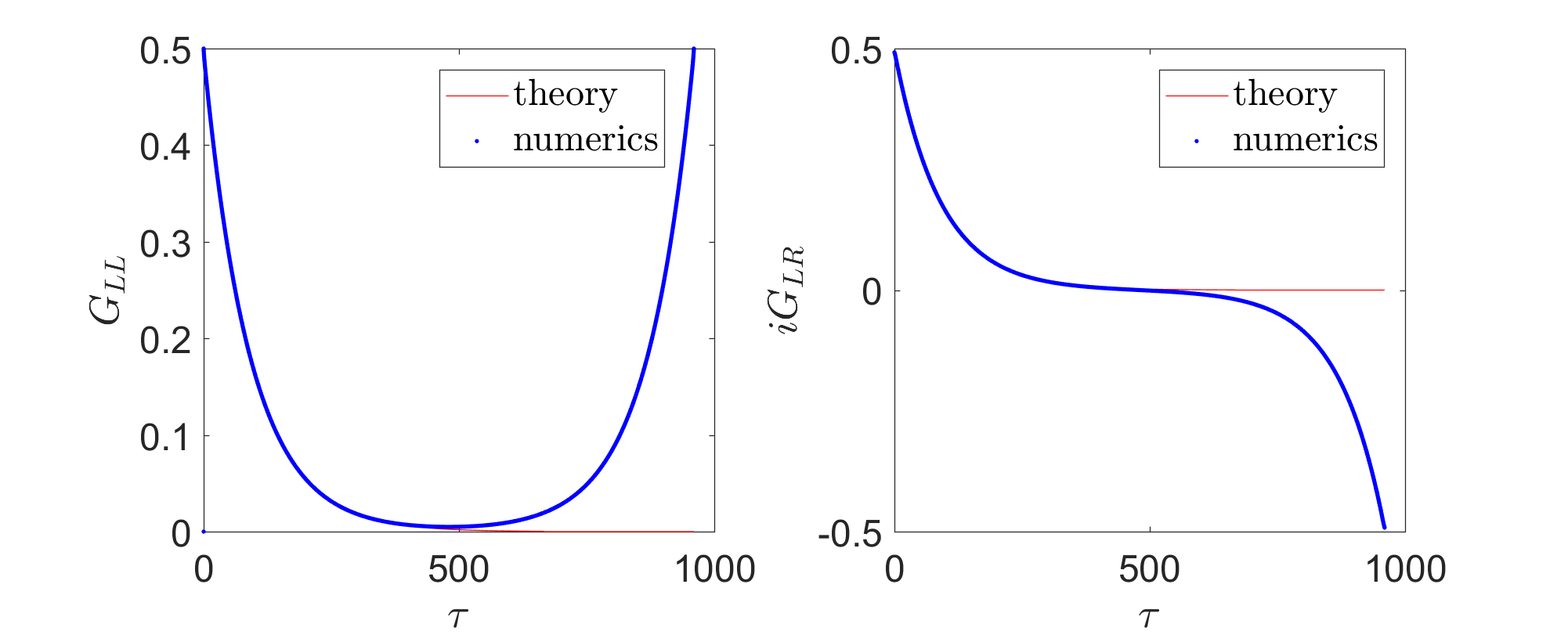}
~~~~~~~~~~~~~(a)~~~~~~~~~~~~~~~~~~~~~~~~~~~~~~~~~~~~~~~~~~(b) 
\caption{The comparison of analytic solution (\ref{gLLgLR}) with numerics for $q=96$ at a low temperature $T=0.001$. The deviation for $\tau>\frac{\beta}2$ is expected because the theoretical formula is for zero temperature, while the numerics at finite temperature is symmetric (for $G_{LL}$ or antisymmetric (for $G_{LR}$) with respect to $\tau=\frac{\beta}2$. }
\label{Glargeq}
\end{center}
\end{figure}

The solutions of \nref{LqE} are 
 \be \la{gLLgLR}
 e^{g_{LL}} = { \alpha^2 \over {\cal J}^2 \sinh^2({ \alpha |\tau |+ \gamma } )} ~,~~~~~~~e^{g_{LR}} = { \tilde\alpha^2 \over  {\cal J}^2 \cosh^2({ \tilde \alpha |\tau| +\tilde  \gamma }) } 
 \ee
 Imposing the boundary conditions \nref{BCzT} leads to 
 \bea \la{ParCond}
 &~ & { \alpha \over {\cal J } \sinh \gamma } =1, ~~~~~~ \hat \mu =2  \tilde \alpha \tanh \tilde \gamma ~,~~~~~~
 \tilde  \alpha  = \alpha ~,~~~~\tilde  \gamma = \gamma ~,~~~~~~~
 \\
 & ~&  \la{SolPar}
 \alpha = J \sinh \gamma ~,~~~~~~~ \tanh^2 \gamma = { \epsilon \over 2} (  \sqrt{ 4 + \epsilon^2 } - \epsilon ) ~,~~~~~\epsilon = { \hat \mu \over 2  {\cal J } } 
 \eea
 Fig. \ref{Glargeq} shows a comparison of the analytic solution with direct numerical solution of the large $N$ equations 
 \nref{SDCoup}, which agrees except that the numerical solution is periodic in imaginary time so it is a periodic identification of the zero temperature solution in the low temperature limit. (Finite temperature effects will be discussed in next subsection.)
 
As an example, we can use these formulas to find an expression for the ground state energy using \nref{EGen} to obtain 
\be \la{ETzLq}
{ E \over N } = { \hat \mu \over q^2 } \left[ - { q \over 2 } + 1 - \log \tanh \gamma - { 1 \over \tanh^2 \gamma } 
\right] 
\ee
with $\tanh \gamma$ given by \nref{SolPar}. When ${\cal J } \to 0$ we get  $E =- N \mu/2$ which is indeed the energy for the ground state of $H_{\rm int}$ in 
\nref{Hint}. 

 In the small $\epsilon$ limit,  we can compare this answer with the general low energy discussion in  section \ref{LowSec}.
  We get agreement once we use the large $q$ expression for 
 $\alpha_S = { 1 \over 4 q^2}$,   note that ${ dt\over d u } =  2  \alpha $, use \nref{tPrime}, 
  and take the large $q$ limit of some of the terms.  The final large $q$ expression for small $\hat \mu \ll {\cal J }$ is 
\be \la{LeLq}
 { d t \over d u }  = 2 \alpha =   \sqrt{ 2 \hat \mu {\cal J } }  ~,~~~~~ \alpha_S = { 1 \over 4 q^2 } ~,~~~~{\rm for }~~~ 1 \ll q  ~,~~~~~ \hat \mu \ll {\cal J } 
 \ee
  Taking the small 
 $\hat \mu$ limit of \nref{ETzLq} we  get 
 \be
  { E \over N } = - { \hat \mu \over 2 q } + { 1 \over q^2 } \left[ - 2 {\cal J } +  { \hat \mu\over 2 } ( 1 - \log { \hat \mu \over 2 {\cal J } } ) \right] 
  \ee
 The term linear in ${\cal J}$ reflects the ``ground state energy''  of the two SYK models at large $q$. This ``ground state energy''  
 is not included in \nref{fLfRres}.  The rest of the terms agree 
 with   the large $q$ (or small $\Delta$)  limit of 
 \nref{GrstEn}, after we use  use \nref{LeLq}. 
 
   
 We can continue  \nref{gLLgLR} to  Lorentzian time to find the left right Lorentzian correlator\footnote{In principle, we can only trust the first term in the $1/q$ expansion 
 of \nref{LRCoq} and \nref{LRCoqmax}. We are using this form because the analysis of higher order corrections (for a single SYK) 
  in \cite{Tarnopolsky:2018env} showed that they are small in the exponential. }
 \be \la{LRCoq}
 \langle \psi_L(t) \psi_R(0) \rangle \sim  { i \over 2 }  e^{g_{LR}/q } =  { i \over 2} \left[ { \sinh \gamma  \over \cos(  t \alpha - i  \tilde \gamma ) } \right]^{2\over q } 
 \ee
  In contrast to \nref{CorrAdS}, now  the correlator is regular at $t =\pi/(2\alpha)$, with a value of 
  \be \la{LRCoqmax} \langle \psi_L(t= { \pi \over 2 \alpha} ) \psi_R(0) \rangle \sim
  { i \over 2} e^{ -i \pi /q} \left[ { \sinh \gamma \over \sinh \tilde \gamma } \right]^{2\over q } 
  \ee
  At zero temperature $\tilde \gamma =\gamma$ and we get 
  the maximal value we could have for the correlators of such operators. This is saying that we are having a perfect information transfer  between the 
 two sides, in this limit.  The factor of $e^{ - i \pi /q}$ seems related to the proper time experienced 
  by the bulk particles as they go from the left boundary to the right boundary. We can define it in a more physical way by comparing this factor for operators of different dimensions, say replacing $\psi_R^{i} \to \psi_R^i(0) \psi^j_R(0)$, and similarly for $\psi_L$, we get a higher dimension $\tilde \Delta$ 
  and a correspondingly extra factor of $e^{ - i \pi \tilde \Delta}$.

  An additional remark is that we can also compute the commutator by using the other operator ordering. The other operator ordering amounts to changing 
  $-i \tilde \gamma \to + i \tilde \gamma $. 
  We can then compute the anticommutator
  \be
 \langle \{\psi_L(t) , \psi_R(0) \}  \rangle =     { i \over 2} \left\{ 
 \left[ { \sinh \gamma  \over \cos(  t \alpha - i  \tilde \gamma ) } \right]^{2\over q } - \left[ { \sinh \gamma  \over \cos(  t \alpha + i  \tilde \gamma ) } \right]^{2\over q }
 \right\} 
 \ee
  For small $\tilde \gamma$ we see that this commutator is small for $ |t | \leq { \pi \over 2 \alpha }$, but it becomes larger for larger times, where we lose the
  extra suppression by $\tilde \gamma$. More explicitly 
  \bea
  \langle\{\psi_L(t) , \psi_R(0) \}  \rangle  
 & \propto &  \left\{ \begin{array}{l}  {2\tilde \gamma    \over q } \tan{ t \alpha } \left[ \cos  \alpha t    \right]^{-2/q }  ~,~~~~~~~~~ 0 <  t\alpha < \pi/2 \cr
 \sin{ 2\pi \over q} \left[ -\cos \alpha t  \right]^{-2/q} ~,~~~~~~~ \pi/2 < t\alpha < 3 \pi/2
 \end{array}\right. \la{anticom}
 \eea
 We see that for small ${\hat \mu \over {\cal J }}$ the anticommutator is suppressed for times less than $ \pi/( 2 \alpha)$ relative to the values it has for later times. 
 This is consistent with what we discussed for the commutator for general $q$ and small ${\hat \mu \over {\cal J } }$ around \nref{CommZe} \nref{CommEta}. 
     
 \subsection{Large $q$ at finite temperature } 
 \la{FinTLq}
 
 We now consider the coupled system in large $q$ limit at finite temperature. As we have discussed in Sec. \ref{sec:numerics}, for $q=4$ we observed a Hawking-Page type first order phase transition numerically. Physically, the transition is between a low temperature phase of global AdS$_2$ geometry and a high temperature phase of two black holes. Numerically, we find that the phase transition exists for all $q$ where the computation can be carried, as long as $\hat{\mu}/\mathcal{J}$ is not too large. At large $q$ limit, there is actually an analytic way to understand the phase transition, which we will discuss in this subsection. 
 
   
 Interestingly, the analytic analysis at large $q$ reveals that the  two minima of free energy corresponding to the two phases  are continuously connected by a saddle point. This continuous connection can be  physically
 explored by considering the theory in the microcannonical ensemble, as opposed to the cannonical ensemble. In other words, the theory in the microcannonical
 ensemble goes continously between these two phases.

  
 
In the following we describe this analysis in more detail, and also analyze other finite temperature properties of the large $q$ problem. For more details see appendix \ref{LargeqApp}. We imagine taking $q$ large 
 holding $\hat \mu \equiv  \mu q $ and $\cal J $ fixed. 
Not surprisingly,  the particular form of the solution depends on how 
 the temperature scales with $q$. 
 Notice that the solution in \nref{gLLgLR} is reasonable only at times that are parametrically less than $q$, otherwise it is not reasonable to expand 
 $G$ as in \nref{GqExp}. In fact, for very large times we can consider a different approximation to the equations \nref{SDCoup}. For that purpose we notice that 
 $\Sigma_{LL}$ and $\Sigma_{LR}$ vary over a relatively short time scale, which is of order one, as compared to the time scale where $G$ varies, which is of order 
 $q$. Therefore, at very long times, we can approximate the convolutions in \nref{SDCoup} as follows. 
 First, let us consider the convolution with $\Sigma_{LR}$. Up to the overall power of $i$, this is a positive function with a non-zero integral. 
 Therefore, we can approximate 
 it as a delta function 
 \be \la{nudef}
 \Sigma_{LR}(\tau) \sim - i \nu \delta(\tau) ~,~~~~~~ \nu \equiv  i \int_{-\infty}^{\infty} d \tau  \Sigma_{LR} = { 2 \tilde \alpha  \over q }  =  {    \mu \over \tanh \tilde \gamma } 
~,~~~~~~~ \mu = {\hat \mu \over q } \ee
 where we used the short time expression for $\Sigma_{LR}$ to evaluate the constant $\nu$. 
 $\Sigma_{LL}$ is an odd function of $\tau$, therefore it leads to a $\delta'(\tau)$, or to a derivative. However the equation \nref{SDCoup} already contains a 
 $\partial_\tau G$ term, while the term coming from $\Sigma_{LL}$ has a $1/q$ suppression and therefore we can ignore it. 
 The conclusion is that, at very long times,  the equations \nref{SDCoup} can be approximated as  
 \be
 \partial_\tau G_{LL} + i \nu G_{RL} = 0 ~,~~~~~~ \partial_\tau G_{LR} + i \nu G_{RR} =0  \la{LargeTG}
 \ee
 and we can use $G_{LL}(\tau) = G_{RR}(\tau)$ and $G_{RL}(\tau) = -G_{LR}(\tau)$ to close the equations. Notice that the equations become local in time. 
 These are  the equations for the correlators of a fermionic harmonic oscillator (in Euclidean signature) with solutions 
 $G_{LL} \propto e^{ \pm \nu \tau } $. Notice that $\nu$ (defined in \nref{nudef}) is setting 
  the long time decay, which can be viewed as the actual energy gap of the system. 
 Notice that it is {\it not} equal to $\mu$. In fact, it is rescaled by the $1/\tanh \tilde \gamma $ factor. So typically, it is larger than $\mu$ itself. 
 In the very low temperature solution, and for small $\mu$, this gap actually goes over to $E_{\rm gap} =  \nu \sim { d t\over du }  \Delta $, which is what we expect from the low energy
 analysis \nref{ConfEn} \nref{LeLq}.    The fact that this goes like $1/q$ means that we start getting deviations from the vacuum at inverse 
 temperatures of order $q$. 
 
 For   $\beta = \infty$ we impose that both functions go to zero at large times to obtain the long time solution 
 \be \la{LBet}
 G_{LL} = - i G_{LR}  = A e^{ - \nu \tau } ~,~~~~{\rm for} ~~~~~~~~\beta = \infty ~,~~~\tau \gg 1/\alpha
 \ee
 This solution was used to set the boundary condition in \nref{BCzT} by matching to the short time behavior of \nref{LBet}.
 More generaly, we can write 
 \be G_{LL} = A \cosh[ \nu (\beta/2 - \tau )] ~,~~~~~~G_{LR} = i A \sinh[ \nu (\beta/2 -\tau) ] \la{Long}
 \ee
 where $A$ should be determined by matching to the shorter time region. 
 
Now we will describe the equations in several consecutive ranges temperature ranges,
 which display various physical phenomena. We will not discuss the derivation of the formulas. 
They are  derived in appendix \ref{LargeqApp}.

\begin{figure}[h]
\begin{center}
\includegraphics[scale=.6]{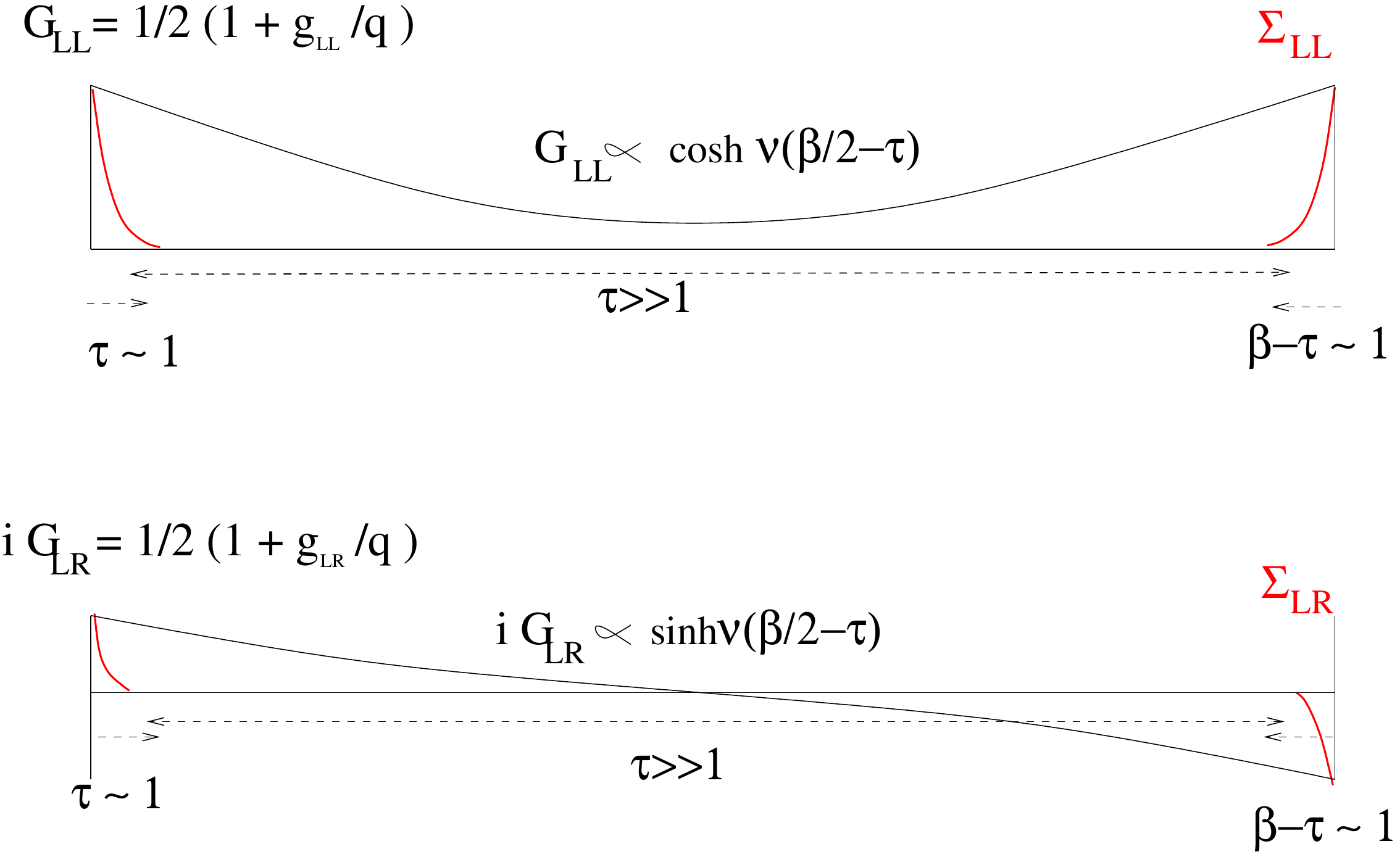} 
\caption{  This is a sketch of the Euclidean coorrelation functions at large $q$ for termperatures of order $q$. The $G_{LL}$ and $G_{LR}$ correlators vary slowly, 
they remain close to maximal for relatively short times, but decay at longer times. In contrast the self energies $\Sigma_{LL}$ and $\Sigma_{LR}$ decay very 
quickly, in times of order one and can be completely neglected for times of order $q$.   }
\label{RegionsG}
\end{center}
\end{figure}

 {\bf Inverse temperatures of order $\beta = q \log q $ } 
 
 In this situation, we can define $\sigma = q e^{ - \beta \nu } $, with $\nu$ as in \nref{nudef}. We take the large $q$ limit while holding $\sigma$ fixed, such that $\beta=\frac1{\nu}\log\frac{q}{\sigma}$. The ansatz for short times is the same as \nref{gLLgLR} and at long
 times \nref{Long}.  Setting the boundary conditions   we get  
 \bea \la{BCTh}
 \tilde \alpha &=&\alpha ~,~~~~ \alpha = {\cal J} \sinh \gamma ~,~~~~~\tilde \gamma = \gamma + \sigma ~,~~~~ \hat \mu = 2 \alpha \tanh \tilde \gamma  ~,~~~~
 \\
 \log (q/\sigma) &=& { \beta \mu \over \tanh \tilde \gamma }  \la{TempE}
 \eea
 where we also listed the definition of $\sigma$. 
 
\begin{figure}[h]
\begin{center}
\includegraphics[scale=.3]{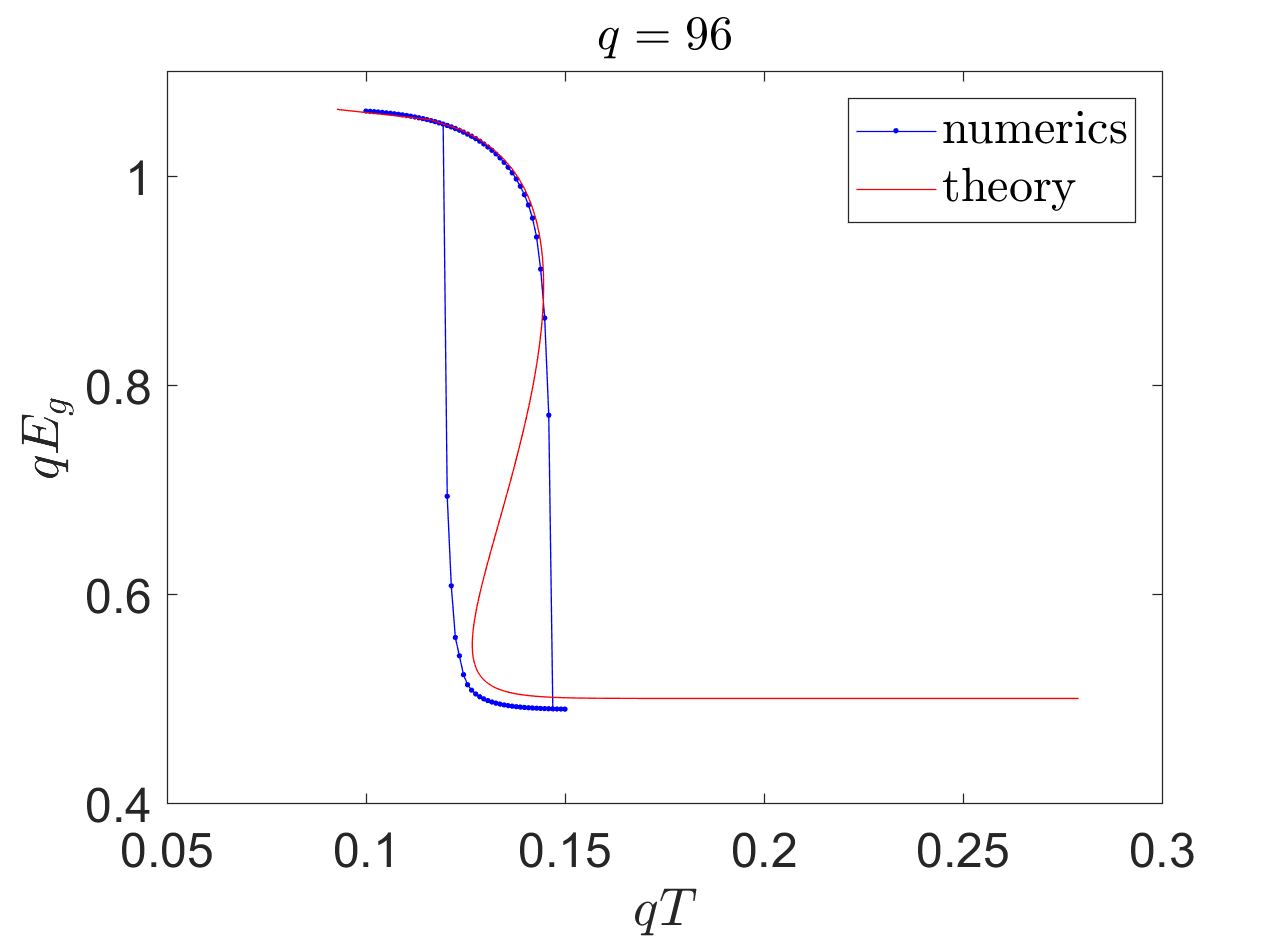} 
\caption{ The relation of energy gap and temperature according to the analytic result (\ref{BCTh}) and (\ref{TempE}) (red line), and the numerical result for $q=96$ (blue line with dots). Note that the numerics can only find the minima, and jumps between the two minima at critical temperature.}
\label{gapvsT}
\end{center}
\end{figure}

 Eq. (\ref{BCTh}) and (\ref{TempE}) determines temperature $\beta$ and parameters $\alpha,\gamma$ in the solution as functions of $\sigma$. The low temperature limit corresponds to $\sigma\rightarrow 0$, in which case Eq. (\ref{BCTh}) reduces to the zero temperature equation (\ref{ParCond}) and (\ref{SolPar}). However, it turns out that the temperature $T(\sigma)=\beta^{-1}(\sigma)$ is not a monotonous function of $\sigma$. Fig. \ref{gapvsT} shows the  relation of the effective energy gap $E_g=\frac{2\alpha}{q}$ and temperature $T$, together with a comparison with numerical results. Due to the non-monotonicity of $T$, there is a temperature window $T_1<T<T_2$ in which there are three solutions with different $E_g$ for a given temperature. They correspond to three saddle points of the free energy, including two minima and one saddle point.  
 
\begin{figure}[h]
\begin{center}
\includegraphics[scale=.6]{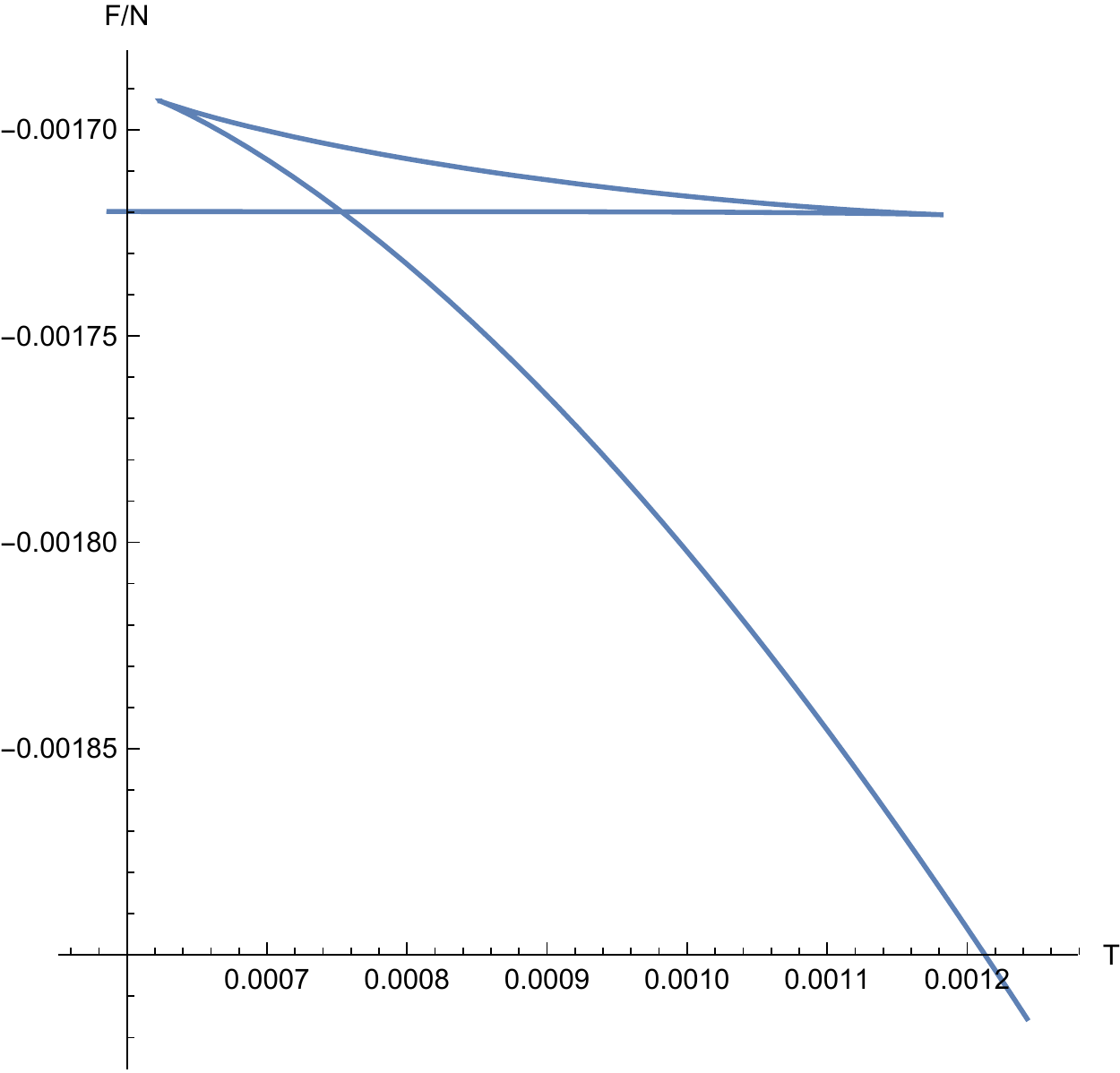}~~~~~~\includegraphics[scale=.6]{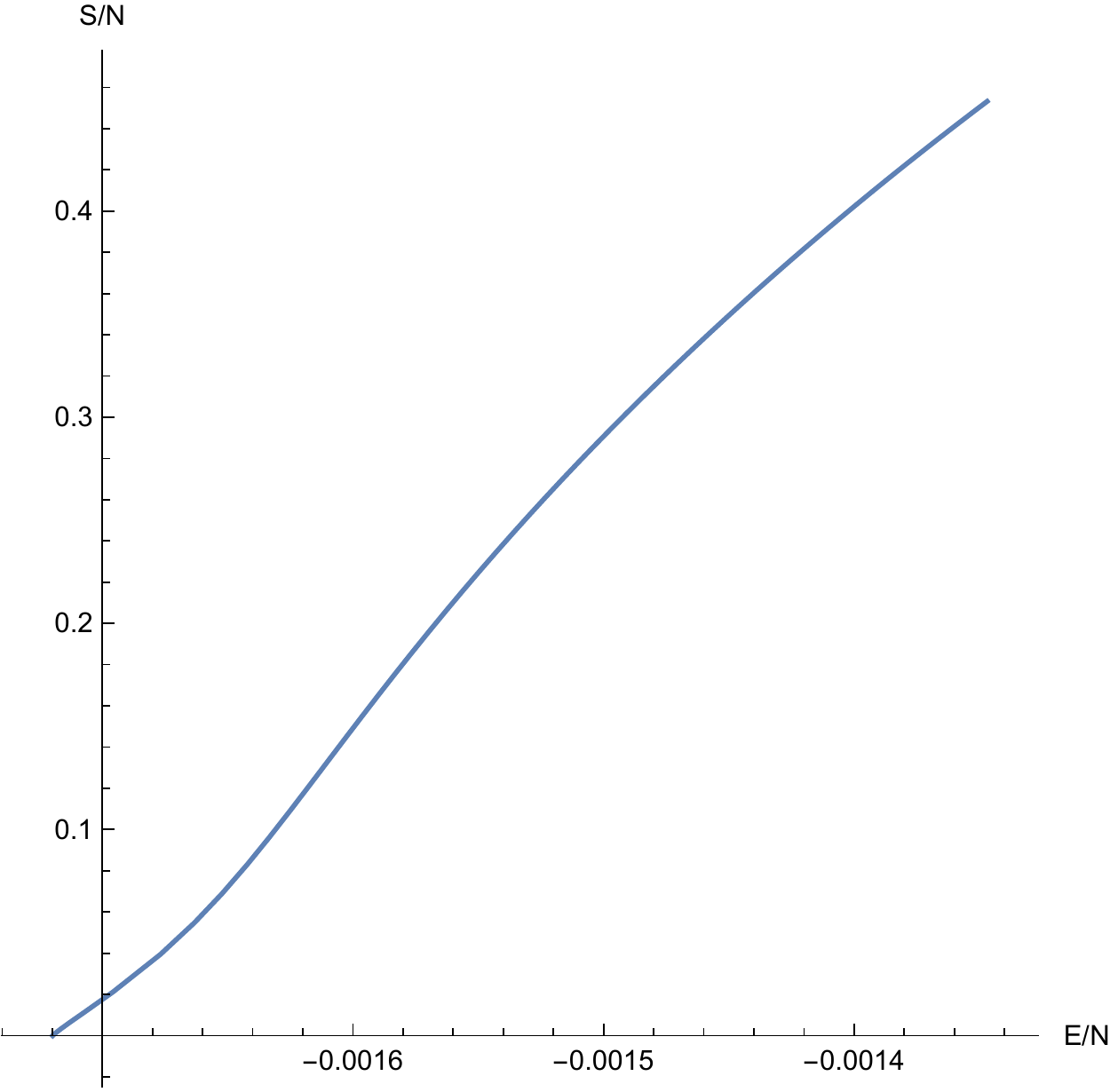} 
~~~~~~~~~~~~~~~~(a)~~~~~~~~~~~~~~~~~~~~~~~~~~~~~~~~~~~~~~~~~~~~~~~~~~~~~~(b)
\caption{ Here we consider a specific case with $q=50$, ${\cal J}=1$, $\hat \mu = .1$ and we plot the free energy as a function of the temperature and
 the entropy as a 
function of the energy, according to \nref{Freeqlog}. The left plot displays a first order phase transition. The top line is unstable in the cannonical ensemble. 
The right  plot shows that in the microcannonical ensemble we have a continuous behavior of the entropy as a function of the energy.
 The changing slope  is related to the change in the temperature as the energy increases, which is displayed more explicitly in figure \ref{Temperature}}
\label{FreeEnergyLq}
\end{center}
\end{figure}

 We can compute the energy from \nref{EGen} and also the free energy, see appendix \ref{LargeqApp} for a derivation.
  We find 
 \bea
 { E \over N } &=& { \hat \mu \over q^2 } \left[ - { q \over 2 } + 1 - { 1 \over \tanh \gamma \tanh \tilde \gamma } - \log { \sinh \gamma \over \cosh \tilde \gamma } 
 \right] 
 \cr \la{Freeqlog}
 - {\beta F \over N} &=& { \beta \hat \mu \over q^2 } \left[  { q \over 2 } - 1 + { 1 \over \tanh \gamma \tanh \tilde \gamma } + \log { \sinh \gamma \over \cosh \tilde \gamma } + { \sigma \over \tanh \tilde \gamma } \right] + { \sigma \over q } 
 \cr
 { S \over N } &=& { \sigma \over q} \left[ 1 + \log { q\over \sigma } \right] = e^{ - \beta \nu} \left[ 1 + \beta \nu \right] 
 \eea
For $\sigma \to 0$ we recover the zero temperature case, see \nref{ETzLq}. For large $\sigma$ we are at a relatively high temperature, where it looks like we start exciting the harmonic oscillators with frequency $\nu$, which in this regime is $\nu \sim \mu$.  In the intermediate temperature range,  the temperature is not a single valued function of $\sigma$, so that 
in the cannonical ensemble we have different branches and a first order transition, see figure \ref{gapvsT} and figure \ref{FreeEnergyLq}(a). 
  All functions are smooth functions of $\sigma$. This can be compared with figure \ref{SDEfig4}(a). 
On the other hand, in the microcannonical ensemble we have a monotonous function $S(E)$ displayed in figure \ref{FreeEnergyLq}(b).
 This shows more explicitly that we go over smoothly from the low temperature to the relatively high\footnote{ We say ``relatively high'' because it is  
  at the higher end of the $\beta \sim q \log q$ window of temperatures.  We still have a few windows to go before we get to really high temperatures!.   } temperature. 
As we increase the energy we encounter a region with negative specific heat, corresponding to the unstable upper curve in figure \ref{FreeEnergyLq}(a). In some
physical systems this could lead to phase separation within the system (as in a mixture of water and ice). However, this seems unlikely to happen in the SYK model 
with its all to all interactions. We have not proven the stability of this region, but we suspect that it is stable in the microcannonical ensemble.

This region was also found in the Schwarzian limit in section \ref{LowerTemp}. In fact, for small $\hat \mu/{\cal J}$, we can make contact between the equations 
\nref{BCTh} and the ones in \nref{finTequl}.
 We approximate \nref{BCTh} as ${ \hat \mu \over \cal J } = 2 \gamma (\gamma + \sigma) $, with $\gamma , \sigma \ll 1$.  These the three terms correspond 
to the three terms in \nref{finTequl}, with $ \gamma \propto t'$ and $\sigma/q = e^{ -\beta 2 J \gamma \over q } $, after using the equation.

       \begin{figure}[h]
\begin{center}
 \includegraphics[scale=.7]{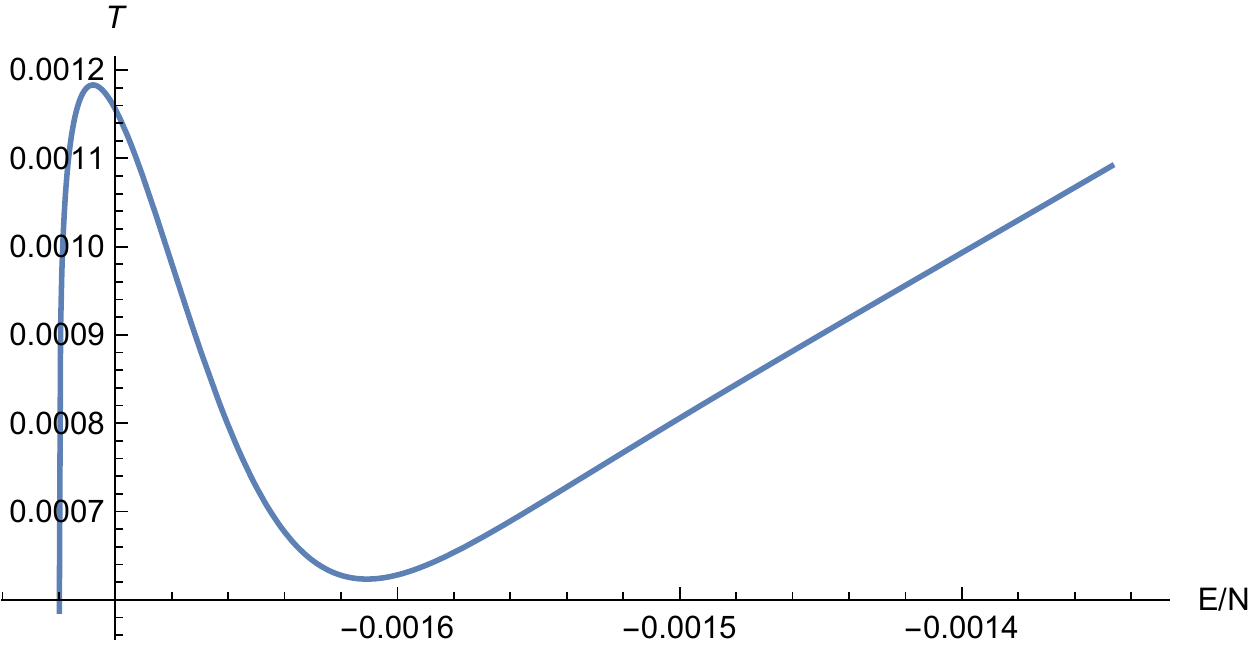}
\caption{ Here we consider a specific case with $q=50$, ${\cal J}=1$, $\hat \mu = .1$ and we plot the temperature as a function of the energy, 
according to \nref{Freeqlog}   }
\label{Temperature}
\end{center}
\end{figure}

  It is interesting to study the Lorentzian correlators in this case. They are given by the analytic continuation of the short time expressions and given by 
 \nref{LRCoq} \nref{LRCoqmax}. 
At finite temperature we have that $\tilde \gamma > \gamma $. As $\sigma = \tilde \gamma - \gamma$ becomes larger, 
the left-right correlator decreases, and the energy 
increases. We also find that the gap $E_{\rm gap} = \nu$ decreases.
 We can interpret this as saying that we are producing bulk excitations, or excitations of the conformal fields that are hindering the transfer of information 
from one side to the other.
We can also say that the positive energy of these excitations is decreasing the total amount of negative energy available to produce the wormhole, and for this reason the 
wormhole is deeper. 

   In this regime
as the temperature becomes progressively higher, or more precisely, as the energy becomes higher, 
 we get $\sigma   \to \infty$ and the value of the left right correlator becomes very small. Therefore we see that the wormhole is closing, or at least is not as ``transparent'' and easy to cross as it was at zero temperature. 
  
\begin{figure}[h]
\begin{center}
 \includegraphics[scale=.3]{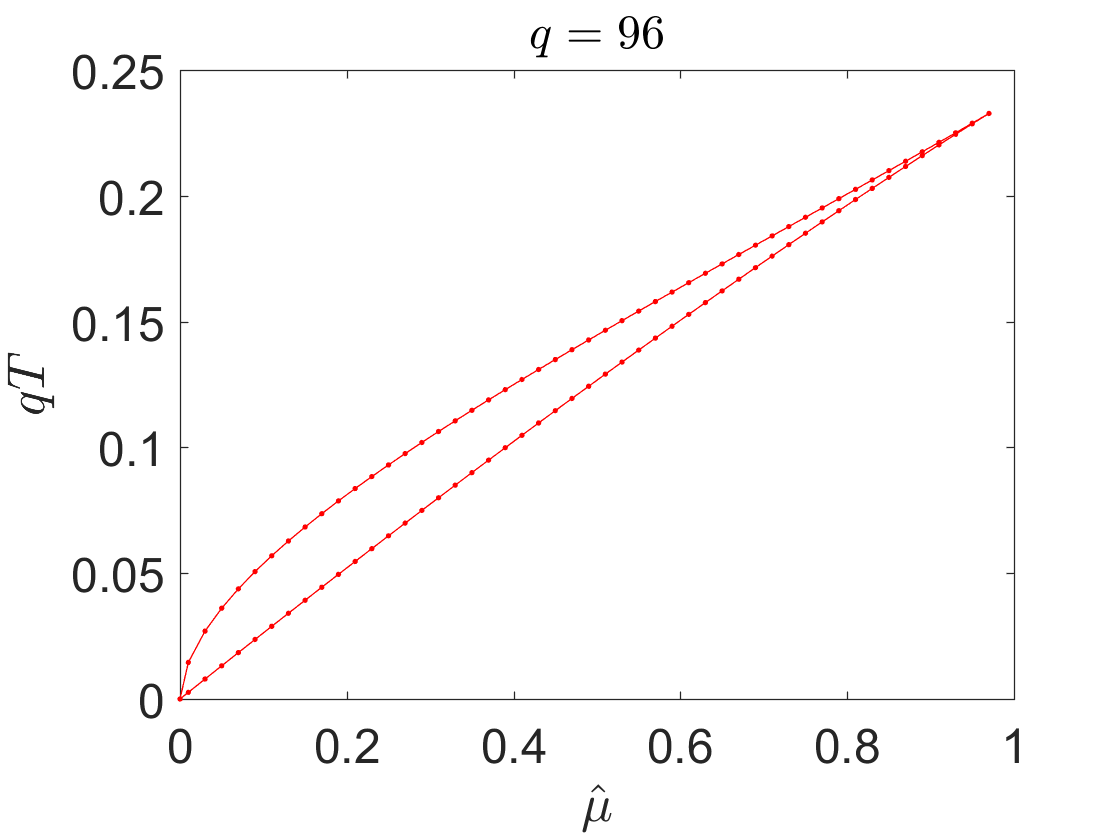}
\caption{ The phase diagram at large $q$. The vertical axis displays $qT$ and the horizonal one $\hat \mu$.    }
\label{largeqPD}
\end{center}
\end{figure}

  The plot in  figure \ref{Temperature} holds for ${ \hat \mu \over {\cal J }} \ll 1$. On the other hand, if ${ \hat \mu \over {\cal J }} \gg 1$, we get from 
  \nref{BCTh} that $\gamma \gg 1$, which also implies that $\tilde \gamma \gg 1$ so that the temperature is approximately given by 
  $\beta \mu \sim \log(q/\sigma)$ which is now a monotonic function of $\sigma$. In this regime, there is no phase transition in the cannonical ensemble. 
 The precise value where the transition disappears  is  around ${\hat \mu \over {\cal J } } \sim 1 $.  The large $q$ phase diagram is shown in Fig. \ref{largeqPD}, where the critical temperatures $T_1,T_2$ are determined by numerically finding the extrema of $T(\sigma)$ curve.  
 
{\bf Inverse temperatures of order $\beta \sim q $ } 

In this regime the function $G_{LR}$ becomes smaller than one everywhere. So we can approximate $G^{q-1}_{LR} =0$ and $\Sigma_{LR}$ contains only the $\mu$ term. 
This implies that we can still make the long time ansatz \nref{Long} but with $\nu = \mu$. Matching to the short time behavior we find 
\be 
2 \alpha = \hat \mu \tanh{ \mu \beta \over 2} ~,~~~~~\sinh \gamma = {\alpha \over {\cal J } } = { \hat \mu \over 2 {\cal J } } \tanh { \mu \beta \over 2 } 
\ee
The free energy is now 
\bea \la{Freeq}
- { \beta F \over N } &=& \log[ 2 \cosh { \beta \mu \over 2 } ] + {  \beta \mu \over q } \tanh{ \beta \mu \over 2 } \left[ \log(2 \sinh \gamma)+ { 1 \over \tanh \gamma } - \gamma 
         -1 \right] 
 \eea
As expected,  the large $\mu \beta$ limit of this expression matches the high temperature limit of \nref{Freeqlog}, so that the two expressions match in their overlapping
range. The most notable feature of \nref{Freeq} is the first term, which is signalling a rise of the entropy, from the relatively low entropy of \nref{Freeqlog} to the 
high entropy of order $S  \sim N \log 2$ that we find in the limit that $\beta \mu $ becomes small. The excitations responsible for this rise look like the free 
 fermionic 
oscillators we would have if our Hamiltonian was given only by the inteaction term $H_{\rm int}$ in \nref{Hint}. 

The main phenomenon that happens in this regime is that the entropy rises to close to its maximal value. We can think that in this regime we are creating lots of bulk 
excitations that make the entropy rise. The final entropy is similar to that of the two separate SYK models. 

{\bf Inverse temperatures of order $\beta \sim \sqrt{q} $ } 

In this regime we can further approximate $G_{LR}$ as 
\be
G_{LR} = { i \over 2 }  \mu  ( { \beta \over 2 } - t ) 
\ee
and this is of order $1/\sqrt{q}$. On the other hand $G_{LL}$ is  close to maximal during the whole range of euclidean times. 
In the equation for $G_{LL}$ we approximate $\Sigma_{LR}$ by just the $\mu \delta(\tau)$ term which then leads to the 
following equation for the variable $g_{LL}$ 
\be
\partial_\tau^2 g_{LL} - 2 {\cal J}^2 e^{ g_{LL}} - { \hat \mu^2 \over q } =0 
\ee
After a rescaling of time by $\beta$ we can remove the $q$ dependence and get a simple equation, which is derived and anlyzed  in the appendix 
\ref{SecSqrtq}. 

The main physical phenomenon that happens within this regime is the growth of the chaos exponent which goes from very small to maximal as the temperature increases within the range we are describing here, see figure \ref{ChaosExponent}. See  appendix \ref{SecSqrtq} for details. 

The expression for the free energy in this regime is given by 
 \be \la{FreeSrtqM}
 - \beta F/N= \log 2 + { ( \beta \mu)^2 \over 8 } + {2  \beta {\cal J } \over   q^2 } + { (\mu \beta)^2 \over 2  q  }  \log( { \cal J }\beta )
+ { h[ q (\mu \beta)^2 ] \over q^2  }  
\ee
where $h$ is a function that we have not determined.

\begin{figure}[h]
\begin{center}
\includegraphics[scale=.7]{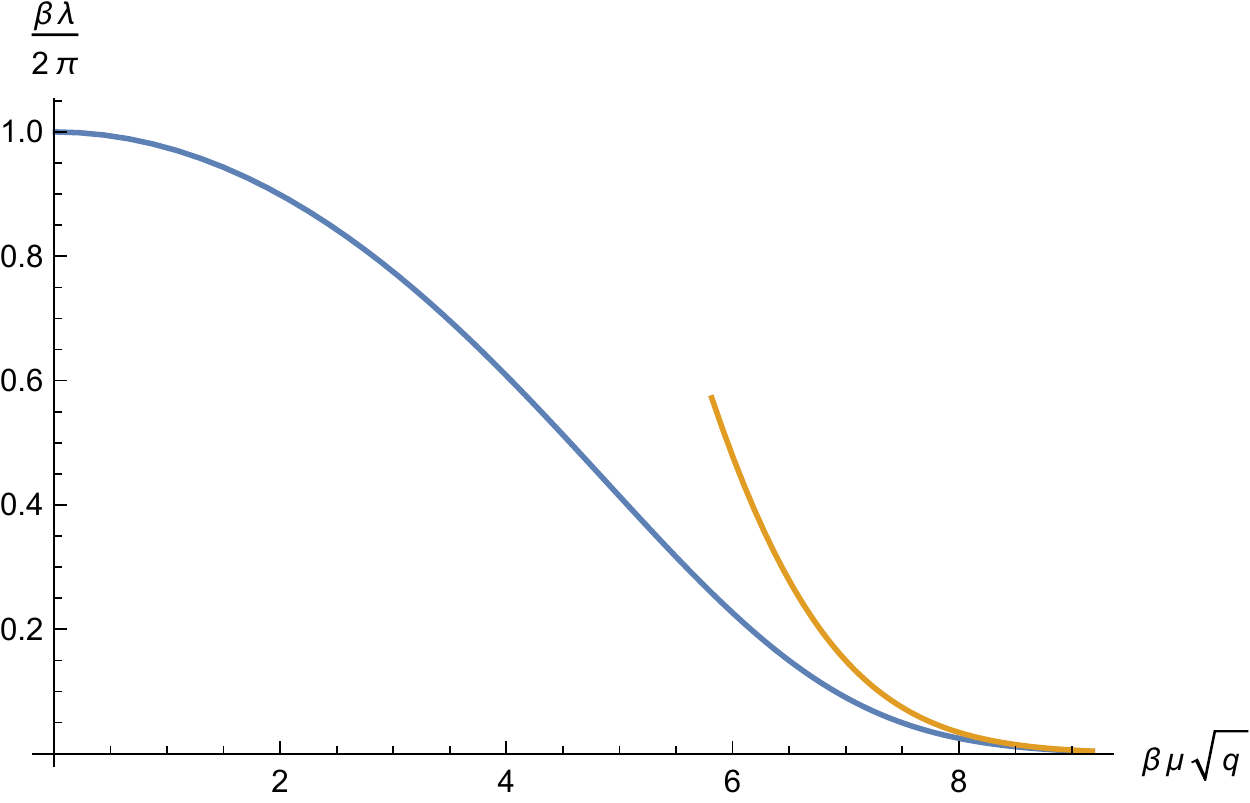} 
\caption{ Here we have plotted the ratio of the chaos exponent to the maximal one as a function of $\sqrt{q}  \beta \mu $.   We see that as the temperature rises we saturate the chaos  bound. The orange curve corresponds to the low temperature analytic  estimate in \nref{LkEstim}.     }
\label{ChaosExponent}
\end{center}
\end{figure}

{\bf Inverse  temperatures of order one }

 Finally, at temperatures of order one we can set $\mu =0$ in the computation of $g_{LL}$ and we recover the single copy, 
 large $q$, SYK result \cite{Maldacena:2016hyu}. And the chaos exponent decreases again from the maximal value to $\lambda = 2 {\cal J }$ for $\beta \to 0$. 
 The free energy can be written as 
 \be
 { F \over N } =   \left. {  F \over N } \right|_{\mu =0 } - {\beta \mu^2  \over 8 } 
 \ee
 Here the second term is a small correction, and we are in the regime described in general by \nref{HigTg}. 
 As we hinted  after \nref{Nft}, and confirmed here, 
  for large $q$ the phase transition does not happen in this regime, it happens at a lower temperature (the regime $\beta \sim q \log q $).

 \subsection{ Overlap of the ground state and the TFD at large $q$ } 
 \la{sec:overlap}
    
At large $q$,   we can also study the overlap between the coupled ground state and TFD state. The overlap $\langle TFD|G\rangle$ can be computed by an Euclidean path integral. The state $|G\rangle$ can be prepared by $\lim_{\tau\rightarrow +\infty}e^{-\tau H}|0\rangle$ with $H$ the coupled Hamiltonian. The initial state $|0\rangle$ only changes the normalization constant, which we should divide out. Therefore 
\be 
\langle TFD|G\rangle=\langle I|e^{-\frac\beta 4H_0-\tau H}|0\rangle= \langle I|e^{-\frac\beta 4 (H_L + H_R) -\tau (H_L+H_R + H_{\rm int} ) }|0\rangle
\ee
which can be written as a Euclidean path integral. Carrying out the random average, one obtains a path integral over the collective fields $\Sigma_{ab}(\tau_1,\tau_2)$ and $G_{ab}(\tau_1,\tau_2)$. The effective action has the same form as Eq. (\ref{EfAct}) except that time runs in the range $\tau\in\left[-\frac\beta 4,+\infty\right)$, and the bare term in self-energy $\mu\delta(\tau_1-\tau_2)$ is replaced by a time dependent term
\bea
\sigma(\tau_1,\tau_2)=\mu \theta(\tau_1)\delta(\tau_1-\tau_2)
\eea
$\theta(\tau_1)$ is the step function which is $1$ for $\tau_1>0$ and zero otherwise. Since there is no time translation symmetry, $G_{ab}$ and $\Sigma_{ab}$ are functions of two time variables $\tau_{1,2}$. 
At $\tau=-\frac\beta 4$ the two sites L and R are smoothly connected, which leads to the boundary condition
\begin{align}
G_{LL}\left(\tau_1,-\frac\beta 4\right)&= - i G_{LR}\left(\tau_1,-\frac\beta 4\right)\nonumber\\
\lim_{\tau_2\rightarrow -\frac\beta 4}\partial_{\tau_2}G_{LL}(\tau_1,\tau_2)&=i \lim_{\tau_2\rightarrow -\frac\beta 4}\partial_{\tau_2}G_{LR}(\tau_1,\tau_2)
\label{BCtwotime}\end{align}
and the same for $\Sigma_{ab}$. The boundary condition at $\tau\rightarrow +\infty$ is set by requiring $G_{ab}(\tau_1,\tau_2)$ to approach the ground state solution $G_{ab}(\tau_1-\tau_2)$ when $\tau_1,\tau_2\rightarrow \infty$. 

\begin{figure}[h]
\begin{center}
\includegraphics[scale=.5]{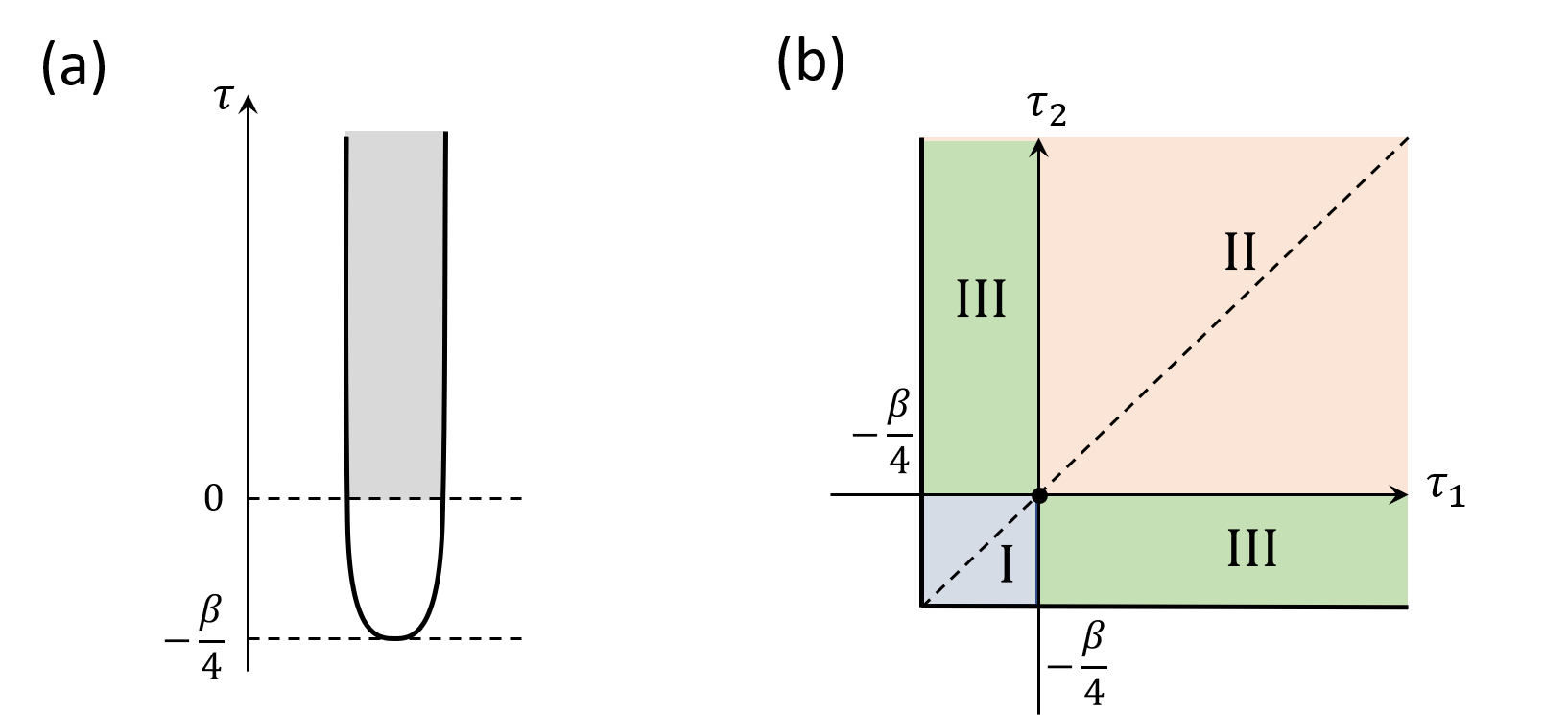}
\caption{(a) Illustration of the Euclidean time path integral that calculates the overlap $\langle TFD|G\rangle$. The initial state at $\tau=-\frac\beta 4$ is a maximally entangled state $|I\rangle$ of the two sites, and the time evolution in $\tau\in[-\frac\beta 4,0]$ with decoupled SYK Hamiltonian prepares the TFD state. The backwards time evolution in the interval $\tau\in(0,+\infty)$ with coupled SYK Hamiltonian prepared the coupled ground state $\langle G|$.  (b) The $\tau_1,\tau_2$ quarter plane on which the two-point functions $G_{ab}(\tau_1,\tau_2)$ is defined. In the large $q$ limit, the Schwinger-Dyson equation is reduced to Liouville equations (\ref{LqEtwotime}) of $g_{ab}(\tau_1,\tau_2)$, with boundary conditions at $\tau_{1,2}=-\frac\beta 4$, $\tau_{1,2}\rightarrow +\infty$ and $\tau_1=\tau_2$.}
\label{figtwotime}
\end{center}
\end{figure}

In the large $q$ limit, one obtains the same Liouville effective action (\ref{LiouvilleAction}) with the boundary condition of $g_{LR}$ changing at $\tau=0$. Since time translation symmetry is absent, the Green's functions are functions of two time variables:
\begin{align}
G_{LL}(\tau_1,\tau_2)&=\frac12{\rm sgn}(\tau_{1}-\tau_2)\left(1+\frac1q g_{LL}(\tau_1,\tau_2)\right), ~
G_{LR}(\tau_1,\tau_2)=\frac i2\left(1+\frac1q g_{LR}(\tau_1,\tau_2)\right)
\end{align}
$g_{LL}$ and $g_{LR}$ satisfy the Liouville equation similar to Eq. (\ref{LqE}), but now with two time variables:
\be 
\frac{\partial^2 g_{LL}}{\partial \tau_1\partial\tau_2}=-2\mathcal{J}^2e^{g_{LL}},~\frac{\partial^2 g_{LR}}{\partial \tau_1\partial\tau_2}=2\mathcal{J}^2e^{g_{LR}}\label{LqEtwotime}
\ee
The equation applies to the quarter plane $\tau_1,\tau_2\in\left[-\frac\beta 4,+\infty\right)$ except the diagonal line $\tau_1=\tau_2$, where the following boundary condition needs to be imposed:
\be 
g_{LL}(\tau,\tau)=0,~\left.\left(\partial_{\tau_1}-\partial_{\tau_2}\right) g_{LR}(\tau_1,\tau_2)\right|_{\tau_2\rightarrow \tau_1}=2\hat{\mu}\theta(\tau_1)\label{BCtwotime2}
\ee
Eq. (\ref{LqEtwotime}) has a general solution (see {\it e.g.} \cite{eberlein2017quantum})
\be
e^{g_{LL}}=\frac{h_1'(\tau_1)h_2'(\tau_2)}{\mathcal{J}^2\left({h_1(\tau_1)-h_2(\tau_2)}\right)^2}~,~~~~~~e^{g_{LR}}=-\frac{f_1'(\tau_1)f_2'(\tau_2)}{\mathcal{J}^2\left({f_1(\tau_1)-f_2(\tau_2)}\right)^2}\label{generalsol2T}
\ee
with functions $h_{1,2}(\tau)$ and $f_{1,2}(\tau)$ determined by the boundary condition. \footnote{An $SL(2,R)$ transformation $h_{1,2}(\tau)\rightarrow \frac{ah_{1,2}+b}{ch_{1,2}+d}$ (with $ad-bc=1$) preserves $g_{LL}$, and similar for $f_1,f_2$. } Due to the symmetry $g_{ab}(\tau_1,\tau_2)=g_{ab}(\tau_2,\tau_1)$, we only need to write the solution for the region $\tau_1>\tau_2$. 

In general, the solution $g_{LL}$ and $g_{LR}$ in the overlap calculation is not related to the saddle point solution for $\langle TFD|TFD\rangle$ and $\langle G|G\rangle$. However, we found that a solution can be found by matching the TFD and coupled SYK saddle points when $\beta$ and $\mu$ satisfies a matching condition. This special case maximizes the overlap $\langle TFD|G\rangle$, and the matching condition determines the effective inverse temperature $\beta(\mu)$ for the reduced density matrix of each SYK site in the coupled ground state. In the following we will provide a summary of the solution, and leave more detailed discussion to Appendix \ref{app:twotime}. 

\begin{figure}[htbp]
\begin{center}
\includegraphics[scale=.27]{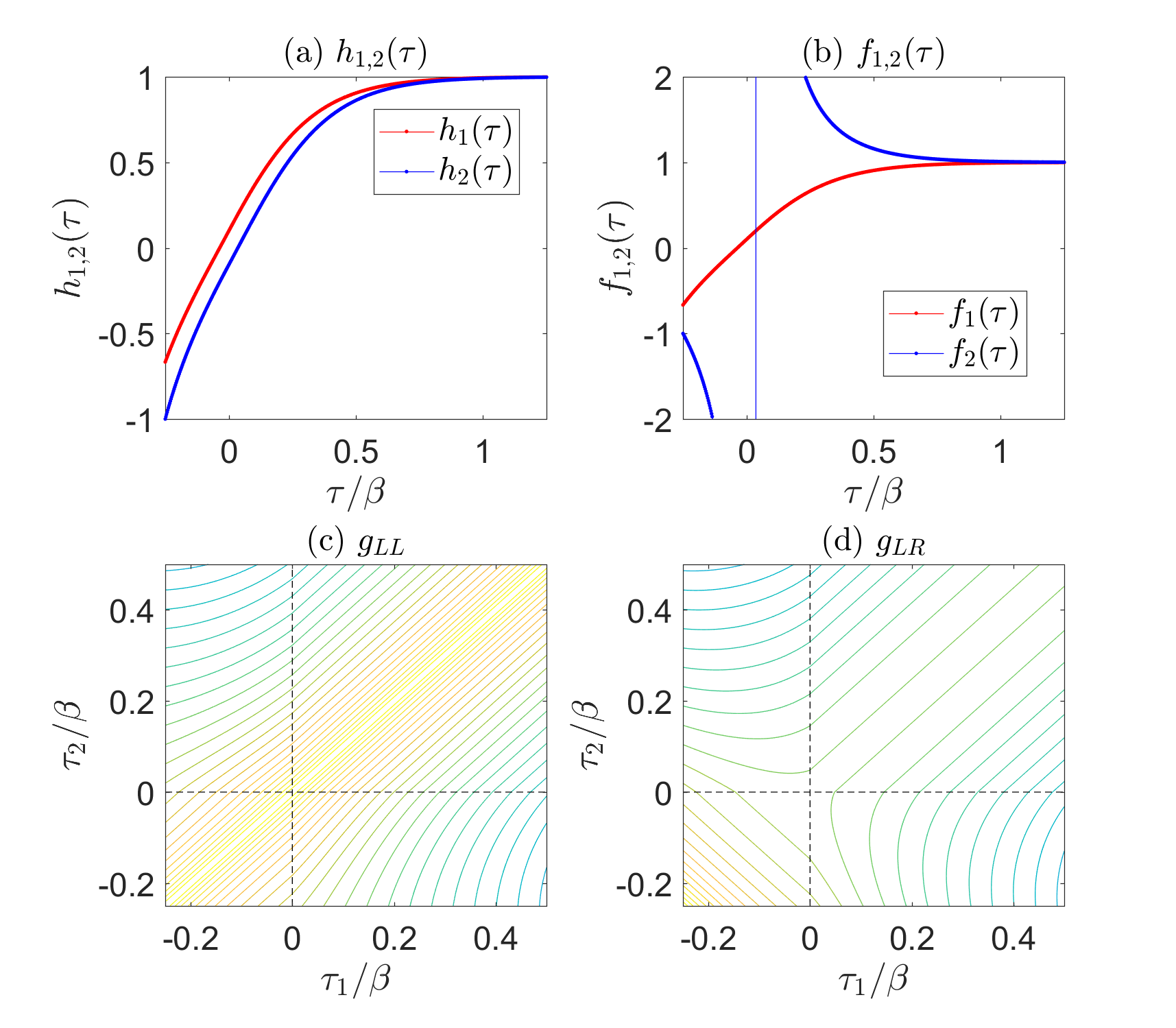}
\caption{(a) and (b) illustrates the functions $h_{1,2}(\tau)$ and $f_{1,2}(\tau)$ that defines the two-point function solution in Eq. (\ref{generalsol2T}). (c) and (d) shows the contour plot of corresponding $g_{LL},g_{LR}$. From the contour plot one can see that $g_{LL}$ is only a function of $\tau_1-\tau_2$ in region I and II of Fig. \ref{figtwotime} (b), while $g_{LR}$ is only a function of $\tau_1+\tau_2$ in region I, and a function of $\tau_1-\tau_2$ in region II. The parameter used in this plot is $\frac{\alpha}{J}=0.2$. }
\label{functwotime}
\end{center}
\end{figure}

The functions $h_{1,2}$ and $f_{1,2}$ are defined as
\begin{align}
h_1(\tau)&=\left\{\begin{array}{cc}\tan\left(\check{\alpha}\tau+\frac12\check{\gamma}\right),&\tau\in\left[-\frac\beta 4,0\right]\\
\tanh\left(\alpha\tau+\frac12\gamma\right),&\tau>0\end{array}\right.,~h_2(\tau)=\left\{\begin{array}{cc}\tan\left(\check{\alpha}\tau-\frac12\check{\gamma}\right),&\tau\in\left[-\frac\beta 4,0\right]\\
\tanh\left(\alpha\tau-\frac12\gamma\right),&\tau>0\end{array}\right.\label{Eqh12}\\
f_1(\tau)&=h_1(\tau),~f_2(\tau)=\left\{\begin{array}{cc}\cot\left(\check{\alpha}\tau-\frac12\check{\gamma}\right),&\tau\in\left[-\frac\beta 4,0\right]\\
\coth\left(\alpha\tau-\frac12\gamma\right),&\tau>0\end{array}\right.
\end{align}
where $\alpha$ and $\gamma$ are determined by the coupled solution in Eq. (\ref{SolPar}).\footnote{$f_2(\tau)$ has a divergence at $\tau=\tilde{\gamma}/2\tilde{\alpha}$ but the correlation function stays regular. } The solution in the region $\tau_{1,2}\in\left[-\frac\beta 4,0\right]$ is a thermal field double solution if $\check{\alpha}$ and $\check{\gamma}$ satisfy
\be 
\check{\alpha} \beta +2 \check{\gamma}= \pi\label{thermalBC2}
\ee
 $\check{\alpha}$ and $\check{\gamma}$ are determined by continuity of $g_{ab}$ and its first derivative at $\tau=0$, which then determines $\beta$ through Eq. (\ref{thermalBC2}). Here we will reserve more details of the derivation to Appendix \ref{app:twotime}, and only write the final result of $\beta$ as a function of $\alpha$:
\be
\beta=\frac{2}{\alpha}\sqrt{1+\left(\frac{\alpha}{\mathcal{J}}\right)^2}{\arctan}\frac{\mathcal{J}}\alpha \label{effectiveT}
\ee

In the weak coupling limit $\hat{\mu}\ll \mathcal{J}$, $\alpha\ll \mathcal{J}$, the formula reduces to $\beta\simeq \frac{\pi}{\alpha}$ as expected, see
\nref{TemMa}. For general value of $\hat{\mu}/\mathcal{J}$, the gap $\alpha$ of the coupled SYK model is different from $\frac{\pi}{\beta}$. In the limit $\hat{\mu}\gg \mathcal{J}$, $\alpha\simeq \hat{\mu}/2$, and $\beta\simeq \frac{2}{\alpha}=\frac{4}{\hat{\mu}}$. In the limit $\hat{\mu}/\mathcal{J}\rightarrow \infty$, $\beta\rightarrow 0$, which means the coupled ground state approaches the infinite temperature TFD state, {\it i.e.} the maximally entangled state $\ket{I}$. 

Now we compute the overlap $\langle TFD|G\rangle$ corresponding to this special solution. To the order of $\frac1 {q^2}$, the overlap is given by the classical saddle point value of the Liouville action (\ref{LiouvilleAction}). Denote $S$ as the saddle point action of the overlap solution, and $S_{TFD}$ and $S_G$ as that of the TFD state and the coupled ground state, respectively, the normalized overlap is
\be 
\left|\langle TFD|G\rangle\right|=\exp\left[-N\left(S-\frac12S_{TFD}-\frac12S_G\right)\right]
\ee
To compute the overlap, we first write the Liouville action (\ref{LiouvilleAction}) in dimensionless couplings $\beta\mathcal{J}$ and $\beta\mu$:
\begin{eqnarray}
\frac1NS&=&\frac1{4q^2}\int_{-\frac{\pi}2}^{+\infty}d\theta_1\int_{\theta_1}^{+\infty}d\theta_2 \left[\partial_{\theta_1}g_{LL}\partial_{\theta_2}g_{LL}-\partial_{\theta_1}g_{LR}\partial_{\theta_2}g_{LR}\right.\nonumber\\
& &\left.-\frac{(\beta\mathcal{J})^2}{\pi^2}\left(e^{g_{LL}}+e^{g_{LR}}\right)\right]-\frac{\beta\hat{\mu}}{2\pi q^2}\int_{-\frac{\pi}2}^{+\infty}d\theta g_{LR}(\theta,\theta)\label{LiouvilleAction2}
\end{eqnarray}
with $\theta_{1,2}=2\pi\frac{\tau_{1,2}}{\beta}$. Now we consider the derivative of $S$ over $\mathcal{J}$ with $\hat{\mu}$ fixed, and $\beta=\beta(\mathcal{J},\hat{\mu})$ determined by Eq. (\ref{effectiveT}). 
\begin{align}
\left.\frac1N\frac{\partial S}{\partial\mathcal{J}}\right|_{\hat{\mu}}&=-\frac1{4\pi^2q^2}\frac{\partial (\beta\mathcal{J})^2}{\partial \mathcal{J}}
\int_{-\frac{\pi}2}^{+\infty}d\theta_1\int_{\theta_1}^{+\infty}d\theta_2\left(e^{g_{LL}}+e^{g_{LR}}\right)-\frac{\partial(\beta\hat{\mu})}{\partial \mathcal{J}}\frac1{2\pi q^2}\int_{-\frac{\pi}2}^{+\infty}d\theta g_{LR}(\theta,\theta)\nonumber\\
&=\frac1{\beta\mathcal{J}q^2}\frac{\partial (\beta\mathcal{J})}{\partial \mathcal{J}}\int_{-\frac\beta 4}^{+\infty} d\tau_1\int_{\tau_1}^{+\infty}d\tau_2\partial_1\partial_2\left(g_{LL}-g_{LR}\right)-\frac1{\beta q^2}\frac{\partial (\beta\hat{\mu})}{\partial\mathcal{J}}\int_{-\frac\beta 4}^{+\infty}d\tau g_{LR}(\tau,\tau)
\end{align}
In the second equality, we used the Liouville equation. \footnote{ Since the term $e^{g_{LL}}+e^{g_{LR}}$ decays in a finite time scale $\sim \alpha$, the Liouville equation of motion is an accurate approximation of the Schwinger Dyson equation. }
Using the fact $\partial_1(g_{LL}-g_{LR})\rightarrow 0$ for $\tau_2\rightarrow+ \infty$, we can integrate over $\tau_2$ in the first term and obtain
\begin{eqnarray}
\frac1N\frac{\partial S}{\partial \mathcal{J}}&=&-\frac1{\beta\mathcal{J}q^2}\frac{\partial (\beta\mathcal{J})}{\partial \mathcal{J}}\int_{-\frac\beta 4}^{+\infty}d\tau_1\left.\left(\partial_1g_{LL}(\tau_1,\tau_2)-
\partial_1g_{LR}(\tau_1,\tau_2)\right)\right|_{\tau_2\rightarrow \tau_1^+}\nonumber\\
& &-\frac1{\beta q^2}\frac{\partial (\beta\hat{\mu})}{\partial\mathcal{J}}\int_{-\frac\beta 4}^{+\infty}d\tau g_{LR}(\tau,\tau)\label{OverlapIntegral}
\end{eqnarray}
The key point of Eq. (\ref{OverlapIntegral}) is that the saddle point value of $S$ is transformed into a single integral over $\tau_1$, and it only depends on the derivative of $g_{ab}$ at the $\tau_1=\tau_2$ line. Since the solution we construct is identical to the TFD solution in region I and is identical to the ground state solution in region II, we obtain 
\begin{align}
\frac{\partial}{\partial \mathcal{J}}\left[S-\frac12\left(S_{TFD}+S_{G}\right)\right]=0
\end{align}
for all $\mathcal{J}$. For $\mathcal{J}=0$, one can explicitly verify $|G\rangle_{\mathcal{J}=0}=|TFD\rangle_{\beta=0}$ is the infinite temperature TFD state. Therefore we can integrate the equation above to conclude $ \left|\langle TFD|G\rangle\right|=1$ for general $\mathcal{J}$ and $\hat{\mu}$. 

Since the Liouville action is only an approximate effective action up to the second order of $\frac1q$, it is possible that the higher order terms contribute a nontrivial suppression to the overlap.
In fact, the computation in  \nref{LogF} shows that there is a non-zero correction that goes like $1/q^3$, obtained after analyzing the small $\Delta$ limit of 
\nref{LogF}, together with \nref{tPrime} \nref{RelPar}. 
 


 As a side remark, we have computed the overlap $\langle TFD|G\rangle$ for small $N$ systems in exact diagonalization. There the temperature $\beta$ in TFD is tuned to a value $\beta(\mu)$ which makes the SYK energy of each site for the two states the same:
\be  
\langle TFD|H_L|TFD\rangle=\langle G|H_L|G\rangle
\ee 
This is a necessary condition for the two states to be approximately the same. The numerics shows that the overlap is always quite close to $1$, although we do not have enough data to do a finite $N$ scaling.  

\begin{figure}[htbp]
\begin{center}
\includegraphics[scale=.27]{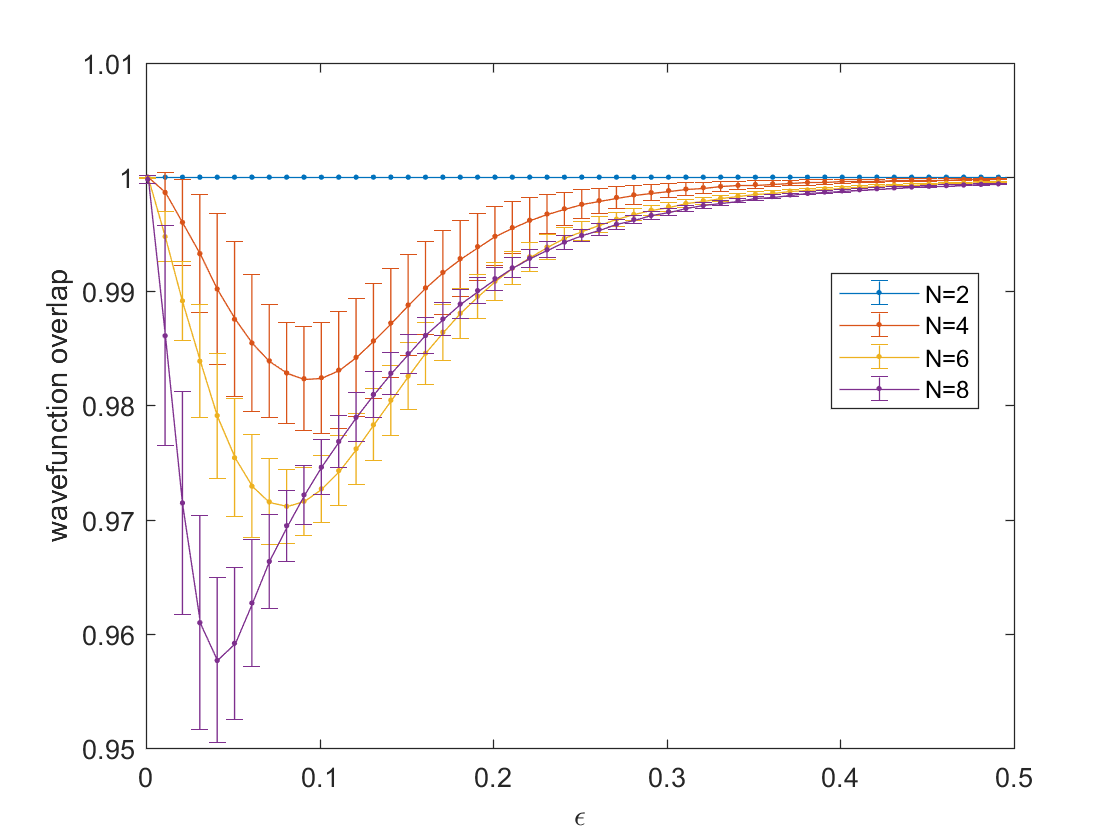}
\caption{The overlap $|\langle TFD|G\rangle|$ for $N=4,8,12,16$ Majorana fermions per site. }
\label{overlapED}
\end{center}
\end{figure}

 \subsection{ Different microscopic couplings } 
 \la{DiffCoupl}
 
 Throughout this paper we have assumed that $H_L$ and $H_R$ have the same couplings, for each random 
 choice of the couplings. We could imagine a different situation where the correlations between the two couplings are not 
 perfect. For example, we could assume that 
 \be
  \langle( J^L_{j_1 \cdots j_q} )^2 \rangle =\langle( J^R_{j_1 \cdots j_q} )^2  \rangle =  { {\cal J}^2 \over \tilde {\cal J}^2 } 
  \langle J^R_{j_1 \cdots j_q} J^L_{j_1 \cdots j_q} \rangle 
 \ee
 
 The large $N$ action in this case has the same form as \nref{EfAct} but with ${\cal J}^2 \to \tilde {\cal J}^2$ in the terms involving 
 $G_{LR}$. 
 
We can easily analyze the equations in the large $q$ limit, where we get the same as in 
 \nref{LqE}, but with ${\cal J } \to \tilde { \cal J }$ in the equation that involves $g_{LR}$. The solutions are again 
 like the ones in \nref{gLLgLR} but with ${\cal  J} \to \tilde {\cal J} $ in the solution for $g_{LR}$. 
 The zero temperature equation imposes again the boundary conditions in \nref{BCzT} which now imply 
 \be \la{ParDC}
 { \alpha \over {\cal J } \sinh \gamma } =1 ~,~~~ \hat  \mu = 2 \tilde \alpha \tanh \tilde \gamma ~,~~~ \tilde \alpha = \alpha 
 ~,~~~~ \tilde \gamma - \gamma = \sigma =\log {  {\cal J } \over \tilde {\cal  J} } 
 \ee
 These are similar to \nref{ParCond}. The only difference is the last equality. In fact, they are the same as the ones 
 we had in \nref{BCTh}, but with a  new  definition for $\sigma$. 
 We see that if the couplings $J_{j_1 \cdots j_q}$ are real, then we always have that $\tilde {\cal J } \leq {\cal J }$. 
 This means that decorrelating the couplings will decrease the physical energy gap, which is 
 \be \la{GapDC}
 \nu = { \hat \mu \over q \tanh \tilde \gamma } = { \mu \over \tanh \tilde \gamma } 
 \ee
 It will also decrease the left-right correlation functions. Curiously, the same sided correlators are not changed
 drastically, in the sense that they also return to their values at $t=0$ at the time $ t = \pi/\alpha$, in Lorentzian time. 

Notice that in the limit that the couplings of the two sides are uncorrelated, we have $\tilde {\cal J} =0$. Then  $\sigma $ in 
\nref{ParDC} goes to infinity. In that case the gap \nref{GapDC} goes down to the naively expected value equal to $\mu$. 
  
 Notice the following point. When discussing  of the thermofield double state and its wormhole dual, one might 
 be left with the impression that in order to have a wormhole one needs to have perfectly identical systems and a 
 perfect matching of energy levels. This does not seem necessary. In fact, a model where we change a bit the 
 couplings between the left and right systems will have energy levels that are shifted much more than their spacing. 
 Nevertheless, the system continues to display a whormhole like behavior. 
 
 In principle, we could also study this problem with different couplings for finite $q$.
 When the couplings are different, even for small $\mu$, we do not expect to be able to approximate the problem 
 using the conformal limit as we did in \nref{fLfRres}. The reason is that the functions $G_{LL}$ and $G_{LR}$ are not given by their 
 conformal field theory limit as a first approximation. 

{\bf  Euclidean wormholes  } 

It is interesting to note that for  $\tilde {\cal J} > {\cal J}$, then we can have a solution of \nref{ParDC} with 
$\mu =\tilde \gamma =0$ and $\gamma >0$. 
Of course, this is possible only if the couplings $J_{ijkl}$ are complex, 
 so that we can arrange that the correlators between left right couplings are larger than the correlator of left-left or right-right couplings.  
 While complex couplings do not  give rise to a sensible Lorentzian theory, they do 
  make sense as a Euclidean theory, in a statistical mechanics context. In this case, the relative (Euclidean) 
   time translation symmetry
between the two sides is spontaneously broken, and the configuration would be associated to a geometry looking 
 like  a  Euclidean wormhole. 
It would be interesting to study this further and see whether it holds some lessons for Euclidean wormholes in general.

\section{Conclusions and discussion }

In this paper we analyzed two closely related problems.
First we considered the theory of gravity that describes nearly extremal black holes, called nearly-$AdS_2$ gravity. 
To this gravity theory we added matter fields with an interaction that looks non-local from the bulk point of view. This interaction connects the two
separated boundaries of $AdS_2$. 
 The interaction creates negative energy in the bulk 
and produces a geometry that is closely related to that of global $AdS_2$, but with just a global time translation isometry. 
The second problem involves two copies of SYK models coupled by a simple interaction. 
  The system becomes gapped at leading order in $N$. Neverthess it displays natural remnants of the nearly conformal symmetry of 
  a single SYK model. In fact, the spectrum is largely controlled by the broken SL(2) symmetry. 
 
 Both systems share a universal subsector   described by a common action \nref{fLfRres}. In the gravity side, this subsector encodes the 
 gravitational interactions of the system. 

The dynamics of the two coupled  SYK systems
 looks like that of a traversable wormhole. Supose that we insert excitations on one of the systems. 
 From the point of view of this system, 
the object becomes more complex, but then, after a while,  the simple objects reasembles on the other side. The dynamics of the simple excitation looks complicated
from the boundary point of view but it is simple in the bulk. The object sails smoothly from one side to the other. It goes through a wormhole. 
 This does not violate any causality constraint since we are adding direct interactions between the two sides.

       \begin{figure}[h]
\begin{center}
 \includegraphics[scale=.5]{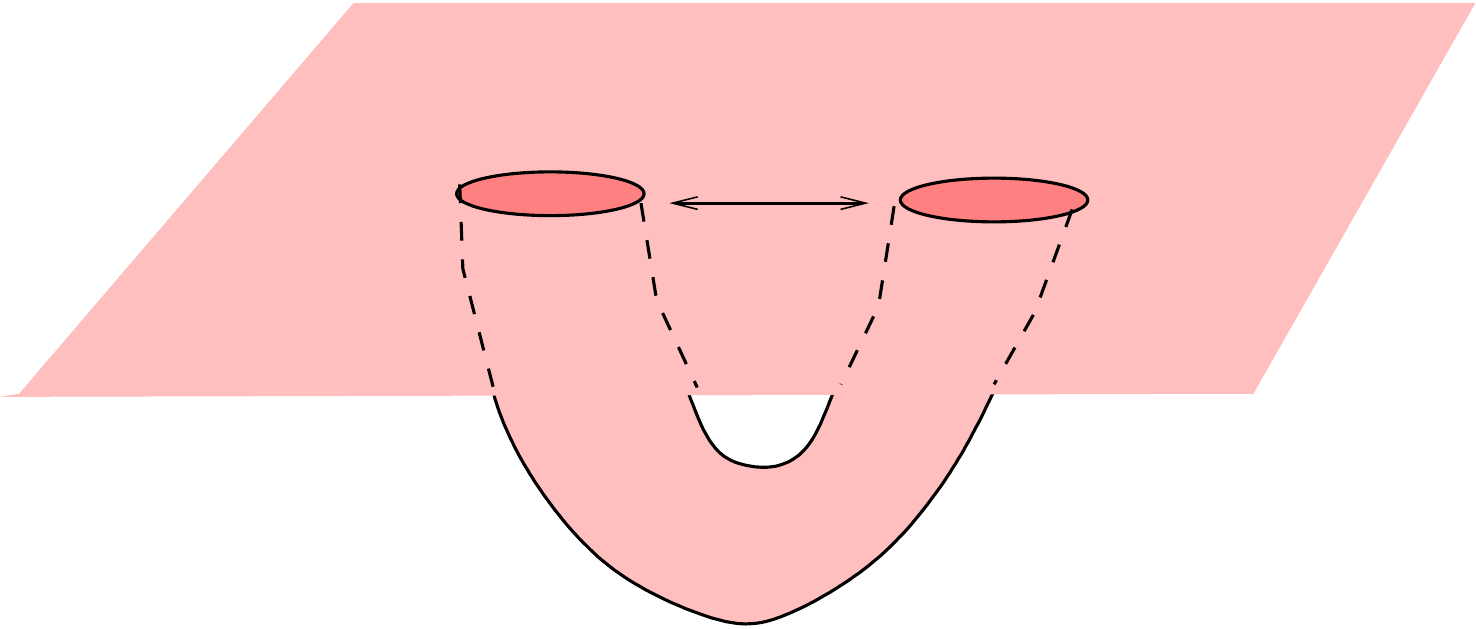}
\caption{ This is a sketch of an idea for producing solutions with non-trivial topology. First we imagine setting up a situation where two near 
extermal black holes are close to each other, without falling into each other. This would require background fields holding the black holes in place. 
 Integrating out the fields modes in the ambient space gives rise to
an interaction between the fields at the top end of the nearly $AdS_2$ throats. This interaction can produce a throat that connecting the two sides so that the
final geometry is a horizonless eternal traversable wormhole. The throat is supported by negative null energy (energy that contributes negatively to the integrated null energy) produced by quantum effects.   }
\label{TwoBlackHoles}
\end{center}
\end{figure}

   It is interesting to ask whether the interaction we  considered in \nref{OOInt} could arise from a higher dimensional set up. 
   In fact, we know that nearly-$AdS_2$ gravity arises universaly in the near horizon region of nearly extremal black holes (see e.g. \cite{Nayak:2018qej}). 
   So, one can imagine a situation where we take two nearly extremal black holes that are relatively near each other, see figure \ref{TwoBlackHoles}. 
    They are far enough that we can think of them as two 
   separate black holes, but close enough that the fields that propagate in that background have correlated fluctuations near the two black holes. 
 Then  \nref{OOInt} can arise by integrating out the field modes that have  energies larger than some small number that is smaller
   than the inverse distance between the two black holes. If the resulting operator of the form \nref{OOInt} has positive sign, then 
   we expect that a traversable wormhole, of the kind discussed here, will form. The full geometry will not contain a horizon, but will have non-trivial topology in the 
   ambient space. 
   There are a few challenges that one has to address. The first is that such near extremal black holes will attract each other by classical forces that are larger than the quantum effects we are discussing. Even if they are orbiting each other, they would be emitting gravitational waves, possibly faster than they can establish the wormhole. 
   Furthermore, the discussion in the present paper was reasonable when the number of 
   fields is large. We can put this large number of fields by hand, but it would be nicer to understand better whether it can be done with a small number of fields.  
   Furthermore, we also need that the fields in the $AdS_2$ regions should be quantized with alternate boundary conditions, see the end of section \ref{SecQuantum}.
   This should be arranged too.

   Another interesting question is the following. Imagine that we couple two systems of this form, which are initially   in a mixed thermal state, entangled with other systems (but not with each other).  Then initially we expect the energy of the combined system will be positive, relative to the ``ground state energy'' of the two SYK models. The expectation value of the interaction term in \nref{OOInt} or \nref{Hint} is zero because the fields are uncorrelated. By further weakly coupling this 
    to another very cold and large system we can let our system of interest cool down and eventually find its ground state, which will be the wormhole like configuration
    described here.  This is  equivalent to saying that we let the black holes evaporate and let them find the ground state.  
    As the system cools down and its energy decreases it will follow the curve of the microcannonical ensemble described in  figure
    \ref{FreeEnergyLq}(b) (for large $q$).     This suggests that the energy decreases smoothly and there is no phase transition between the state containing a pair of black holes vs the state containing the single wormhole.
    This seems surprising from the geometric point of view, since there is a discrete transition between the two geometries. It seems that one should necessarily go through a non-geometric phase. In fact, the phase diagram in figure \ref{FreeEnergyLq}(a) is reminiscent of what is found for black holes in higher dimensional 
    $AdS_D$, $D> 3$ \cite{Witten:1998zw}. The curve sloping down to the right are the large black holes, the left cusp is the lowest temperature black hole, and the upper curve joining the two cusps is similar  to the small black holes in $AdS_D$. The horizontal curve is similar to global Euclidean $AdS_D$ with an identification 
    of the euclidean time coordinate.
    In pure gravity the small black hole and the thermal-$AdS_D$ line meet at infnite temperature (in the  classical limit). In string theory, they  meet at the Hagedorn temperature.
    In this two dimensional  case,
     the unstable phase corresponds to slight excitations on the the thermal $AdS_2$ branch, see figure \ref{TwoPhases}(c), rather than a ``small'' black hole. So the region near the right cusp of figure \ref{FreeEnergyLq} is described simply, as in section \ref{LowerTemp}. On the other hand, we do not have a simple picture from the 
     low energy Schwarzian description, or $AdS_2$ gravity, for the left cusp in figure \ref{FreeEnergyLq}. This would be interesting to find since it seems to involve 
     topology change. Surprisingly, in the SYK model this seems to be a smooth transition (at least at large $q$). 
     Returning to the higher dimensional case, one could also wonder whether there is a smooth transition between the small black hole and thermal $AdS_D$. 
      In classical string theory these two solutions have a different order parameter, the expectation value 
     of the winding tachyon field on the thermal  circle (or Polyakov loop in the boundary gauge theory description) \cite{Witten:1998zw}. Nevertheless it has been suggested that in the microcannonical ensemble, one might have a smooth transition between a gas of strings, or a highely excited string and a small black hole of string size 
     \cite{Horowitz:1997jc}.
    
   Notice that the above procedure is a relatively practical  and physical way for constructing a pair of  SYK models in the thermofield double state. 
   In other words, first we produced the two coupled SYK models and then we let the system cool down to find its ground state. After it has done this, we can 
   turn off the coupling. If the coupling between the two SYK models is small, then the state that we produce is close to the thermofield double, as discussed in section 
   \ref{MatchingTFD}, \ref{SecOver}. 
   Note that, due to the comments in section \ref{DiffCoupl}, the individual couplings do not need to be fine tuned with very high precision.

{\bf Acknowledgements } 

We thank A. Almheiri, Y. Gu, D. Harlow, A. Kitaev, D. Marolf, S. Shenker, D. Stanford, E. Witten, Z. Yang for discussions. 
J.M. is supported in part by U.S. Department of Energy grant
de-sc0009988 and the It from Qubit grant from the Simons foundation. XLQ is supported by the National Science Foundation grant 1720504, and the David and Lucile Packard foundation.

\appendix

\section{Thermodynamics of the two coupled SYK models at large $q$}

  \la{LargeqApp}

In this appendix we explore the thermodynamics of the model at large $q$. 
Writing $\mu = \hat \mu/q$, we are interested in the large $q$ limit for fixed $\hat \mu$ and ${\cal J }$. 
Another parameter is  $\beta$, the inverse temperature. We can scale the inverse temperature $\beta$ in various ways with $q$. 

We   divide the inverse temperature range into four windows, inverse temperatures of order $q \log q$,   $q$,  
$\sqrt{q}$ and 1. 
In each of these temperature ranges we will make different approximations for solving the equations for $G_{ab}(\tau)$, \nref{SDCoup}. 

\subsection{ Inverse temperature of order $q \log q $ } 

Here we make two distinct approximations for the equations. For euclidean times that are small compared to $q$ we approximate $G_{LL}$ and $G_{LR}$ as in 
\nref{GqExp}, leading to equations \nref{gLLgLR}. 
On the other hand, at times of order $q$ or larger we approximate the equations as indicated around \nref{nudef} leading to the equations 
\nref{LargeTG}, with solutions \nref{Long}. 
For convenience we summarize here the solutions in these two regimes 
\bea
 e^{g_{LL}}&=& { \alpha^2 \over {\cal J}^2 \sinh^2({ \alpha |\tau |+ \gamma } )} ~,~~~~~~~e^{g_{LR}} = { \tilde \alpha^2 \over  {\cal J}^2 \cosh^2({ \tilde \alpha |\tau| +\tilde  \gamma }) } ~,~~~~~~~~ \tau \ll q \la{shortAp} 
 \\ 
  G_{LL} &=& A \cosh[ \nu ({\beta\over 2 }  - \tau )] ~,~~~G_{LR} = i A \sinh[ \nu ({\beta\over 2 }  -\tau) ]  ~,~~ \nu = { \mu \over \tanh \tilde \gamma } ~,~~ 1 \ll \tau 
  ~~~~~~ \la{LongAp} 
 \eea
 In this temperature regime we  scale the temperature so that 
 \be \la{SigDef}
 { \sigma \over q } = e^{ -\beta \nu } ~,~~~~~~~ \nu \equiv { \mu \over \tanh \tilde \gamma } 
 \ee
 with constant $\sigma$. 
 
We now equate the long time expansion of \nref{shortAp} to the short time expansion of \nref{LongAp} to obtain 
\bea
 G_{LL} &\sim & \half ( 1 + { g_{LL} \over q }   ) \sim  \half -  { 1 \over q } ( \log{J \over \alpha } + \gamma + \alpha \tau ) = A \cosh { \beta \nu \over 2 } - \tau \nu A \sinh { \beta \nu \over 2 }  
\cr
-i G_{LR} & \sim  & \half ( 1 + { g_{LR} \over q }   )\sim  \half -  { 1 \over q } ( \log{J \over \tilde \alpha } +\tilde \gamma + \tilde \alpha \tau ) = A\sinh{ \beta \nu \over 2 } - \tau \nu 
A \cosh{ \beta \nu \over 2}  ~~~ 
\eea
These equations, together with \nref{SigDef}, imply that 
\be
\tilde \alpha = \alpha ~,~~~~~~~ \tilde \gamma = \gamma + \sigma 
\ee
In addition, from $g_{LL}(0)=0$ and the proper discontinuity of the derivative of $g_{LR}$ at the origin we get 
\be
\alpha = {\cal J } \sinh \gamma ~,~~~~~~~~ \hat \mu = 2 \tilde \alpha \tanh \tilde \gamma = 2 {\cal J } \sinh \gamma \tanh \tilde \gamma
\ee
These equations, together with \nref{SigDef} give us $\gamma$ and $\sigma$ as a function of the physical parameters $\hat \mu , ~{\cal J },~ \beta$. 
Alternatively, we can first write 
\be \la{gasig}
{ \hat \mu \over 2 {\cal J } } =  \sinh \gamma \tanh \tilde \gamma = \sinh \gamma \tanh(\gamma + \sigma) ~,~~~~~~ \beta \mu = \tanh \tilde \gamma \log( q/\sigma)
\ee
and solve the first equation to find $\gamma(\sigma)$  and then use the second equation 
 to find $\beta \mu $ as a function of $\sigma$ (all for a fixed ratio $\hat \mu/{\cal J }$). 
Below we will write expressions for the energy, free energy and entropy, in terms of $\gamma$ and $\sigma$, which can then be viewed as functions of 
the temperature by solving the equations as indicated here. 

We can   find the energy using  \nref{EGen} to obtain 
\be \la{EnAp}
{ E \over N } = { \hat \mu \over q^2 } \left[ - { q \over 2 } + 1 - { 1 \over \tanh \gamma \tanh \tilde \gamma } - \log\left( { \sinh \gamma \over \cosh \tilde \gamma } \right)
 \right] = - \partial_\beta \ell  ~,~~~~\ell \equiv \log Z/N
 \ee
where we denoted by  $\ell = \log Z/N$ the logarithm of the partition function, then we know that $E/N = -\partial_\beta \ell $. In principle, we could integrate this 
 equation to find $\ell$. The problem is that this derivative is taken with fixed values of $\hat \mu $ and $\cal J$. 
 
 Instead we will notice that from the effective action \nref{EfAct} we can write 
 \bea 
 {\cal J } \partial_{\cal J } \ell &=&  \beta \int_0^\beta  d\tau { \cal J }^2 ( e^{g_{LL} } + e^{g_{LR} } ) = 
 { \beta \hat \mu \over q^2 } \left[ { 1 \over \tanh \gamma \tanh \tilde \gamma } - 1 \right] 
 \cr
 \mu \partial_\mu \ell &=& - i \beta \mu G_{LR}(0)  = {\beta  \hat \mu \over q^2 } \left[  { q \over 2 } + \log \left( {\sinh \gamma \over \cosh \tilde \gamma }\right)  \right]  \la{FreenAp}
 \eea
 where we used that $G_{ab}$ obeys the equations of motion to conclude that only the explicit dependence on $J$ and $\mu$ 
  of the action \nref{EfAct} contributes to its derivatives. 
 These expressions are consistent with \nref{EnAp} being $E/N = -\partial_\beta \ell $, and with $\ell$ being a function 
 $\ell( \beta \hat \mu , \beta {\cal J } )$ of dimensionless variables.  We can use this fact, together with
  \nref{FreenAp} to write the derivatives of $\ell$ with respect to $\gamma$ and $\sigma$, with $\ell$ viewed as a function of $\gamma $ and $\sigma$. 
  Using \nref{gasig} and \nref{SigDef} we can compute 
 \bea \la{Conv}
    & ~&  { \partial_\gamma \mu \over \mu } - { \partial_\gamma {\cal J } \over {\cal J } } =  { 1 \over \tanh \gamma } + { 
     1 \over \sinh \tilde \gamma \cosh \tilde \gamma } ~,~~~~
     { \partial_\gamma ( \beta \mu ) \over \beta \mu } = { 1 \over \sinh \tilde \gamma \cosh \tilde \gamma } 
    \cr
    & ~&  { \partial_\sigma  \mu \over \mu } - { \partial_\sigma {\cal J } \over {\cal J } } =  { 
     1 \over \sinh \tilde \gamma \cosh \tilde \gamma } ~,~~~~~ \partial_\sigma \log (\beta \mu) = { 1 \over \sinh \tilde \gamma \cosh \tilde \gamma }  - { 1\over \sigma \log(q/\sigma) } 
     \eea
    We now view $\ell$ as a function of $\ell(\gamma , \sigma)$ with the physical ratios of parameters $\hat \mu, ~\cal J ,~1/\beta$ determined as in 
    \nref{gasig}. We then write 
    \bea \la{siggader}
    \partial_\gamma \ell &=& {\partial_\gamma (\beta \hat \mu ) \over \beta \hat \mu } \beta \partial_\beta \ell  + ({\partial_\gamma {\cal J } \over {\cal J } } - 
     { \partial_\gamma \mu \over \mu } ) { \cal J } \partial_{\cal J } \ell 
     \cr
     \partial_\sigma \ell & =&  {\partial_\gamma (\beta \hat \mu ) \over \beta \hat \mu } \beta \partial_\beta \ell +  ({\partial_\sigma {\cal J } \over {\cal J } } - 
     { \partial_\sigma \mu \over \mu } ) { \cal J } \partial_{\cal J } \ell 
     \eea
       where we used that $(\mu \partial_\mu + {\cal J } \partial_{\cal J } - \beta \partial_\beta ) \ell =0$  by dimensional analysis. Using \nref{gasig}, \nref{Conv}, 
       \nref{EnAp}  we can 
       express the right hand sides of \nref{siggader} in terms of $\sigma, ~\gamma$ and then integrate the two equations to find the expression for 
       $\ell$ quoted in \nref{Freeqlog} and reproduced here for convenience 
       \be
       \ell =   {   \tanh \tilde \gamma  \log(q/\sigma) \over q  } \left[  { q \over 2 } - 1 + { 1 \over \tanh \gamma \tanh \tilde \gamma } + \log { \sinh \gamma \over \cosh \tilde \gamma } + { \sigma \over \tanh \tilde \gamma } \right] + { \sigma \over q } 
       \ee
   Using the energy \nref{EnAp} we can also write the entropy 
   \be
    S/N = \ell - \beta \partial_\beta \ell =  { \sigma \over q } \left( 1 + \log{q\over \sigma } \right)  = e^{ - \nu \beta } ( 1+ \beta \nu ) 
    \ee
    It is curious  that the entropy looks exactly like that of a slightly excited oscillators of energy $\nu$. 
  
  In figure \ref{FreeEnergyLq}(a) we plot the resulting free energy as a function of the temperature. We see that we have a first order transition. This is despite the fact that 
  the free energy is a smooth function of $\sigma$ (for fixed $\mu/{\cal J }$). Figure \ref{FreeEnergyLq}(b)  shows that in the microcannonical ensemble we have a completely smooth 
  behavior. We also find that the energy is a monotonic function of $\sigma$. Finally, in figure \ref{Temperature} 
  we have ploted the temperature as a function of the energy to show how it increases, 
  decreases and then increases again.

                 When we take the limit $\sigma \to 0$ we get $\tilde \gamma = \gamma$ and reproduce the   zero temperature solution \nref{SolPar}. On the other hand, 
           when $\sigma \to \infty$, we get that $\tilde \gamma \to \infty$ and 
           \bea 
           2 \alpha &=& \hat \mu , ~~~~~ \sinh \gamma = { \hat \mu \over 2 {\cal J } } ~,~~~~~~{\rm for } ~~~ \sigma \gg 1 
           \cr
           \ell & \sim & { \beta \mu \over 2  } + e^{ - \beta \mu }  +  { \beta \mu \over 2 } \left[ \log(2 \sinh \gamma)+ { 1 \over \tanh \gamma } - \gamma 
         -1\right] ~,~ ~~~ \sigma \gg 1 \la{FreeLlq}
              \eea
           Note that in this regime $\nu$ is becoming independent of the temperature and equal to $\mu$, which is 
            the naively expected gap for the Hamiltonian $H_{\rm int}$. 
           We also see that since $\tilde \gamma $ is becoming large, then $ e^{g_{LR}}$ is becoming very small everywhere, a fact that will be important below.  
           

      \subsection{Inverse temperature of order $\beta \sim q $ }
      
      We have seen that in the  hight temperature region  of the previous range the function $e^{g_{LR}}$ is becoming very small. What is happening is that 
      $- 2i G_{LR}$ is becoming less than one everywhere, so that we can neglect the term $G_{LR}^{q-1}$ in the expression for the self energy. 
      Therefore, in this regime,    the left-right  self energy becomes $\Sigma_{LR}(\tau) =  -i \mu \delta(\tau)$. 
      This also  implies  that $\nu = \mu$ in the long time approximation to the functions 
      \bea  \la{Gbetq}
      G_{LL} &=& \half { \cosh \mu ( { \beta \over 2} - \tau ) \over \cosh { \mu \beta \over 2 } }  + o(1/q)  
      \cr
      G_{LR} &=& { i \over 2}  { \sinh \mu ( { \beta \over 2} - \tau ) \over \cosh { \mu \beta \over 2 } }   + o(1/q) ~,~~~~~~~ {\rm for} ~~ \tau\gg 1
      \eea
      where we have imposed the appropriate periodicity conditions on the thermal circle.   Notice that in 
      this regime $\beta \mu$ and  $\beta \tau $  are of order one. 
      We will now consider a short time solution $g_{LL}$ given by the usual expression \nref{gLLgLR}. Matching its long time behavior to the short time behavior of 
      \nref{Gbetq} we obtain 
      \be \la{algambeq}
         2 \alpha = \hat \mu \tanh { \mu \beta \over 2}  ~,~~~~ \sinh \gamma = { \hat \mu \over 2 {\cal J } } \tanh {\mu \beta \over 2 }      
          ~,~~~\mu = {\hat \mu \over q} 
         \ee
         On the other hand, the short time behavior of $G_{LR}$ is still given by \nref{Gbetq}.  
         Notice that at short times 
         $- 2 i G_{LR} (0) \sim \tanh { \mu \beta \over 2 } $ which is less than one in the present regime, consistent with our assumptions. 
         We can now evaluate various derivatives of the free energy in a simple way, as it was done above. 
         We find 
         \bea
         {\cal J } \partial_{\cal J } \ell &=& {   \beta \mu  \over q }\tanh {\beta \mu \over 2 } \left[ { 1 \over \tanh \gamma} -1 \right] 
         \cr
         \mu \partial_\mu \ell & = & { \beta \mu \over 2 } \tanh { \mu \beta \over 2 }   + o(1/q) 
         \eea 
         This can be integrated to 
         \be \la{Freebetq}
         \ell = \log[ 2 \cosh { \beta \mu \over 2 } ] + {  \beta \mu \over q } \tanh{ \beta \mu \over 2 } \left[ \log(2 \sinh \gamma)+ { 1 \over \tanh \gamma } - \gamma 
         -1\right] 
         \ee
         where we matched the integration constant by comparing with the expected answer at ${\cal J}=0$ (or $\gamma = \infty$). 
          Notice that the first term in \nref{Freebetq}
           is the leading order piece of the free energy and it looks like the partition function of $N$ free fermionic oscillators of frequency $\mu$. 
          The second term is a $1/q$ correction. For relatively low temperature, where $\beta \mu \gg 1$, these expressions match \nref{FreeLlq}, 
           which is the high temperature limit of the previous temperature range.  
          It is also interesting to take the large temperature limit of \nref{algambeq} and \nref{Freebetq}  to obtain 
          \be \la{FbeqLa}
          \gamma \sim { (\mu \beta)^2 \over 4 q {\cal J } } ~,~~~~~~ \ell \sim \log 2 + { (\beta \mu)^2 \over 8 }   +
          { 2 \beta {\cal J } \over     q^2 } + { (\beta \mu)^2 \over 2 q } \log\left( { (\mu \beta )^2 \over 4 q {\cal J } } \right)  + \cdots  
          \ee
          
          One important point about this regime is that when the temperature increases from  
           low to  high values  within this regime, the entropy rises from close to zero to close to $\log 2$, which is close to the twice the ground state entropy of two separate SYK systems at large temperature. 
         However, despite the entropy  being large, there are still some other properties that are still different from the high temperature regime. 
         For example the Liapunov exponent is still far from maximal. We will discuss this in more detail  as we  examine the next regime. 
         
         \subsection{  Inverse temperatures of the form $\beta \sim \sqrt{q} $ } 
         \la{SecSqrtq}

         In this regime we can approximate $G_{LL} = \half (1 + g_{LL}/q) $ in the whole temperature range.
         As before, we can approximate $\Sigma_{LR}(\tau) = -  i \mu \delta(\tau)$.  From \nref{SDCoup}  
         we get 
         \be \la{GLRHT}
         0=\partial_\tau   G_{LR} - \Sigma_{LR} * G_{LR} +   i \mu G_{RR} \sim \partial_\tau G_{LR} + i { \mu \over 2 }  \longrightarrow G_{LR} =  
         { i \over 2 } \mu ( { \beta \over 2 } - \tau ) = -G_{RL}
         \ee
         where we used that $\Sigma_{LL}$ goes as $1/q$ and can be neglected, as well as the fact that $G_{RR} = \half $ to leading order. 
          We have also used the condition that  $G_{LR}({\beta\over 2 } + x) = - G_{LR}( {\beta \over 2 } - x ) $. 
          This solution also agrees with the large temperature approximation (small $\mu \beta$) of \nref{Gbetq}. 
           We now use this equation and insert it into the time derivative of the first equation in \nref{SDCoup} to get 
           \be \la{eqnlin}
  0=\partial_\tau \left[ \partial_\tau G_{LL} - \Sigma_{LL} * G_{LL} + i \mu G_{RL} \right] ~~\longrightarrow~~ 0=
              \partial_\tau^2 g_{LL} - 2 {\cal J}^2 e^{ g_{LL} } - {q  \mu^2 }  
           \ee
           where we used \nref{GLRHT}. 
           Now, naively, we could neglect the last term since it goes like $\hat \mu^2/q$. However, for large times, of order $\sqrt{q}$,  the first terms are also similar. 
           This can be seen most clearly by defining rescaled variables $x$ and $e^{\hat g}$ via 
           \be
           x = { \tau - \beta/2 \over \beta }   ~,~~~~ x \in [ -\half, \half ]~,~~~~~ e^{\hat g } = {      (\beta  {\cal J })^2 e^{   g_{LL} }  } 
            \ee
           Now \nref{eqnlin} becomes 
            \be \la{eqnghat}
            \partial_x^2 \hat g - 2 e^{\hat g } - 2k =0 ~,~~~~~~~k \equiv  { q \beta^2 \mu^2 \over 2 } 
            \ee
            We should then look for solutions to this equation where $\hat g $ diverges at $x =\pm \half $. The first integral of 
            \nref{eqnghat} is $ {({\hat g}')}^2 - 4 e^{\hat g } - 4 \hat g k  =$constant. We can then further integrate this as 
            \be \la{IntF}
  2  x = \int_{\hat g_0}^{\hat g}  { d g \over \sqrt{ e^{ g} - e^{\hat g_0 } + k( g - \hat g_0 ) } }
  \ee
  where $\hat g_0$ is the value of $\hat g$ at $x=0$. This is determined by solving the equation 
  \be
  1 = \int_{\hat g_0}^{\infty}  { d g \over \sqrt{ e^{ g} - e^{\hat g_0 } +  k( g - \hat g_0 ) } } ~,~~~{\rm or }~~~~ e^{  \hat g_0/2} = 
  \int_0^\infty  { d g \over \sqrt{ e^g - 1 + \tilde k g} }
 \la{Cond}
  \ee
  where $\tilde k = k e^{-\hat g_0 }$. The 
   second expression is a bit more useful for numerical evaluations. 
 
   When $k=0$, we have $e^{\hat g_0} = \pi^2$, $ e^{\hat g} = { \pi^2 \over \cos^2 \pi x } $. 
  We can then consider the regime when $k$ is very large.  One can check numerically that $\hat g_0$ decreases as $k$ increases. At $k=0$ we have the same
  equation as for a single copy of a large $q$ SYK model \cite{Maldacena:2016hyu}, but the boundary conditions we are imposing here are slightly different here due
  to the way we rescaled the coordinates\footnote{Namely, 
  in this subsection we are approximating the boundary condition $e^{\hat g(x=\pm 1/2)} = ( {\cal J \beta } )^2 $ by infinity. }. 
 
 The equations discussed here have also appeared in \cite{Garcia-Garcia:2017bkg}, for a closely related problem.

 The free energy has the form 
 \be \la{FreeSrtq}
 \ell = \log 2 + { ( \beta \mu)^2 \over 8 } + {2  \beta {\cal J } \over   q^2 } + { (\mu \beta)^2 \over 2  q  }  \log( { \cal J }\beta )
+ { h[ q (\mu \beta)^2 ] \over q^2  }  \ee
 We used that  
 \be 
  - 2 e^{ \hat g_0/2 } + \int_{\hat g_0}^\infty d g \left[ {e^g \over \sqrt{ e^g - e^{\hat g_0} + k ( g - \hat g_0 ) } }  - e^{ g/2} \right] = k = { q (\beta \mu)^2 \over 2 } 
 \ee
 thanks to \nref{Cond}. Interestingly, the coefficient of the $(\beta \mu)^2$ 
  term in \nref{FreeSrtq} is the same as the one in 
  \nref{FbeqLa} throughout the temperature regime of this subsection.
   In order to determine the logarithmic term in \nref{FreeSrtq} we used the expression for 
  ${\cal J } \partial_{\cal J } \ell ={ 1 \over q^2}  \int_{-1/2}^{1/2} dx e^{ \hat g} $, and took into account that the divergence near the end points is regularized and gives a factor of ${\cal J }$, with the rest being ${\cal J}$ independent. This method does not
  determine the $\cal \mu$ dependence at this order. For that reason we could not fix the function $h$ in 
  \nref{FreeSrtq}. The leading $\mu$ dependence comes from the expression for the $\mu$ derivative, 
  $\partial_\mu \ell = - i \beta G_{LR}(0) = \beta^2 \mu/4$. 
 
 {\bf Chaos exponent } 
 
 We now look at the chaos exponent. This can be done by computing the retarded kernel in the thermofield double configuration. 
 Here we are thinking about the thermofield double of the already doubled physical system. This thermofield double contains four copies 
 of the SYK model. To go to the second side of the
  thermofield double we can analytically continue after a shift by  $\beta/2$ along the Euclidean time direction. 
  The retarded Kernel obeys the equation 
  \be
  \partial_{t_1} \partial_{t_2 } K(t_1,t_2;t_3 ,t_4) = 2 q  \Sigma_{LL}( \beta/2 + i t_{34} ) 
  \ee
 Here we have neglected $\Sigma_{LR}( \beta/2 + i t_{34})$. Here we think of $t_1,~t_3$ as living in the original system while $t_2,~t_4$ live in the second copy of the thermofield double. 
  We are interested in eigenfunctions of this Kernel of the form $K \psi = \psi$, with $\psi = e^{\lambda (t_1 + t_2)/2} \chi(t_1 -t_2)$. 
  Then $\chi$ obeys the equation 
  \be  \la{lameq}
   - { (\lambda \beta)^2 \over 4 } \chi = [ - \partial_y^2 - 2 e^{ \hat g_l } ]\chi(y) 
   \ee
   where $\hat g_l(y) = \hat g( i y )$ is the analytic continuation of the Euclidean funtion $\hat g(x)$ (note that $x=0=y$ corresponds to $\tau = \beta/2$). 
   $\hat g_l(y)$ has a maximum at $y=0$ and then it decreases as $y \to \pm \infty$. This implies that \nref{lameq} is a Schroedinger problem 
   with an attractive potential that asymptotes to zero at infinity, and we are looking for its bound states. 
   There is exactly one bound state for the following reason. We can show that 
   $\chi \sim \partial_y \hat g_l$ is a zero energy state by  taking the 
   $\partial_y$ derivative of the equation that $\hat g_l(y)$ obeys (which is the analytic continuation of  \nref{eqnghat}). 
   Furthermore $\partial_y \hat g_l$ crosses zero only once. Therefore there is a single state with energy less than zero in the Schroedinger problem \nref{lameq}.
   
    We can analyze extreme limits analytically. For $k=0$ we recover the usual expression $\lambda = 2\pi/\beta$. For large $k$, $k\gg 1$, 
     the potential can be approximated
    by a $\delta$ function and we can estimate 
    \be
    \la{LkEstim} 
     { \lambda \beta \over 2 \pi }  \sim { k^{3/2} e^{ - k/4}  \over \sqrt{\pi } } ~, ~~~~~~~k = { q \mu^2 \beta^2 \over  2 } 
     \ee
      which is very small for large $k$. 
    So we see that as the temperature increases, and we go from large $k$ to small $k$, the chaos exponent goes from being very small to being maximal.

  To study the problem numerically and produce figure \ref{ChaosExponent} we found it convenient to do the following. 
  First we defined a shifted function $\tilde g = \hat g - \hat g_0$ and rescaled variables so that now we have
  \be
   2 \tilde y = \int_{\tilde g } ^{0} { dg \over \sqrt{ 1 - e^g - \tilde k g } }~,~~~~~~ \tilde y = y e^ {  \hat g_0/2}   ~,~~~~\tilde k = k e^{ - \hat g_0 } 
   \ee
   The eigenvalue equation now looks 
   \be \la{tlaDe}
    - {\tilde \lambda}^2 \chi = [ - \partial_{\tilde y}^2 - 2 e^{ \tilde g } ] \chi(\tilde y ) ~,~~~~~ \tilde \lambda = { \lambda \beta \over 2 } e^{-\hat g_0/2 } 
  \ee
  Then we changed variables\footnote{We thank Zhenbin Yang for this suggestion.} 
   from $\tilde y$ to $\tilde g$ so that the eigenvalue equation is 
  \be
  -{ \tilde \lambda}^2 \chi = \left[  - 4 (1 - e^{\tilde g } - \tilde k \tilde g ) \partial_{\tilde g}^2  + 2 ( e^{ \tilde g } + \tilde k ) \partial_{\tilde g } - 2 e^{ \tilde g } 
  \right] \chi 
  \ee
  We want to solve this equation for $g\in [ -\infty, 0]$ with the boundary condition that $\chi$ is regular at $\tilde g =0$ and that it decays as $\tilde g \to - \infty$. 
  In this way we find $\tilde \lambda(\tilde k)$. We can then find $\hat g_0$ as a function of $\tilde k$ from \nref{Cond}, and then finally compute $k(\tilde k)$ and 
  $\lambda$ from \nref{tlaDe} \nref{eqnghat}.

    \subsection{Inverse temperatures of order one } 
    
    In this regime we can set $k=0$ in \nref{eqnghat} and impose $e^{\hat g (x=\pm \half) } = (\beta {\cal J } )^2$. 
     We recover the results we had for the two decoupled SYK models. 
    We will not do this in detail since this was discussed in \cite{Maldacena:2016hyu}. The free energy is twice the result in \cite{Maldacena:2016hyu}
    \be
    \ell_{\mu =0} = \log 2 + { 4 \over q^2 } \hat v \left[ \tan \hat v  - {\hat v \over 2  } \right]  ~,~~~~~~{\beta \cal J } = { 2 \hat v \over \cos\hat v } 
    \ee
    What we should note is that the leading deviation from the $\mu=0$ result is 
    \be
     \ell = \ell_{\mu =0} +  { ( \beta \mu)^2 \over 8 } 
     \ee
  This term can be directly computed by using $\partial_\mu \ell = - i \beta G_{LR}(0) $ and the expression \nref{GLRHT}
    which continues to be valid in this regime.  It is also given by iterating twice the interaction Hamiltonian and using the leading order answer for the correlator 
    \be
    \ell - \ell_{\mu =0} =   {\mu^2 \over 2 } \int d\tau_1 d\tau_2 G_{LL}(\tau_{12}) G_{RR}(\tau_{12}) = { ( \beta \mu)^2 \over 8 } 
    \ee
    
    The chaos exponent is maximal in  the region  $\sqrt{q} \ll \beta {\cal J } \ll 1$, as
    we had found in the large temperature range of the previous subsection. It then becomes less than maximal when  
    $\beta {\cal J } \sim 1$, becoming   $\lambda = 2 {\cal J }$ at very high temperatures \cite{Maldacena:2016hyu}. 
    
    \section{ SL(2) charges} 
    \la{SLTwo} 
    
    In this appendix we spell out the general form of the SL(2) charges for the action \ref{fLfRres}. Here we set $N=1$.  
    Using the Noether procedure for the transformation 
      \be  \la{GTapp}
     \delta t_l = \epsilon^0 + \epsilon^+ e^{ i t_l} + \epsilon^-  e^{ - i t_l } ~,~~~~~~~\delta t_r = \epsilon^0  - \epsilon^+ e^{ i t_r} - \epsilon^- e^{ - i t_r }
     \ee
    we find 
    \bea \la{CharG}
   Q_0/N &=& Q^S_0[t_l] + Q_0^S[t_r] + \left( { 1 \over t'_l} + { 1 \over t'_r}  \right) F
   \cr
   Q_+/N &=&  Q^S_+[t_l] - Q_+^S[t_r] + \left ( { e^{ i t_l }  \over t'_l} - { e^{ i t_r}  \over t'_r}  \right) F
   \cr
   Q_- /N&=&  Q^S_-[t_l] - Q_-^S[t_r]  +  \left ( { e^{- i  t_l  }  \over t'_l} - { e^{- i t_r}  \over t'_r} \right ) F
   \cr
   F &\equiv&    \Delta \eta \left[  { t'_l t'_r \over \cos^2{ t_l - t_r \over 2 } } \right]^{\Delta }
   \cr
   Q_0^S[t] &=&- {t'} + { { t''}^2 \over { t'}^3 } - { t''' \over { t'}^2 } 
   \cr
   Q_+^S[t] & = & e^{ it} \left(     i {t'' \over t' } + { {t''}^2 \over { t'}^3 } - { t''' \over { t'}^2 } \right) 
   \cr
   Q_-^S[t] & = & e^{ - i t} \left( -  i {t'' \over t' } + { {t''}^2 \over { t'}^3 } - { t''' \over { t'}^2 } \right) 
   \eea
   
We see that if $t_l(\tilde u) = t_r(\tilde u)$, then $Q_\pm =0$ automatically and $Q_0$ gives 
\be
Q_0/N  = 2 e^{ -\varphi } \left[ - \varphi''   - e^{ 2 \varphi } + \eta \Delta e^{ 2 \Delta \varphi }  \right] ~,~~~~~~{\rm with} ~~~\varphi = \log t' 
\ee

 Let us imagine we  look for solutions with no matter. 
 In this case we need to set the above charges to zero. We now want to argue that any such solution can be gauge transformed to a solution with 
$t_l(\tilde u) = t_r(\tilde u)$. The argument is the following. 
 First imagine we have a general solution with $t_l(\tilde u) \not = t_r(\tilde u)$. Then, by a global SL(2) transformation we can set the two times and their derivatives
 equal at $\tilde u=0$, namely $t_l(0)=t_r(0)$ and $t'_l(0) = t'_r(0)$. 
 Then by imposing that $Q_- - Q_+ =0$ we get that $t''_l(0) = t''_r(0)$. Finally imposing that $Q_- + Q_- =0$ we conclude that $t'''_l(0) = t'''_r(0)$. 
 Since the equations of motion are of fourth order, this implies that they are equal for all later times. 
 This shows that   solutions obeying the constraints can be gauge transformed to solutions obeying $t_l(\tilde u) = t_r(\tilde u)$. 
 
 When we have matter in the bulk (or in the SYK language we excite the  approximately conformal invariant degrees of freedom) 
 we have an extra matter contribution, $q^{M}_a$ to the SL(2) charges. 
 
 It is also interesting to compute the expression for the charges, after we expand around the solution we discussed around \nref{CandSol} \nref{tPrime}. 
 We write 
 \be
 t_l = t' \tilde u + \chi_+(\tilde u) + \chi_-(\tilde u) ~,~~~~~~~t_r = t' \tilde u + \chi_+(\tilde u) - \chi_-(\tilde u) 
 \ee
  The action of the SL(2) gauge transformations \nref{GTapp} is 
  \be
  \delta \chi_+ = \epsilon^0 ~,~~~~~~ \delta \chi_- = \epsilon^+ e^{ i t } + \epsilon^- e^{ - i t } ~,~~~t \equiv t' u 
  \ee
  where we rescaled the time so that now we work in terms of $t$, which is natural if we are expanding around the vacuum solution. 
  Assuming that $t'$ obeys \nref{tPrime} 
  we then find 
  \be \la{CHLin}
  Q_0/N  = - 2 t'  d_t [ d_t^2 + 2 ( 1 - \Delta) ] \chi_+ ~,~~~~~ Q_\pm/N = -2 t' e^{ \pm i t }   (d_t  \mp i ) (d_t^2 + 1) \chi_- 
   \ee
   It is also instructive to write down the expression for the total energy \nref{EnGenN}, which is 
   \be \la{EnLinCh}
   E_{\tilde u}  = t'^2N  \left[ - { 1 -\Delta   \over \Delta }   - 2 \left\{ d_t^3 \chi_+ + 2 ( 1 - \Delta) d_t \chi_+  \right\} \right]  = 
   -  t'^2 N  { 1 -\Delta  \over \Delta } - t' q_0 
  \ee
  This is the expected expression for the energy at this order in the $1/N$ expansion. Namely, the energy is just given by the bulk energy\footnote{
  The energy is minus the quantity $q_0$ which is the Noether charge under time translations.}  in global time, up to a 
  rescaling factor of $t'$. 
  Now, here $q_a$ are the charges of the matter theory. We have used the expression for $Q_0$ in \nref{CHLin} and the constraint
  \be \la{Cos}
  Q_a + q_a =0 
  \ee
  Notice that the $\pm$ components of these constraint equations \nref{Cos} simply determine $\chi_-$ from $q_\pm$ and we obtain 
  \be
  Q_\pm + q_\pm =0 ~~\longrightarrow ~~~~\chi_- =-  { 1 \over 8 N  t' } \left(  t e^{ -it } q_+ + t e^{   i t } q_-  \right) 
  \ee
  Where we used that $q_\pm$ are just constants. We can solve these even as operator equations. 
  The expression for $\chi$ is of order $1/N$, for $q_\pm $ of order one, so that the small $\chi$ approximation was justified. 
  Notice that $\chi_\pm$ do not appear in the expression for the
  energy \nref{EnLinCh} to this order in the $1/N$ expansion. 
 
     It is also interesting to note that 
     \bea \la{BoostG}
     H_R - H_L &=& { d\tilde u \over d u } \left[ -\{ \tan{ t_r(\tilde u)  \over 2 } , \tilde u \} + \{ \tan{ t_l(\tilde u)  \over 2 } , \tilde u \} \right] = 2 N  { d \tilde u \over u } 
     t'^2  \left[ d_t^2 \chi_-+ d_t \chi_- \right] =
     \cr
     &=&{ dt \over d u } \left[ \cos t { ( q_+ + q_-) \over 2}  + \sin t { ( q_+ - q_-) \over  2 i } \right] 
     \eea
     Note that here $H_R$ and $H_L$ should be the Hamiltonians of the right or left SYK models, without the interaction term. In the gravity picture this is the difference of
     the left and right ADM 
     mass operators. 
     Now, this equation has an interesting interpretation. It is saying that at $t=0$, the operator $H_R - H_L$ can be identified with $(q_+ + q_-)/2$ which turns out to
     be the boost generator around the origin in $AdS_2$. 
     Finally, the same operator, but at $t= \pi/2$, can be identified as the third generator of the bulk SL(2) transformations. Near the origin of $AdS_2$, it acts as
     a spatial translation. By origin we mean $t=0$, $\sigma = \pi/2$ in the coordinates in \nref{GlobCoord}.  
     
     Notice that in the SYK model we can still define the generators $q_a$ as acting on the conformal invariant degrees of freedom, the excitations which are 
     not described by the $t_r, ~t_l$ variables. 
     
     Note that the operator in \nref{BoostG} is time dependent. Indeed, this operator is not conserved due to the interaction term in the full Halmitonian of the coupled 
     system. If we turn off the coupling at $t=0$, then $H_R - H_L$ will be conserved and equal to the boost generator in the bulk field theory, consistent with the
     picture in section \ref{MatchingTFD}. 
     
\section{ Boundary conditions and negative energy for a CFT in the bulk } 

In this section we imagine we start with $AdS_2$ and we consider a conformal field theory on $AdS_2$. 
Since it is a conformal field theory, we can forget about the overall scale factor of the metric in \nref{GlobCoord}  and consider it on 
a strip, $ds^2 = - dt^2 + d\sigma^2$. 
In this appendix we discuss a couple of issues about this. With standard boundary conditions on the strip, we will ask 
whether this setup gives rist to negative energy or not.  
Then we will consider the specific case of free fermions and work out the state for various value of the boundary coupling. 

\subsection{Negative energy on the strip goes to zero energy in $AdS_2$} 
 \la{GenBulkCFT}

It is well known that the energy of a conformal field theory on a strip is given by 
\be  \la{CardyEn}
E = - { c \over 24 } { \pi \over L } ~,~~~~~~~{\rm or } ~~~~~~ E = - { c \over 24 } ~,~~{\rm for}~~L= \pi 
\ee
Here we ask whether this is enough to give rise to negative null energy in the sense discussed around \nref{NegEn}. The conclusion will be no. 
We have to realize that we are talking about two different notions of energy. When we think about the flat strip, we imagine renormalizing the energy with a constant cutoff on the flat metric of the strip. When we talk about $AdS_2$ we imagine doing the same with a cutoff that is constant in $AdS_2$ proper length. The difference is just a scale factor. We can work out the difference by looking at the general form of the conformal anomaly (in Lorentzian signature)
\be
 Z[ g = e^{ 2 \omega } \hat g ] = \exp\left\{ i { c \over 2 4 \pi  } \int d^2 x \sqrt{\hat g } [ \hat R \omega + ( \hat \nabla \omega )^2 ] \right\}  Z[ \hat g ] 
 \ee
 This means that the stress tensor computed by taking derivatives with respect to each metric is given by 
 \bea \la{Tghat}
 T_{\mu \nu}^{g} & =& T_{\mu \nu}^{\hat g } - { c \over 12 \pi  }   \left[ 
  \partial_\mu \omega  \partial_\nu \omega - \half \hat g_{\mu \nu } ( \hat \nabla \omega )^2 -  \hat \nabla_\nu \hat \nabla_\mu \omega +
   \hat g_{\mu \nu} \hat \nabla^2 \omega
\right] 
\cr 
 T_{\mu \nu}^{g}&=& T_{\mu \nu}^{\hat g } + { c \over 24 \pi } \left( \begin{array}{cc} 1 & 0 \\ 0 & 1 \end{array} \right)  - { c \over 24 \pi} 
 g_{\mu \nu} 
\eea
We see that the second term cancels precisely the negative energy we had in $T^{\hat g}_{\mu \nu}$. The last term 
is proportional to $g_{\mu \nu}$ (not  $\hat g_{\mu \nu}$) and it is the piece that will contribute to the conformal 
anomaly in $AdS_2$, $T^{g , \, \mu}_{\mu } = { c \over 24 } R $. 

We conclude that $T^{g}_{++}$ vanishes. The piece that is proportional to the metric in the stress tensor is a contribution proportional to $\sqrt{g} $ in the action, and it can absorbed by a shift of the dilaton (and also a shift of the coefficient of
the topological term $\phi_0 \int \sqrt{g} R $ in the action). 

  One comment, is that in a CFT we have have various boundary conditions. In this discussion we have assumed that
  the boundary condition on the left side is the ``same'' as the one on the right side. By ``same'' here we mean the 
  CPT conjugate one. It is the one we get by taking a boundary condition along the real line and then mapping the 
  upper half plane to the strip by a conformal transformation. With this pair of boundary conditions we have the minimal energy $E=- { c \over 24 }$. For other possible combinations we have higher energy, again consistent with the null energy condition in $AdS_2$. 

\subsection{Negative energy for a free fermion $AdS_2$ plus a boundary interaction }
\la{FBF}

In this subsection we consider a free real 
fermion theory on a flat strip. We start with the usual boundary conditions 
\be 
\psi_+= \psi_- |_{\sigma =0} ~,~~~~~~ \psi_+ = - \psi_- |_{\sigma = \pi }  ~~~{\rm for ~all~~}t
\ee
After we add the interaction that relates left and right, we end up modifying the boundary conditions to 
\be  \la{BCFer}
\psi_+ |_{\sigma =0} = \cos \pi  \epsilon \psi_-|_{\sigma =0}  - \sin \pi  \epsilon \psi_+ |_{\sigma = \pi } 
 ~,~~~~~~~~  \psi_-|_{\sigma =\pi } = - \cos \pi \epsilon \psi_-|_{\sigma = \pi }  - \sin \pi  \epsilon \psi_- |_{\sigma =\pi } 
\ee
We can get the first relation if we thinking of the reflection and transmision of the incoming fermion. The reflection and transmision factors should be real since the fermion fields are real. In addition, consistency with commutation relations (or unitarity) implies that we can write them as a sine and cosine. We get the second relation in \nref{BCFer} by a rotation (or CPT transformation)
 to the other boundary.  Here  $\epsilon$ is a parameter that is related to the coupling between the left and the right sides. We expect that $\epsilon \propto \eta $, and in comparing with the general discussion, this case corresponds to 
 $\Delta =\half$. 
  Writing $\psi_+(t)  \propto  b_\omega  e^{ - i \omega t } $ and inserting in \nref{BCFer} we get the condition 
  \be \la{Freqep}
  \omega = \pm \left[  \half + \epsilon  + 2 n \right] ~,~~~~~~~~~{\rm or }~~~ \cos \pi \omega = - \sin \pi \epsilon 
  \ee 
  Assuming $-\half  < \epsilon < \half$, we see that the positive energies are $\half + \epsilon + 2 n $, $n \geq 0$ and 
  $ -\half - \epsilon +  2n $, $n\geq 1$. For $\epsilon =0$ these two two towers collapse to $ \half + m $, $m\geq 0$. 
  We can compute the ground state energy by adding all the zero point energies of these fermionic oscillators (suitably) 
  regularized to obtain 
  \be  \la{GsV}
  E = - { 1 \over 48 } \left[ 1 + 12 \epsilon ( 1 -  \epsilon ) \right] ~,~~~~~~ -\half \leq \epsilon \leq \half 
  \ee
  Special values are 
  \be
  E( \epsilon =0) = - { 1 \over 48 } ~,~~~~~~E(\epsilon = \half ) = - { 1 \over 12 } ~,~~~~~~ E( \epsilon = - \half) = { 1\over 6 } 
  \ee
  In the first case we recover the energy expected from the general formula \nref{CardyEn}, where for a single 
  real fermion we have $c=\half$. 
  The second case corresponds to a fermion on a circle of length $L=\pi$ 
   with antiperiodic boundary conditions. For a circle the 
  ground state formula analogous to \nref{CardyEn} is  $E = - { c  \over 12 } { 2 \pi \over L } $, where $L$ is the size of the circle. The last case also corresponds to a fermion on a circle of length $\pi$, but with periodic boundary conditions. 
  So we see that \nref{GsV} interpolates between these last two extreme situations.
  
  Notice that, for any $CFT$ on the strip, we can couple the two sides such that we end up with ``transparent'' boundary conditions. Namely, we end up with the boundary conditions for a CFT on a circle of radius $\pi$. In such a case, 
  the final energy can be computed in general as $E =- { c \over 6}$, which is more negative than the strip result 
  \nref{CardyEn}. 
   
  The analysis we made in the bulk of the paper we always assumed that $\eta \propto \epsilon$ was small. 
  For this case, we see we can also take it to be large. However, there is a maximum effective value which is 
  $\epsilon \pm \half$. If we continue beyond that value we get the same physics \nref{Freqep}.

\subsection{Profile for the bulk dilaton } 
\la{DilProf} 

Here we will compute the profile for the dilaton for the negative energy states discussed above. 

In Lorentzian signature, the final form of the stress tensor on the flat strip  is then  $T_{t\sigma}=0$, 
  $T_{tt} = T_{\sigma \sigma }  = E/\pi $ ,with $E$ as in \nref{GsV} and $T_{t\sigma} =0$. 
  When we transform this into an $AdS_2$ stress tensor, as in \nref{Tghat}  that still has $T_{t\sigma}=0$, but now
  the piece that is not proportional to the metric is 
  \be
 { 1 \over 2} (  T^g_{tt} + T^g_{\sigma \sigma} ) = - { 1 \over 4 \pi } \epsilon (1 - \epsilon ) 
\ee
Comparing this with the contribution to the energy of the interaction Hamiltonian in section \ref{LowSec} we find that they 
agree if  
get $\epsilon/4 = \eta $, for small $\epsilon$. 
We can then make an ansatz $\phi = \phi(\sigma)$ and then from \nref{NegEn} we find 
\be \la{phiso}
- \partial_\sigma ( \sin^2 \sigma \partial_\sigma \phi) = - { N \over 2 \pi } \epsilon (1- \epsilon ) \sin^2 \sigma ~~~\longrightarrow ~~~ \phi = N { \epsilon (1-\epsilon) \over 4 \pi }  \left[ { ( { \pi \over 2 } - \sigma ) \over \tan \sigma }   + 1 \right]  + { c \over 24 \pi } 
\ee
where we have set the integration constants appropriately. Actually, to fix the additive constant we also need to impose the equation for the dilaton that comes from 
the trace variation of the metric, and use the full stress tensor \nref{Tghat}. Here $N$ is the total number of Majorana fermions, and $c=N/2$.  The additive constant
in \nref{phiso} that is proportional to $c$ comes from the term in the stress tensor \nref{Tghat} that is propotional to the metric. It is easy to see from the form of the
original action \nref{ActJT} that such a term can be removed by a shift of the dilaton, up to an overall topological term in the action. 

We see that we still get that $\phi \propto 1/\sigma$ near $\sigma =0$ and similarly near $\sigma = \pi$. 
Even though the stress tensor components are constant in these coordinates, their invariant values are becoming 
small. In some sense, the negative stress tensor is concentrated away from the boundaries. 

The methods discussed in section \ref{LowSec} reproduce the physics we obtain from the small $\epsilon$ limit 
of this case. 

\section{Derivation of the Liouville effective action}\label{app:Liouville}

In this appendix, we derive the Liouville effective action from the large $q$ limit of the $G,\Sigma$ action (\ref{EfAct}). 

\begin{eqnarray}
 -S_E/N&=&\frac12{\rm Tr}\log\left(\partial_\tau\delta_{ab}-\Sigma_{ab}\right)-\frac12\int d\tau_1d\tau_2\sum_{a,b}\left[\Sigma_{ab}(\tau_1,\tau_2)G_{ab}(\tau_1,\tau_2)-s_{ab}\frac{ {\cal J}^2}{2 q^2}[ 2 G_{ab}(\tau_1,\tau_2)]^q\right] + \nonumber\\
 & &+ { i \hat{\mu} \over {2q}}  \int d\tau_1 \left[ - G_{LR}(\tau_1,\tau_1) + G_{RL}(\tau_1,\tau_1) \right] 
 \end{eqnarray}
 Define
 \be 
 G_{ab}(\tau_1,\tau_2)=G_{0ab}(\tau_1,\tau_2)\left(1+\frac1 qg_{ab}(\tau_1,\tau_2)\right)
 \ee
 with $G_{0LL}(\tau_1,\tau_2)=\frac12{\rm sgn}(\tau_1-\tau_2),~G_{0LR}=\frac i2{\rm sgn}(\hat{\mu})$. $\left[G_0^{-1}\right]_{ab}=\delta_{ab}\partial_\tau$. $G_0$ is chosen to be the two-point function of free fermion with a Hamiltonian $H=i\mu \sum_i\chi_{iL}\chi_{iR}$ with $\mu\rightarrow 0$. In large $q$ limit, the self energy at the saddle point also scales with $\frac 1q$, so that we can expand the determinant term as
 \be
{\rm Tr}\log\left(\partial_\tau\delta_{ab}-\Sigma_{ab}\right)={\rm Tr}\log\left(G_0^{-1}\right)-{\rm Tr}\left(G_0*\Sigma\right)-\frac12{\rm Tr}\left(G_0*\Sigma*G_0*\Sigma\right)+...
 \ee
 Omitting the constant terms ${\rm Tr}\log \left(G_0^{-1}\right)$ and $G_{0LR}-G_{0RL}$ in the action, we obtain
 \begin{eqnarray}
 \frac1{N}S_E&\simeq &\frac14{\rm Tr}\left(G_0*\Sigma*G_0*\Sigma\right)+\frac1{2q}\int d\tau_1 d\tau_2\sum_{ab}\Sigma_{ab}(\tau_1,\tau_2)G_{0ab}(\tau_1,\tau_2)g_{ab}
 (\tau_1,\tau_2)\nonumber\\
 & &-\frac{\mathcal{J}^2}{4q^2}\int d\tau_1d\tau_2e^{g_{ab}(\tau_1,\tau_2)}-\frac{\left|\hat{\mu}\right|}{q^2}\int d\tau g_{LR}(\tau,\tau)
 \end{eqnarray}
 By integrating out $\Sigma$, we will obtain an action of $g_{ab}$. To do that it is helpful to introduce 
 \bea
 \Phi_{ab}(\tau_1,\tau_2)=\left[G_0*\Sigma(\tau_1,\tau_2)\right]_{ab}=\int d\tau G_{0ac}(\tau_1,\tau)\Sigma_{cb}(\tau,\tau_2)
 \eea
 Thus by construction
 \bea
 \Sigma_{ab}(\tau_1,\tau_2)=\partial_{\tau_1}\Phi_{ab}(\tau_1,\tau_2)
 \eea
 
 The effective action can be written as
 \begin{eqnarray}
 \frac1{N}S_E&\simeq &\frac14{\rm Tr}\left(\Phi*\Phi\right)+\frac1{2q}\int d\tau_1 d\tau_2\sum_{ab}\partial_{\tau_1}\Phi_{ab}(\tau_1,\tau_2)G_{0ab}(\tau_1,\tau_2)g_{ab}
 (\tau_1,\tau_2)\nonumber\\
 & &-\frac{\mathcal{J}^2}{4q^2}\int d\tau_1d\tau_2e^{g_{ab}(\tau_1,\tau_2)}-\frac{\left|\hat{\mu}\right|}{q^2}\int d\tau g_{LR}(\tau,\tau)\nonumber\\
 &\simeq &\frac12{\rm Tr}\left(\Phi*\Phi\right)-\frac1{2q}\int d\tau_1 d\tau_2\sum_{ab}\Phi_{ab}(\tau_1,\tau_2)\partial_{\tau_1}\left(G_{0ab}(\tau_1,\tau_2)g_{ab}
 (\tau_1,\tau_2)\right)\nonumber\\
& & -\frac{\mathcal{J}^2}{4q^2}\int d\tau_1d\tau_2e^{g_{ab}(\tau_1,\tau_2)}-\frac{\left|\hat{\mu}\right|}{q^2}\int d\tau g_{LR}(\tau,\tau)
 \end{eqnarray}
 where we have done an integration by part in the second term. Integrating over $\Phi_{ab}$ we obtain the effective action
 \begin{eqnarray}
\frac1NS_{\rm eff}&=&\frac1{4q^2}\int d\tau_1 d\tau_2\sum_{ab}\partial_{\tau_1}\left(G_{0ab}(\tau_1,\tau_2)g_{ab}(\tau_1,\tau_2)\right)\partial_{\tau_2}\left(G_{0ab}(\tau_1,\tau_2)g_{ab}(\tau_1,\tau_2)\right)\nonumber\\
& &-\frac{\mathcal{J}^2}{4q^2}\int d\tau_1d\tau_2\sum_{ab}e^{g_{ab}(\tau_1,\tau_2)}-\frac{\left|\hat{\mu}\right|}{q^2}\int d\tau g_{LR}(\tau,\tau)
 \end{eqnarray}
It should be noted that there is a nontrivial Jacobian relating the path integral of $\Sigma_{ab}$ to that of $\Phi_{ab}$, but the Jacobian is independent from $g_{ab}$ which only contributes a constant term to the effective action.

Since $G_{0ab}$ is a constant except $G_{0LL}$ at $\tau_1=\tau_2$, and $g_{0LL}$ satisfies the boundary condition $g_{0LL}(\tau_1,\tau_1)=0$, the effective action can be simplified to
 \begin{eqnarray}
\frac1NS_{\rm eff}&=&\frac1{8q^2}\int d\tau_1 d\tau_2\left(\partial_{\tau_1}g_{LL}(\tau_1,\tau_2)\partial_{\tau_2}g_{LL}(\tau_1,\tau_2)-\partial_{\tau_1}g_{LR}(\tau_1,\tau_2)\partial_{\tau_2}g_{LR}(\tau_1,\tau_2)\right)\nonumber\\
& &-\frac{\mathcal{J}^2}{2q^2}\int d\tau_1d\tau_2\left(e^{g_{LL}(\tau_1,\tau_2)}+e^{g_{LR}(\tau_1,\tau_2)}\right)-\frac{\left|\hat{\mu}\right|}{q^2}\int d\tau g_{LR}(\tau,\tau)
 \end{eqnarray}
The role of the last term is imposing a boundary condition for $g_{LR}$ at $\tau_1=\tau_2$. Using the symmetry $g_{ab}(\tau_1,\tau_2)=g_{ab}(\tau_2,\tau_1)$ one can consider the action as the Liouville action defined on the half plane $\tau_1>\tau_2$:
  \begin{eqnarray}
\frac1NS_{\rm eff}&=&\frac1{4q^2}\int_{\tau_1>\tau_2} d\tau_1 d\tau_2\left(\partial_{\tau_1}g_{LL}(\tau_1,\tau_2)\partial_{\tau_2}g_{LL}(\tau_1,\tau_2)-\partial_{\tau_1}g_{LR}(\tau_1,\tau_2)\partial_{\tau_2}g_{LR}(\tau_1,\tau_2)\right)\nonumber\\
& &-\frac{\mathcal{J}^2}{q^2}\int_{\tau_1>\tau_2} d\tau_1d\tau_2\left(e^{g_{LL}(\tau_1,\tau_2)}+e^{g_{LR}(\tau_1,\tau_2)}\right)-\frac{\left|\hat{\mu}\right|}{q^2}\int d\tau g_{LR}(\tau,\tau)
\end{eqnarray}
with the boundary condition
\be
g_{LL}(\tau,\tau)=0,~\left.\left(\partial_{\tau_1}-\partial_{\tau_2}\right)g_{LR}(\tau_1,\tau_2)\right|_{\tau_2=\tau_1}=2\left|\hat{\mu}\right| 
\ee
In addition to the boundary condition at $\tau_1=\tau_2$ line, other boundary conditions depend on the problem we are considering. For finite temperature thermal ensemble, periodic boundary condition in $\tau_1,\tau_2$ are imposed. (In that case, we choose $G_0$ is anti-periodic in $\tau_1\rightarrow \tau_1+\beta$ or $\tau_2\rightarrow \tau_2+\beta$, so that $g_{ab}$ is periodic in both directions.)

\section{The two-time solution}\label{app:twotime}

In this appendix we provide more details on finding the interpolating solution we discussed in Sec. \ref{sec:overlap}. The starting point is the general solution to Liouville equations (\ref{generalsol2T}), which we copy here:
\be
e^{g_{LL}}=\frac{h_1'(\tau_1)h_2'(\tau_2)}{\mathcal{J}^2\left({h_1(\tau_1)-h_2(\tau_2)}\right)^2},~e^{g_{LR}}=-\frac{f_1'(\tau_1)f_2'(\tau_2)}{\mathcal{J}^2\left({f_1(\tau_1)-f_2(\tau_2)}\right)^2}
\ee
We start from the two time-translation-invariant solutions, the TFD solution and the coupled ground state solution. The TFD solution is obtained from the thermal solution of a single SYK site\cite{Maldacena:2016hyu}:
\be
e^{g(\tau_1,\tau_2)}=\frac{\check{\alpha}^2}{\mathcal{J}^2\sin^2(\check{\alpha}\left|\tau_1-\tau_2\right|+\check{\gamma})}
\ee
with $\check{\alpha}$ and $\check{\gamma}$ determined by the boundary conditions
\be
\check{\alpha}=\mathcal{J}\sin\check{\gamma},~\check{\alpha}\frac{\beta}2+\check{\gamma}=\frac\pi2 \label{thermalBC}
\ee
By viewing the thermal circle as a doubled system, we can write the thermal solution above in the form of thermal double solution, with $\tau_1,\tau_2\in\left[-\frac\beta4,\frac\beta 4\right]$:
\be
e^{g_{LL}}=\frac{\check{\alpha}^2}{\mathcal{J}^2\sin\left(\check{\alpha}|\tau_1-\tau_2|+\check{\gamma}\right)^2},~ e^{g_{LR}}=\frac{\check{\alpha}^2}{\mathcal{J}^2\sin\left(\check{\alpha}(\tau_1+\tau_2+\frac\beta2)+\check{\gamma}\right)^2}\equiv \frac{\check{\alpha}^2}{\mathcal{J}^2\cos^2\left(\check{\alpha}(\tau_1+\tau_2)\right)}
\ee
The boundary condition requires 
\be
g_{LL}(\tau,-\frac\beta 4)=g_{LR}(\tau,-\frac\beta 4),~\lim_{\tau_2=-\frac\beta 4}\partial_{\tau_2}g_{LL}(\tau_1,\tau_2)=-\lim_{\tau_2=-\frac\beta 4}\partial_{\tau_2}g_{LR}(\tau_1,\tau_2) 
\ee
and similar at $\tau=\frac\beta 4$. 
This solution can be realized as the general solution (\ref{generalsol2T}) of Liouville equation, with the choice
\bea
h_1(\tau)&=&\tan\left(\check{\alpha}\tau+\frac{\check{\gamma}}2\right),~h_2(\tau)=\tan\left(\check{\alpha}\tau-\frac{\check{\gamma}}2\right),\nonumber\\
f_1(\tau)&=&\tan\left(\check{\alpha}\tau+\frac{\check{\gamma}}2\right),~f_2(\tau)=\cot\left(\check{\alpha}\tau-\frac{\check{\gamma}}2\right)
\label{thermalhf}
\eea

On the other hand, the ground state solution (\ref{gLLgLR}) corresponds to the choice of functions
\bea
h_1(\tau)&=&\tanh\left(\alpha \tau+\frac{\gamma}2\right),~h_2(\tau)=\tanh\left(\alpha\tau-\frac{\gamma}2\right)\nonumber\\
f_1(\tau)&=&\tanh\left(\alpha \tau+\frac{\gamma}2\right),~f_2(\tau)=\coth\left(\alpha\tau-\frac{\gamma}2\right)\label{groundstatehf}
\eea
Now the question is whether we can find a new solution by simply joining the functions in (\ref{thermalhf}) and those in (\ref{groundstatehf}) at $\tau=1$. To get a solution of Liouville equation, we need $h_{1,2},~f_{1,2}$ and $h_{1,2}',~f_{1,2}'$ to be continuous. 

However, one should remember that there is an SL(2,R) gauge symmetry for the choice of functions. The solution is invariant under transformations
\be
h_1(\tau)\rightarrow \frac{ah_1(\tau)+b}{ch_1(\tau)+d},~ h_2(\tau)\rightarrow \frac{ah_2(\tau)+b}{ch_2(\tau)+d},~ad-bc=1
\ee
There are two independent SL(2,R) gauge symmetry, one for $h_{1,2}$ and one for $f_{1,2}$. Therefore when we try to match the two solutions, it is sufficient to find a match up to SL(2,R) transformation. As a gauge fixing, we could fix the functions (\ref{groundstatehf}) and carry SL(2,R) transformation only to the TFD part (\ref{thermalhf}). The SL(2,R) transformed functions can be written as
\bea
h_1(\tau)&=&A\tan\left( \check{\alpha}\tau+\frac{\check{\gamma}}2+B\right)+C,~h_2(\tau)=A\tan\left( \check{\alpha}\tau-\frac{\check{\gamma}}2+B\right)+C,\nonumber\\
f_1(\tau)&=&D\tan\left( \check{\alpha}\tau+\frac{\check{\gamma}}2+E\right)+F,~f_2(\tau)=D\cot\left( \check{\alpha}\tau-\frac{\check{\gamma}}2-E\right)+F
\label{thermalhf2}
\eea
with arbitrary parameters $A,B,C,D,E,F$ labeling $SL(2,R)\times SL(2,R)$. The matching condition for $h_1,h_2$ at $\tau=0$ gives
\bea
A\tan\left(\frac{\check{\gamma}}2+B\right)+C&=&\tanh\left(\frac{\gamma}2\right),~A\tan\left(\frac{-\check{\gamma}}2+B\right)+C=\tanh\left(-\frac{\gamma}2\right),\nonumber\\
A\check{\alpha}\sec^2\left(\frac{\check{\gamma}}2+B\right)&=&\alpha{\rm sech}^2\left(\frac{\gamma}2\right),~A\check{\alpha}\sec^2\left(-\frac{\check{\gamma}}2+B\right)=\alpha{\rm sech}^2\left(-\frac{\gamma}2\right)\label{EqofABC}
\eea
These equations require 
\be 
B=0,~C=0,~	A\tan\frac{\check{\gamma}}2=\tanh\frac{\gamma}2,~A\check{\alpha}\left(1+\tan^2\frac{\check{\gamma}}2\right)=\alpha\left(1-\tanh^2\frac{\gamma}2\right)\label{SolofABC}
\ee
Similarly, the equations for $f_1,f_2$ requires
\bea
D\tan\left(\frac{\check{\gamma}}2+E\right)+F&=&\tanh\frac{\gamma}2,~D\cot\left(-\frac{\check{\gamma}}2-E\right)+F=-\coth\frac\gamma2,\nonumber\\
D\check{\alpha}\sec^2\left(\frac{\check{\gamma}}2+E\right)&=&\alpha{\rm sech}^2\frac{\gamma}2,~-D\check{\alpha}\csc^2\left(\frac{\check{\gamma}}{2}+E\right)=-\alpha{\rm csch}^2\frac{\gamma}2\label{EqofDEF}
\eea
From the second line we get 
\bea
\tan^2\left(\frac{\check{\gamma}}2+E\right)&=&\tanh^2\frac\gamma 2
\eea
Using this equation in Eq. (\ref{EqofDEF}) we obtain 
\bea
D=1,~F=0,~\check{\alpha}=\frac{\alpha}{\cosh\gamma}\label{SolofDEF}
\eea

In addition, from Eq. (\ref{SolofABC}) and Eq. (\ref{EqofDEF}) we see that
\bea
\frac{\tan\frac{\check{\gamma}}2}{\sec^2\frac{\check{\gamma}}2}=
\frac{\tan\left(\frac{\check{\gamma}}2+E\right)}{\sec^2\left(\frac{\check{\gamma}}2+E\right)}=\frac{\check{\alpha}\tanh\frac{\gamma}2}{\alpha{\rm sech}^2\frac\gamma 2}\Rightarrow E=0,~A=1
\eea

In summary, we obtain the following matching condition:
\begin{align} 
A&=1,~B=0,~C=0,~D=1,~E=0,~F=0\nonumber\\
\tan\frac{\check{\gamma}}2&=\tanh\frac\gamma 2,~\check{\alpha}=\frac{\alpha}{\cosh\gamma}\label{matchingcond}
\end{align}
Note that Eq. (\ref{matchingcond}) is consistent with the thermal boundary condition (\ref{thermalBC}) since
\bea
\alpha=\mathcal{J}\sinh\gamma, ~\check{\alpha}=\frac{\alpha}{\cosh\gamma}=\mathcal{J}\tanh\gamma=\mathcal{J}\frac{2\tanh\frac\gamma 2}{1+\tanh^2\frac\gamma 2}=\mathcal{J}\sin\check{\gamma}
\eea
This consistency is because $e^g\rightarrow 1$ in the $\tau_1-\tau_2\rightarrow 0$ limit in both regions. The temperature that matches the ground state solution is determined by the boundary condition (\ref{thermalBC}). $\beta$ can be expressed as a function of $\alpha$:
\bea
\beta=\frac{2}{\check{\alpha}}\left(\frac\pi 2-\check{\gamma}\right)=\frac 2{\alpha}\sqrt{\frac{\alpha^2}{\mathcal{J}^2}+1} \, \arctan\frac{\mathcal{J}}{\alpha}
\eea

\mciteSetMidEndSepPunct{}{\ifmciteBstWouldAddEndPunct.\else\fi}{\relax}
\bibliographystyle{utphys}
\bibliography{GlobalAdSDraft}{}

\end{document}